\documentclass[%
twocolumn,
superscriptaddress,
longbibliography,
nofootinbib,
 amsmath,amssymb,
 aps,
prb,
]{revtex4-2}

\usepackage{lipsum}
\usepackage{caption}
\usepackage{dsfont}
\usepackage{float}
\usepackage{wasysym}
\usepackage{cancel}
\usepackage{braket}
\usepackage{physics}
\usepackage{comment}
\usepackage{xcolor}
\usepackage{appendix}
\usepackage{caption}
\usepackage{subcaption}
\usepackage{graphicx}
\usepackage{dcolumn}
\usepackage[mathscr]{euscript}
\usepackage[scr=boondox]{mathalpha}
\usepackage{bm}
\usepackage[linktocpage]{hyperref}
\hypersetup{colorlinks=true,citecolor=blue,linkcolor=blue, urlcolor=blue, breaklinks=true}
\usepackage{soul}

\makeatletter
\newcommand\xlabel[2][]{\phantomsection\def\@currentlabelname{#1}\label{#2}}
\makeatother

\newcommand{\Ox}{\text{o},x}
\newcommand{\Oy}{\text{o},y}
\newcommand{\SO}{\mathscr{S}_{\text{o}}}
\newcommand{\OO}{\text{o}}
\newcommand{\PO}{\vec{\mathscr{P}}_{\text{o}}}
\renewcommand{\P}{\mathscr{P}}

\newcommand{\Z}{\mathbb{Z}}

\renewcommand{\H}{\mathcal{H}}

\begin{document}

\preprint{APS/123-QED}

\title{
Quantized charge polarization as a many-body invariant \\ in (2+1)D crystalline topological states and Hofstadter butterflies
}

\author{Yuxuan Zhang}
\affiliation{Department of Physics and Joint Quantum Institute, University of Maryland,
College Park, Maryland 20742, USA}
\affiliation{Condensed Matter Theory Center, University of Maryland,
College Park, Maryland 20742, USA}
\author{Naren Manjunath}
\affiliation{Department of Physics and Joint Quantum Institute, University of Maryland,
College Park, Maryland 20742, USA}
\affiliation{Condensed Matter Theory Center, University of Maryland,
College Park, Maryland 20742, USA}
\author{Gautam Nambiar}
\affiliation{Department of Physics and Joint Quantum Institute, University of Maryland,
College Park, Maryland 20742, USA}
\author{Maissam Barkeshli}
\affiliation{Department of Physics and Joint Quantum Institute, University of Maryland,
College Park, Maryland 20742, USA}
\affiliation{Condensed Matter Theory Center, University of Maryland,
College Park, Maryland 20742, USA}

\begin{abstract}
We show how to define a quantized many-body charge polarization $\vec{\mathscr{P}}$ for
(2+1)D topological phases of matter, even in the presence of non-zero Chern number and magnetic field. For invertible topological states, $\vec{\mathscr{P}}$ is a $\Z_2 \times \Z_2$, $\Z_3$, $\Z_2$, or $\Z_1$ topological invariant in the presence of $M = 2$, $3$, $4$, or $6$-fold rotational symmetry, lattice (magnetic) translational symmetry, and charge conservation. $\vec{\mathscr{P}}$ manifests in the bulk of the system as (i) a fractional quantized contribution of $\vec{\mathscr{P}} \cdot \vec{b} \text{ mod 1}$ to the charge bound to lattice disclinations and dislocations with Burgers vector $\vec{b}$, (ii) a linear momentum for magnetic flux, and (iii) an oscillatory system size dependent contribution to the effective 1d polarization on a cylinder.  
We study $\vec{\mathscr{P}}$ in lattice models of spinless free fermions in a magnetic field. We derive predictions from topological field theory, which we match to numerical calculations for the effects (i)-(iii), demonstrating that these can be used to 
extract $\vec{\mathscr{P}}$ from microscopic models in an intrinsically many-body way. We show how, given a high symmetry point $\text{o}$, there is a topological invariant, the discrete shift $\mathscr{S}_{\text{o}}$, such that $\vec{\mathscr{P}}$ specifies the dependence of $\mathscr{S}_{\text{o}}$ on $\text{o}$. We derive colored Hofstadter butterflies, corresponding to the quantized value of $\vec{\mathscr{P}}$, which further refine the colored butterflies from the Chern number and discrete shift. 
\end{abstract}

\maketitle

\tableofcontents

\begin{figure*}[t]
    \centering
    \includegraphics[width=17.6cm]{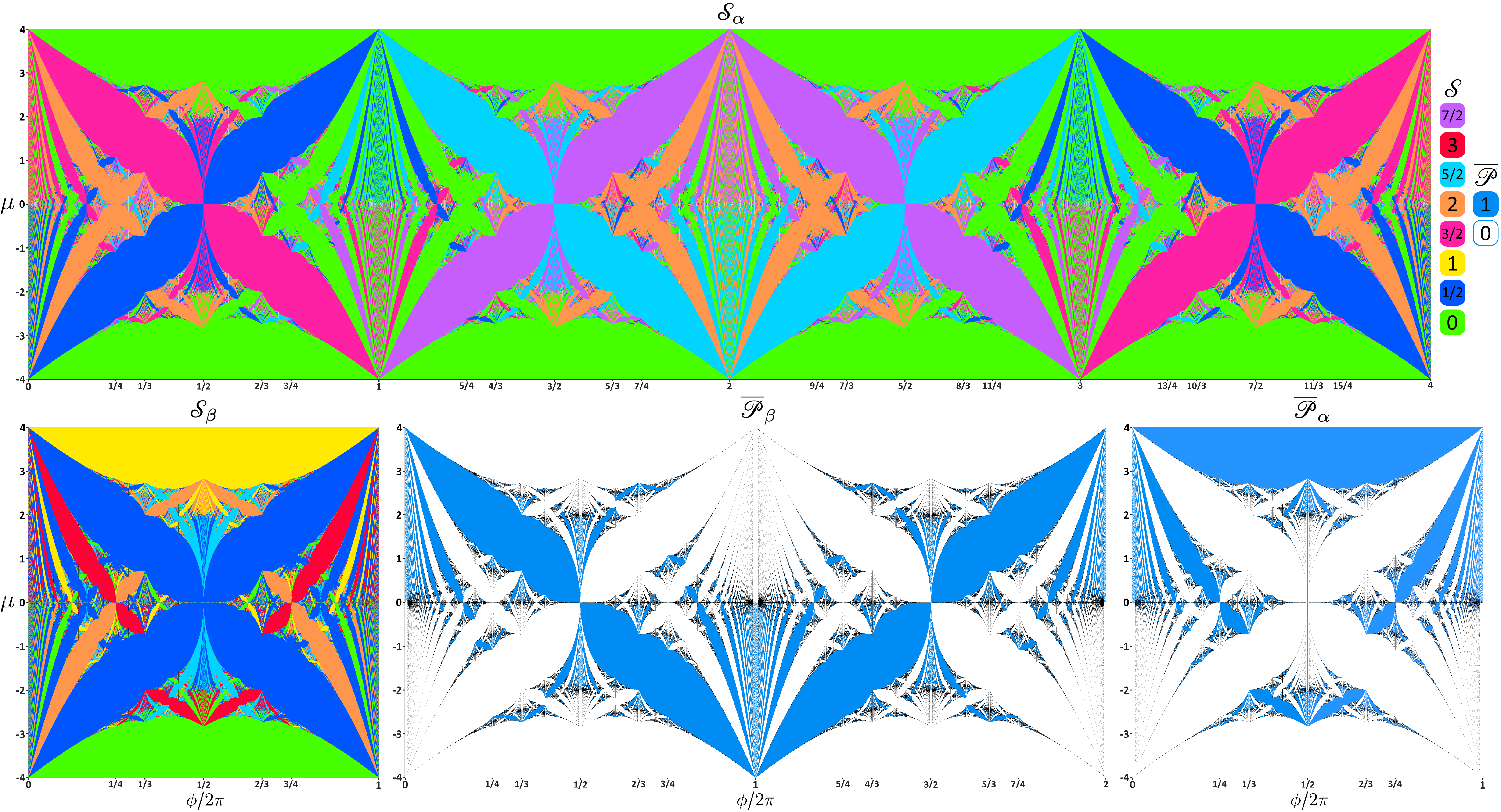}
    \caption{A menagerie of butterflies in the spinless square lattice Hofstadter model.
    $\alpha$ and $\beta$ represent a plaquette center and vertex respectively. For any $C_4$ symmetric origin $\OO$, $\SO$ has a $\Z_4$ classification, while $\overline{\mathscr{P}}_{\text{o}}$ has a $\Z_2$ classification. We empirically find that $\{\mathscr{S}_{\beta},\mathscr{S}_{\alpha},\overline{\mathscr{P}}_{\beta},\overline{\mathscr{P}}_{\alpha}\}$ follow Eqs.\{\eqref{eq:s_formula_b}, \eqref{eq:s_formula_a}, \eqref{eq:t_formula}, \eqref{eq:t_formula2}\} respectively. 
    Note that $\mathscr{S}_{\alpha}$ has period $8\pi$ in $\phi$, $\overline{\mathscr{P}}_{\beta}$ has period $4\pi$ in $\phi$, while $\mathscr{S}_{\beta}$ and $\overline{\mathscr{P}}_{\beta}$ has period $2\pi$ in $\phi$.
    See Sec.~\ref{sec:c4_SO} and \ref{sec:c4_PO} for discussions.}
    \label{fig:SandP}
\end{figure*}

\section{Introduction}

In the presence of symmetry, gapped quantum phases of matter can acquire \it symmetry-protected \rm topological invariants. The paradigmatic example is the quantized Hall conductance, which is specified by the Chern number, and is defined only for systems with a $U(1)$ charge conservation symmetry. Since the discovery of topological insulators and superconductors \cite{hasan2010,qiRMP2011,bernevig2013topological}, there has been spectacular progress in our understanding of symmetry-protected topological invariants for both single-particle free fermion models \cite{Kitaev2009periodic,ryu2010,Chiu2016review} and for interacting many-body systems \cite{Chen2013,kapustin2014SPTbeyond,Gu2014Supercoh,kapustin2015fSPT,senthil2015,Wang2020fSPT,Barkeshli2019,barkeshli2021invertible,bulmashSymmFrac,aasen2021characterization}. Despite these advances, a complete understanding of topological invariants arising from crystalline symmetries is still lacking. 

Recently Refs. \cite{manjunath2021cgt,Manjunath2020fqh} applied ideas from topological quantum field theory (TQFT) and the algebraic theory of symmetry defects \cite{Barkeshli2019}, which can be used to characterize gapped quantum many-body systems, to develop a systematic classification of topological invariants for systems with $U(1)$ charge conservation, discrete (magnetic) translational symmetry and rotational symmetry in two spatial dimensions. In particular, Refs. \cite{manjunath2021cgt,Manjunath2020fqh} showed how TQFT predicts the existence of a quantized many-body polarization in the presence of $2$, $3$, $4$ or $6$-fold rotational symmetry. For invertible topological phases, which do not host anyon excitations, the polarization acquires a $\Z_2 \times \Z_2$, $\Z_3$, $\Z_2$, or $\Z_1$ classification, respectively.

Remarkably, the TQFT prediction of a quantized charge polarization applies also in the presence of a non-zero Chern number and a non-trivial magnetic field. This appears to be in tension with several statements made previously in the literature about whether the polarization is well-defined in the presence of a non-zero Chern number \cite{Fang2012PGS,Song2021polarization}. 

The TQFT not only predicts the presence of the invariant, but also its bulk physical manifestation. This is in terms of a fractional quantized contribution of the charge bound to lattice defects and a dual response, the momentum of the ground state in the presence of additionally inserted magnetic flux.

In contrast, the modern theory of polarization in insulators is based on the Berry-Zak phase of single-particle wave functions in momentum space \cite{resta2007theory,resta1994,watanabe2018,coh2009}. This Berry phase theory of polarization assumes the phase of the single-particle states is globally well-defined throughout the Brillouin zone, which applies only in the case of zero Chern number. For the case of non-zero Chern number, while there has been work showing how one may define a notion of polarization in the single-particle context by fixing an origin in the Brillouin zone \cite{coh2009}, its quantization from crystalline symmetries, the effects of non-zero magnetic field, and its implication for bulk response properties have not been studied. 

In many-body systems with interactions, the single-particle Berry phase formulation breaks down. It can be replaced with a Berry phase theory based on twisted boundary conditions or with an expectation of Resta's exponentiated polarization operator \cite{resta1998eq}. However these apply only to the effective 1d polarization, meaning the system is viewed as an effective one-dimensional system; such a 1d polarization is no longer an intensive quantity in a higher dimensional system.  

In this paper we show how one can indeed define a quantized charge polarization in an intrinsically many-body fashion and in the presence of both non-zero Chern number and non-zero magnetic field. This is not an effective 1d polarization obtained by viewing the system as a 1d system -- rather, this is an intrinsic bulk 2d polarization, which has non-trivial bulk responses mentioned above. 

More specifically, we show that upon fixing a choice of high symmetry point $\text{o}$ in the unit cell, one can define two invariants, $\mathscr{S}_{\text{o}}$ and $\vec{\mathscr{P}}_{\text{o}}$.
$\SO$ is a discrete analog of the Wen-Zee shift \cite{Wen1992shift,haldane2011fqh,Gromov2014,Bradlyn2015,Gromov2015,Biswas2016}, which is an invariant associated to $U(1)$ charge conservation and $SO(2)$ plane rotational symmetry. We refer to $\SO$ as the `discrete shift' because it is a $\Z_M$ invariant, while the Wen-Zee shift is a $\Z$ invariant. $\PO$ denotes the quantized charge polarization. 

We show, through extensive numerical studies, how these invariants can be extracted from bulk response properties of microscopic models in multiple different ways. We show how the predictions of the TQFT, including the bulk response properties and the dependence of $\mathscr{S}_{\text{o}}$ and $\vec{\mathscr{P}}_{\text{o}}$ on $\text{o}$, can be precisely matched to calculations on microscopic models. 

As an application, we show how one can extract the quantized charge polarization for the Hofstadter model \cite{hofstadter1976} of spinless free fermions in a non-zero magnetic field on a lattice. This provides yet another way to color Hofstadter's butterfly (see Fig. \ref{fig:SandP}), extending the recent coloring in Ref. \cite{zhang2022fractional} based on the discrete shift, and the earlier coloring with the Chern (TKNN) number \cite{osadchy2001db,thouless1982}. 

We note that the dependence of the polarization on a choice of origin $\text{o}$ is a well-known property of all definitions of the polarization in electronic systems; it is usually dealt with by considering instead changes in the polarization as an external parameter is tuned, or by using the overall charge-neutrality of the system (for example by taking into account the background positive ions), which removes the origin-dependence \cite{resta2007theory}.
While at first glance it seems unusual that an invariant of a phase of matter could have a dependence on a choice of origin, we will explain it further in subsequent sections.

\subsection{Relation to prior work}

Our work is closely related to several works over the past decade that also study polarization and its physical consequences but all in the context of Chern number $C = 0$. Responses associated to the discrete shift have also been explored in microscopic models in recent works \cite{Liu2019ShiftIns,Li2020disc,zhang2022fractional} and in the context of topological field theory \cite{manjunath2021cgt,Manjunath2020fqh,Han2019}, although the origin-dependence of the discrete shift has not been discussed in prior work. 

Refs.~\cite{Fang2012PGS,Jadaun2013PGS} discuss a quantized charge polarization in free fermion crystalline insulators with different point group symmetries, assuming zero Chern number.

Ref.~\cite{Song2021polarization} showed how, ignoring rotational symmetry, polarization is a `non-quantized' topological response and can be defined for zero Chern number systems in an intrinsically many-body fashion in terms of the momentum of the ground state in the presence of magnetic flux. Ref.~\cite{Song2019_2pi} earlier studied the momentum of magnetic flux and mentioned its quantization by rotational symmetries. We note that the definition of the magnetic translation operator in a magnetic field, which is used to compute the momentum, has a number of ambiguities that were not fully considered in these previous works. 

Refs.~\cite{Fang2012PGS,Song2021polarization} both asserted that the polarization is not well-defined in the presence of non-zero Chern number, which disagrees with our results in the case where we have both translational and rotational symmetry. 

Ref.~\cite{Li2020disc} defines the polarization for systems with $C = 0$ and zero magnetic field via Wannier representation theory, and characterizes it in terms of a fractional charge bound to lattice defects with non-trivial Burgers vector. Ref.~\cite{Miert2018dislocationCharge} also finds that lattice dislocations can have fractionally quantized charges in a rotationally symmetric system; here it appears that $C=0$ is being implicitly assumed. 
We emphasize that our definition of fractional charge of the lattice defects differs from the definition presented in Refs.~\cite{Li2020disc,Miert2018dislocationCharge}.

\subsection{Organization of paper}

The rest of the paper is organized as follows. In Sec.~\ref{sec:ovviewmain} we summarize our main results. In Sec.~\ref{sec:useful_bgd} we review some basic properties of lattice defects. In Secs.~\ref{sec:c4_SO} and~\ref{sec:c4_PO} we present detailed results for $\SO$ and $\PO$ respectively on the square lattice, highlighting the various subtleties that arise in matching the field theory to numerics. Sec.~\ref{sec:general} does the same for $M=2,3,6$. In Sec.~\ref{app:ftrelabel} we discuss the origin dependence of $\SO,\PO$ from a field theory perspective. We then conclude and discuss future directions.

\section{Overview of main results}\label{sec:ovviewmain}

\bgroup
\def\arraystretch{1.8}
\begin{table}
    \centering
\begin{tabular}{ c||l|l  }

\hline

$M$ &
$\mathscr{S}_{\text{o}+\vec{v}}$ &  $\vec{\mathscr{P}}_{\text{o}+\vec{v}}$ \\
\hline
2& $
   \mathscr{S}_{\text{o}} - 4 \vec{v}\cdot\vec{\mathscr{P}}_{\text{o}} + 4\kappa(v_x^2 + v_y^2 + v_x v_y)
$&$\vec{\mathscr{P}}_{\text{o}}+
(-v_y \kappa,  v_x \kappa)$
\\
\hline
4 & $
   \mathscr{S}_{\text{o}} +  2\overline{\mathscr{P}}_{\text{o}} - \kappa
$ & $\vec{\mathscr{P}}_{\text{o}}+(-\frac{1}{2} \kappa,  \frac{1}{2} \kappa)$\\

\hline
3 & $\mathscr{S}_{\text{o}}- 3 v_y \overline{\mathscr{P}}_{\text{o}} -3\kappa (v_x^2 + v_y^2 + v_x v_y)$ & $\vec{\mathscr{P}}_{\text{o}}+(-v_y \kappa,  v_x \kappa)$\\
\hline
6 & $\mathscr{S}_{\text{o}}$ & 0\\
\hline

\end{tabular}
\caption{Transformation of $\mathscr{S}_\text{o}$ and $\vec{\mathscr{P}}_{\text{o}}$ under $\text{o}\rightarrow \text{o}+(v_x,v_y)$, which shifts the origin from one $C_M$ symmetric point to another $C_M$ symmetric point. Note that $\mathscr{S}_{\text{o}}$ and $\vec{\mathscr{P}}_{\text{o}}$ are only defined up to equivalences as described in the main text. For $M=4$ we have taken the unique non-trivial choice $\vec{v}=(\frac{1}{2},\frac{1}{2}).$
}\label{table:shiftingO}

\end{table}
\egroup

We consider a gapped phase of matter with the symmetry group 
\begin{align}
G = U(1) \times_{\phi} [\mathbb{Z}^2 \rtimes \Z_M], 
\end{align}
where $\Z^2$ denotes magnetic lattice translations and $\Z_M$ for $M = 2,3,4,6$ denotes point group rotations.\footnote{More specifically $G$ is the fermionic symmetry group, which acts non-trivially on fermionic operators. It is sometimes written as $U(1)^f \times_{\phi} [\mathbb{Z}^2 \rtimes \Z_M]$. The symbol $U(1)^f$ means that the order 2 element of this group is identified with the fermion parity operation. 
}  The symbol $\times_{\phi}$ implies that the magnetic translation operators, generated by $\tilde{T}_{\bf x}, \tilde{T}_{\bf y}$, obey the algebra $\tilde{T}_{\bf y}^{-1} \tilde{T}_{\bf x}^{-1} \tilde{T}_{\bf y} \tilde{T}_{\bf x} = e^{i \phi \hat{N}}$ where $\hat{N}$ is the total fermion number. The tilde superscript indicates that the definition of the operator involves a $U(1)$ gauge transformation.

The charge conservation and translation symmetries allow us to define a charge per unit cell $\nu$. Each unit cell can be divided into $M$ subcells with equal flux $\phi_{sub}$. The total flux per unit cell is then $\phi = M \phi_{sub}$. Note that for our purposes, depending on the microscopic model we may need to specify the flux within even smaller subregions of the unit cell. Therefore we assume that the 2d system is embedded in a continuum, and that the magnetic field $B$ is specified at each continuum point. This allows us to specify $\phi, \phi_{sub}$ exactly as real numbers, even though the symmetry only requires us to define $\phi \mod 2\pi$. We comment further on this in Sec.~\ref{sec:c4_SO}.

Let $C$ be the Chern number of the system. We then define the integer
\begin{align}\label{eq:k0def}
    \kappa \equiv \nu - C \frac{\phi}{2\pi}. 
\end{align}
$\kappa$ is a $\Z$ topological invariant for the system if $\phi$ is known exactly (and not just modulo $2\pi$).
Fixing $C$ and $\phi$, if the charge per unit cell increases by an integer $l$, then $\kappa \rightarrow \kappa + l$. For further intuition about $\kappa$ and a heuristic derivation of Eq.~\eqref{eq:k0def}, see App.~\ref{app:defkappa}.

For a given high symmetry point $\text{o}$ of the lattice unit cell, we will see that one can define a set of topological invariants 
$\{\mathscr{S}_{\text{o}}, \vec{\mathscr{P}}_{\text{o}}\}$. The transformation of $\{\mathscr{S}_{\text{o}}, \vec{\mathscr{P}}_{\text{o}}\}$ under a change of $\text{o}$ is fully determined if $\text{o}$ is preserved by a $\Z_M$ rotation symmetry group. Therefore it is sufficient to specify $\{\mathscr{S}_{\text{o}}, \vec{\mathscr{P}}_{\text{o}}\}$ for a single such $\text{o}$. 

A subtle point is that the definition of the invariants requires a choice of operators that represent the symmetry group elements. In this work, we take $\tilde{C}_{M, \text{o}}$ to be a `magnetic' rotation operator about $\text{o}$ (a spatial rotation combined with a gauge transformation), such that $(\tilde{C}_{M, \text{o}})^M = 1$. If we change this choice by $\tilde{C}_{M, \text{o}} \rightarrow \tilde{C}_{M, \text{o}} e^{i \frac{2 \pi}{M} \chi \hat{N}}$ for some real number $\chi$, then the invariant $\mathscr{S}_{\text{o}}$ also transforms, $\mathscr{S}_{\text{o}} \rightarrow \mathscr{S}_{\text{o}} + \chi C$. Nevertheless, we will see that there are canonical choices for $\tilde{C}_{M, \text{o}}$ that can be made.

We caution that the vector $\PO$ we refer to throughout the text is different from the standard charge polarization vector $\vec{P}_{\OO}$, which for zero Chern number satisfies ${\bf j} = \partial_t \vec{P}_{\OO}$ where ${\bf j}$ is the induced current. We show in App.~\ref{app:C=0calcs} that 
\begin{equation}\label{eq:P_vs_scrP}
    (\mathscr{P}_{\Ox}, \mathscr{P}_{\Oy}) = (P_{\Oy}, -P_{\Ox}) = \vec{P}_{\OO} \times \hat{z}.
\end{equation}
However, since $\PO$ appears most naturally in our theory, we will work in terms of $\PO$ throughout and refer to it as the polarization, recognizing this as a slight abuse of terminology.

\begin{figure}[t]
    \centering
    \includegraphics[width=8.5cm]{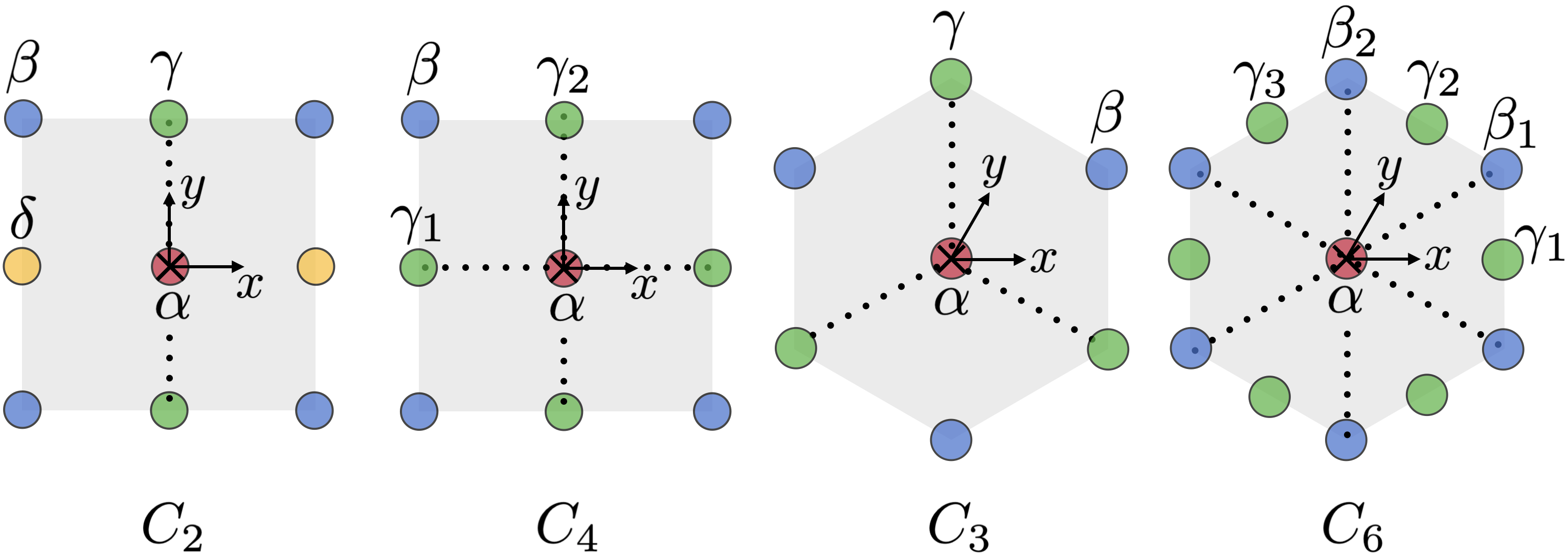}
    \caption{Maximal Wyckoff positions for $C_2$, $C_3$, $C_4$ and $C_6$ symmetries (colored circles). $\times$ marks
    the unit cell center, which we denote as $\alpha$. The high symmetry points $\beta_i$ and $\gamma_i$ each belong to a single maximal Wyckoff position ($\beta$ or $\gamma$), but are all inequivalent under lattice translations. The dotted lines show a possible division of the unit cell into $M$ subcells.}
    \label{fig:wyckoff}
\end{figure}

\subsection{Warmup: $C_4$ symmetric lattice}

We first illustrate our main results for the square lattice. A representative unit cell with high symmetry points (HSPs) $\alpha$ (unit cell center), $\beta$ (unit cell vertices), $\gamma_i$ (edge centers) is shown in Fig.~\ref{fig:wyckoff}. The points $\gamma_1,\gamma_2$ are not translation-equivalent but are related by rotations about $\alpha$. $\alpha,\beta,\gamma$ refer to maximal Wyckoff position (MWPs), which are collections of points related by lattice symmetries; the precise definition of a MWP is given in App.~\ref{sec:bgd}. $\alpha$ and $\beta$ have an order 4 site symmetry group generated by the ``magnetic" rotation operators $\tilde{C}_{4,\alpha},\tilde{C}_{4,\beta}$ (which also include a gauge transformation), with 
$\tilde{C}_{4,\alpha}^4 = \tilde{C}_{4,\beta}^4 = 1$.
The point $\gamma$ has an order 2 site symmetry group generated by the operator $\tilde{C}_{2,\gamma}$, with $\tilde{C}_{2,\gamma}^2 = 1$. We can pick any of these points as our origin $\OO$.

First we define the invariants $\SO,\PO$ for each possible HSP $\OO$ and list their properties; thereafter, we explain how to use them to characterize the topological phase of the given system. 

Suppose $\OO = \alpha$ or $\beta$. Then, $\SO$ is defined mod 4 and can take one of 4 possible values for a fixed Chern number $C$. Next, suppose $\OO = \gamma$, which is a $C_2$ symmetric point. Then $\mathscr{S}_{\gamma}$ is only defined mod 2. In all cases, we have the constraint \cite{zhang2022fractional}
\begin{align}\label{eq:S-C-mod1}
    \SO \mod 1 = 
      \frac{C}{2} \mod 1 . 
\end{align}

We next turn to $\PO$. For $\OO = \alpha,\beta$,
\begin{align}
\PO \in \{ (0,0), (\frac{1}{2},\frac{1}{2}) \}
\end{align}
up to integer vectors. We write 
\begin{align}
\PO = \frac{\overline{\mathscr{P}}_{\OO}}{2}(1,1) 
\end{align}
in this case; $\overline{\mathscr{P}}_{\OO}$ is an integer defined mod 2. For $\OO = \gamma$, there are 4 possible choices:
\begin{align}
\vec{\mathscr{P}}_{\gamma} \in \{(0,0),(\frac{1}{2},0),(0,\frac{1}{2}),(\frac{1}{2},\frac{1}{2})\}
\end{align}
up to integer vectors. 

$\SO,\PO$ for a $C_4$ symmetric point, together with $\kappa$, determines $\mathscr{S}_{\OO'},\vec{\mathscr{P}}_{\OO'}$ for all other $\OO'$. For example, 
\begin{align}\label{eq:SaSbPa}
    \{\mathscr{S}_{\beta}, \overline{\mathscr{P}}_{\beta},\kappa\} &=   \{\mathscr{S}_{\alpha} + 2 \overline{\mathscr{P}}_{\alpha}-\kappa, \overline{\mathscr{P}}_{\alpha}+ \kappa,\kappa\}. 
\end{align}
If we only know $\mathscr{S}_\gamma, \vec{\mathscr{P}}_\gamma,\kappa$, we can determine $\vec{\mathscr{P}}_\alpha$ and $\vec{\mathscr{P}}_{\beta}$ fully, but can only determine $\mathscr{S}_\alpha$ and $\mathscr{S}_\beta$ mod 2 and not mod 4. The relevant formulas are given in Table~\ref{table:shiftingO}. Thus, to fully specify $\SO,\PO$ for each high symmetry point in the unit cell, we need to determine them for some $\OO$ with the largest possible site symmetry group.

In Fig.~\ref{fig:SandP}, we show colored Hofstadter butterflies for two different origins $\alpha,\beta$ extracted for the square lattice Hofstadter model of spinless fermions. We find that $\{\mathscr{S}_{\beta},\mathscr{S}_{\alpha},\overline{\mathscr{P}}_{\beta},\overline{\mathscr{P}}_{\alpha}\}$ follow the empirical equations Eqs.\{\eqref{eq:s_formula_b}, \eqref{eq:s_formula_a}, \eqref{eq:t_formula}, \eqref{eq:t_formula2}\} respectively. In this figure $\beta$ corresponds to a site and $\alpha$ to a plaquette center, where there is no site.

To distinguish two phases of matter based on $\SO,\PO$ (assuming $C, \kappa$ and all other invariants are equal), we first fix a common origin $\OO$ (which must be a $C_4$ symmetric point) and find $\SO,\PO$ for the two systems. If their values are not equal, the two systems cannot be adiabatically connected to each other in a symmetry-preserving manner. It is important to note that comparing the crystalline topological invariants of two phases is only meaningful after fixing a common origin.

\subsection{Basic properties and classification for $\{\mathscr{S}_{\text{o}}, \vec{\mathscr{P}}_{\text{o}}\}$}

\begin{table}[]
\def\arraystretch{1.1}
    \centering
    \begin{tabular}{c||c|c|c|c}
    \hline
    $M$ & 2& 3& 4&6 \\ \hline
     $U(2\pi/M)$ & $\begin{pmatrix}-1 & 0 \\ 0 & -1\end{pmatrix}$ & $\begin{pmatrix}-1 & -1 \\ 1 & 0\end{pmatrix}$ &  $\begin{pmatrix}0 & -1 \\ 1 & 0\end{pmatrix}$ & $\begin{pmatrix}0 & -1 \\ 1 & 1\end{pmatrix}$  \\ \hline
    \end{tabular}
    \caption{Elementary rotation matrices $U(2\pi/M)$, for the coordinate basis shown in Fig.~\ref{fig:wyckoff}.}
    \label{tab:U_mats}
\end{table}
 
We now generalize the above discussion to the case with $M = 2,3,4,6$.\footnote{For the case $M = 1$, there are no quantized discrete shift and polarization invariants \cite{manjunath2021cgt}} In Table \ref{tab:U_mats} we define the $2\times 2$ matrices $U(2\pi/M)$, corresponding to elementary $2\pi/M$ rotations around an origin $\OO$ with $M$-fold rotational symmetry. These describe the action of the rotation operator $\tilde{C}_{M,\OO}$ on space. As above we fix $(\tilde{C}_{M,\OO})^M = +1$. 

Then for any given $\text{o}$ which is fixed under an order-$M$ rotation, $\{\mathscr{S}_{\text{o}}, \vec{\mathscr{P}}_{\text{o}}\}$ 
have a $\Z_M \times K_M$ classification, where $K_M = \{\Z_2 \times \Z_2, \Z_3, \Z_2, \Z_1\}$ for $M = 2,3,4,6$. A derivation is given in App.~\ref{sec:bgd}. More specifically, $\mathscr{S}_{\text{o}}$ is an integer or half-integer defined modulo $M$, and it satisfies Eq.~\eqref{eq:S-C-mod1}.  $\vec{\mathscr{P}}_{\text{o}}$ is a 2-component vector with the following quantization condition and equivalence relation: 
\begin{align}\label{eq:PO_props}
    & (1 - U(2\pi/M)) \vec{\mathscr{P}}_{\text{o}} \in \Z^2, 
    \nonumber \\
    &\vec{\mathscr{P}}_{\text{o}} \sim \vec{\mathscr{P}}_{\text{o}} + \vec{\Lambda}, \;\; \vec{\Lambda} \in \Z^2.
\end{align} 

For 4-fold and 6-fold point groups, it is possible for the HSP $\text{o}$ to only be invariant under a smaller $M'$-fold rotation. For example we can have $M' = 2$ when $M = 4$, or $M' = 2,3$ when $M = 6$. In these cases, the possible values of $\{ \mathscr{S}_{\text{o}}, \vec{\mathscr{P}}_{\text{o}}\}$ have a $\Z_{M'} \times K_{M'}$ classification. The relations defining them will be as above, with $M$ replaced by $M'$.

It will be convenient to parameterize $\vec{\mathscr{P}}_{\text{o}}$ in the following way:
\begin{equation}\label{eq:parametrization}
 \vec{\mathscr{P}}_{\text{o}} =\begin{cases}
     \frac{1}{2} (\overline{\mathscr{P}}_{\text{o},x},\overline{\mathscr{P}}_{\text{o},y}) &\quad M'=2\\
 \overline{\mathscr{P}}_{\text{o}} \frac{1}{2} (1,1) &\quad M'=4\\
 \overline{\mathscr{P}}_{\text{o}} \frac{1}{3} (1,2) &\quad M'=3\\
 0 &\quad M'=6.\\
\end{cases}
\end{equation}
Here $M'$ is the maximal integer such that $\text{o}$ is a fixed point under rotations of order $M'$. $\overline{\mathscr{P}}_{\text{o}}$ can take any integer value according to the $K_{M'}$ classification. For example, when $M'=2$, there are 4 inequivalent choices for $\PO$: $\PO \in \{(0,0),(1/2,0),(0,1/2),(1/2,1/2)\}$. When $M'=3$, there are 3 inequivalent choices: $\PO \in \{(0,0),(1/3,2/3),(2/3,1/3)\}$. This is derived in App.~\ref{app:Pcquant}. 

Next we discuss the origin dependence of $\SO, \PO$. If we shift $\text{o} \rightarrow \text{o}' = \text{o} + (v_x, v_y)$, then we can determine $\{\mathscr{S}_{\text{o}'}, \vec{\mathscr{P}}_{\text{o}'}\}$ from $\{\mathscr{S}_{\text{o}}, \vec{\mathscr{P}}_{\text{o}}\}$ and $\kappa$, as specified in Table \ref{table:shiftingO}. Note that $(v_x,v_y)$ can be fractions of a lattice unit. For $\{\mathscr{S}_{\text{o}'}, \vec{\mathscr{P}}_{\text{o}'}\}$ to be fully specified in terms of $\{\mathscr{S}_{\text{o}}, \vec{\mathscr{P}}_{\text{o}}\}$, the minimal rotation angle which preserves $\text{o}'$ must be a multiple of the minimal rotation angle which preserves $\OO$. In other words, the site symmetry group of $\OO'$ is isomorphic to a subgroup of the site symmetry group of $\OO$. For example, if $\text{o}$ and $\text{o}'$ have site symmetry groups $\Z_4$ and $\Z_2$ respectively, then $\{\mathscr{S}_{\text{o}'}, \vec{\mathscr{P}}_{\text{o}'}\}$ can be determined using the $M=2$ entry of Table~\ref{table:shiftingO}.

Thus, in order to obtain a complete specification for each high symmetry point, we need to know $\{\mathscr{S}_{\text{o}}, \vec{\mathscr{P}}_{\text{o}}\}$ for at least one $\text{o}$ which is invariant under an $M$-fold rotation. Otherwise we will not be able to fully recover $\{\mathscr{S}_{\text{o}'}, \vec{\mathscr{P}}_{\text{o}'}\}$ for each $\text{o}'$. Note, in particular, that the dependence of $\vec{\mathscr{P}}_{\text{o}}$ on $\text{o}$ is completely determined by $\kappa$, and the dependence of  
$\mathscr{S}_{\text{o}}$ on $\text{o}$ is completely determined by $\vec{\mathscr{P}}_{\text{o}}$ and $\kappa$.

Nevertheless, differences
\begin{align}
\Delta \mathscr{S}
&= \mathscr{S}_{\text{o}}(\lambda_1) - \mathscr{S}_{\text{o}}(\lambda_2)
\nonumber \\
\Delta \vec{\mathscr{P}} &= \vec{\mathscr{P}}_{\text{o}}(\lambda_1) - \vec{\mathscr{P}}_{\text{o}}(\lambda_2)
\end{align}
are independent of $\text{o}$. Here $\lambda$ is some tuning parameter in the Hamiltonian, $H[\lambda]$, which keeps the invariant $\kappa$ fixed and preserves the crystalline symmetry. This can be done for example by fixing $\{\nu,\phi_{\text{sub}}\}$. The reason we need to fix $\kappa$ is discussed in Sec.~\ref{sec:trans_PO}. 

Note that if we have a solid state system of electrons with some background positive charge due to ions, then the total polarization of the system will be $\vec{\mathscr{P}}_{tot} =\vec{\mathscr{P}}_{\text{o}} + \vec{\mathscr{P}}_{ion, \text{o}}$. If we assume that the ions have a charge of $\kappa$ per unit cell, then the origin dependence cancels and $\vec{\mathscr{P}}_{tot}$ becomes origin-independent. In realistic systems, the excess charge per unit cell $\nu - \kappa$ will be neutralized by a metallic gate, which we would ignore to compute the total polarization. 

As another example, if we take $M = 3$, we have 3 maximal Wyckoff positions invariant under $3$-fold rotation symmetry: $\alpha$, $\beta$, and $\gamma$, with $\beta=\alpha+(1/3,1/3)$ and $\gamma=\alpha+(-1/3,2/3)$ (see Fig.~\ref{fig:wyckoff}). Then
\begin{align}
    \{\mathscr{S}_{\beta}, \overline{\mathscr{P}}_{\beta},\kappa\} &=   \{\mathscr{S}_{\alpha} -\overline{\mathscr{P}}_{\alpha}-\kappa, \overline{\mathscr{P}}_{\alpha}-\kappa,\kappa\} 
    \nonumber \\
    \{\mathscr{S}_{\gamma}  , \overline{\mathscr{P}}_{\gamma} ,\kappa\} &= \{\mathscr{S}_{\alpha}+\overline{\mathscr{P}}_{\alpha} - \kappa, \overline{\mathscr{P}}_{\alpha}+ \kappa ,\kappa\}. 
\end{align}

\subsection{Extracting $\{\mathscr{S}_{\text{o}}, \vec{\mathscr{P}}_{\text{o}}\}$ from microscopic models}

For a given microscopic model, we can extract $\{\mathscr{S}_{\text{o}}, \vec{\mathscr{P}}_{\text{o}}\}$ in several distinct ways, as summarized in Table~\ref{tab:geneqs}. 
To set up the calculations, we first need to fix a rotation operator $\tilde{C}_{M',\text{o}^*}$, where the high symmetry point $\text{o}^*$ is invariant under $2\pi/M'$ rotations and $M' \leq M$.  
In our examples we will always choose this operator so that 
$(\tilde{C}_{M',\text{o}^*})^{M'} = +1$. We also define translation operators $\tilde{T}_{\textbf{x}}$, $\tilde{T}_{\textbf{{y}}}$ corresponding to the elementary lattice vectors ${\bf x},{\bf y}$, which obey the magnetic translation algebra. 

Our numerical procedure is guided by a topological response theory derived using TQFT ideas \cite{manjunath2021cgt,Manjunath2020fqh,barkeshli2021invertible}. This gives a Lagrangian density in terms of background gauge fields:
\begin{align}\label{eq:chargeresponse}
 \mathcal{L} &=  \frac{C}{4\pi} A \wedge dA + \frac{\mathscr{S}_{\text{o}}}{2\pi} A \wedge d\omega + \frac{\vec{\mathscr{P}}_{\text{o}}}{2\pi} \cdot A \wedge \vec{T} + \frac{\kappa}{2\pi}A \wedge A_{XY} \nonumber \\
  &+ \frac{\tilde{\ell}_s}{4\pi} \omega \wedge d\omega + \frac{\vec{\mathscr{P}}_{s}}{2\pi} \cdot \omega \wedge \vec{T} +\frac{\nu_s}{2\pi} \omega \wedge A_{XY} + \dots
\end{align}
Here, $A$ is a background $U(1)$ gauge field, and is defined so that $\int dA$ represents the full magnetic field (and not just its deviation from some background value) $\omega$ is a background $\Z_M$ `rotation' gauge field, which is treated as a real field with quantized holonomies. $A_{XY}$ and $\vec{T}$ are the area element and torsion 2-form, respectively, which are constructed using translation gauge fields. The notation is described more fully in 
App.~\ref{sec:field}. 

Importantly, the coefficients of these terms are all quantized in specific patterns and defined modulo certain equivalence relations, which can be systematically derived for bosonic systems using group cohomology \cite{manjunath2021cgt}. We will only be concerned with the first four terms, which have the coefficients $C,\SO,\PO,\kappa$.
In this paper, we carry out a derivation of the 
quantization conditions on $\PO$ in the case of fermionic systems using a general theory of invertible fermionic phases developed in Ref.~\cite{barkeshli2021invertible} (see App.~\ref{sec:field} of this paper). We show that the quantization conditions on $\PO$ in invertible fermionic systems (i.e. without fractionalized excitations) are the same as for invertible bosonic systems, in contrast to the Chern number and discrete shift \cite{zhang2022fractional}.

\subsubsection{$\{\mathscr{S}_{\text{o}}, \vec{\mathscr{P}}_{\text{o}}\}$ from fractional charge of lattice disclinations and dislocations}

Given the magnetic rotation operator $\tilde{C}_{M',\text{o}^*}$ about a high symmetry point $\text{o}^*$ and translation operators $\tilde{T}_{\bf{x}}$, $\tilde{T}_{\bf{y}}$, and the Hamiltonian for the clean system with the full crystalline symmetry, $H_{\text{clean}}$, one can define a Hamiltonian in the presence of a lattice disclination or dislocation $H_{\text{defect}}$. This is done through a cut-and-glue procedure described in App.~\ref{sec:dislocationConstruction}. $H_{\text{defect}}$ is uniquely defined up to local operators at the core of the defect. In our numerics we will take $H_{\text{clean}}$ to be a free fermion Hofstadter model, usually with nearest neighbor hopping terms, but our methods conceptually apply more generally, as we discuss in Secs.~\ref{sec:c4_SO},~\ref{sec:general}. In App.~\ref{app:nnnhopping}, we demonstrate that the dislocation charge calculation generalizes naturally to Hamiltonians with next nearest neighbour hopping.

A lattice disclination has a non-zero disclination angle $\Omega$ (Frank angle $-\Omega$), and $\vec{b}_{\text{o}}$ is the Burgers vector. Here, the subscript $\OO$ means that the Burgers vector is measured by the holonomy of a loop encircling the defect, which starts at the point $\OO$. Note that $\OO$ and $\OO^*$ need not be equal in general. As we explain in Sec. \ref{sec:useful_bgd}, for a disclination with $\Omega \neq 0$, the value of $\vec{b}_{\text{o}}$ as defined above depends on $\text{o}$. However, for a lattice dislocation, which has $\Omega = 0$, the value of $\vec{b}_{\text{o}}$ is independent of $\text{o}$.

We can compute the fractional charge in the ground state, $Q_W \mod 1$, in a large region $W$ surrounding a lattice disclination or dislocation. We require that the boundaries of the region $W$ align with the boundaries of the unit cell $\Theta$. The linear size of $W$ must be much larger than the correlation length. We first define the charge in a region $W$:
\begin{align}
    Q_{W} &= \sum_{i \in W} \text{wt}(i) Q_i,
\end{align}
where $\text{wt}(i) = 1$ if $i$ is in the interior of $W$, while if $i$ lies on the boundary $\partial W$, $2\pi \text{wt}(i)$ is the angle subtended by $\partial W$ in the interior of $W$ at $i$. $Q_i$ is the charge on site $i$ in the ground state of $H_{\text{defect}}$. An example is shown in Fig.~\ref{fig:disclinations}.

Using Eq.~\eqref{eq:chargeresponse} along with Eq.~\eqref{eq:k0def}, we then find 

\begin{align}\label{eq:Qwpred}
    Q_{W} &= C \frac{\Phi_{W}}{2\pi} + \mathscr{S}_{\text{o}} \frac{\Omega}{2\pi} + \vec{\mathscr{P}}_{\text{o}} \cdot \vec{b}_{\text{o}} + \kappa n_{W,\text{o}} \mod 1\\
    &= C \frac{\delta \Phi_{W, \text{o}}}{2\pi} + \mathscr{S}_{\text{o}} \frac{\Omega}{2\pi} + \vec{\mathscr{P}}_{\text{o}} \cdot \vec{b}_{\text{o}} + \nu n_{W,\text{o}} \mod 1.
\end{align}
$\Phi_{W}$ is the total flux through the region $W$. $n_{W, \text{o}}$ is the number of unit cells in $W$, and may be fractional. $\Phi_{W}$ has two contributions, $\Phi_{W}=\phi n_{W,\OO}+\delta \Phi_{W, \text{o}}$. Here, $\phi n_{W,\OO}$ is the reference background flux within $W$. $\delta \Phi_{W, \text{o}}$ is the excess magnetic flux in the region $W$ relative to this reference. The precise microscopic definitions of $\delta \Phi_{W, \text{o}}$ and $n_{W, \text{o}}$ are quite subtle and non-trivial and explained in detail in Secs~\ref{sec:c4_SO}, \ref{sec:c4_PO} and~\ref{sec:general}.

Importantly,  $\delta \Phi_{W, \text{o}}$ and $n_{W, \text{o}}$ in general depend on the position of $\text{o}$ relative to the chosen unit cell $\Theta$. Nevertheless, the final results for $\mathscr{S}_{\text{o}}$ and $\vec{\mathscr{P}}_{\text{o}}$ are independent of $\Theta$. This is explained using the \textit{trimming} method developed in App.~\ref{sec:trim}. 

Naively it may seem that the coefficients in Eq.~\eqref{eq:Qwpred} should also depend on $\OO^*$. One reason for our notation is that Eq.~\eqref{eq:Qwpred} comes from a TQFT which is only sensitive to $\OO$. But even in microscopic calculations, we find that neither $\SO$ nor $\PO$ actually depends on $\OO^*$. This is easily seen for $\PO$, which can be defined using pure dislocations, for which $\Omega = 0$ and $\OO^*$ does not appear. To show that $\SO$ is independent of $\OO^*$, we give a theoretical argument when $C=0$, in App.~\ref{sec:ostarindependence}. We also have extensive numerical evidence for this when $C \ne 0$. 

The above discussion implies that we can consider any defect Hamiltonian, and extract $\mathscr{S}_{\text{o}}$ and $\vec{\mathscr{P}}_{\text{o}}$ (which only depend on $\text{o}$) by suitably defining $\delta\Phi_{W,\OO}$ and $n_{W,\OO}$ along with the appropriate Burgers vector. To simplify the disclination charge calculation of $\SO$ we will often choose $\OO = \OO^*$, but this is not a requirement of the theory. 

\subsubsection{$\SO$ from angular momentum}

Alternatively, we can examine the action of rotations and translations on the ground state in order to extract an angular momentum or linear momentum. These dual responses are a valuable consistency check on the value of $\SO$ obtained from the above charge response. Importantly, to compare the values of discrete shift from the disclination charge and angular momentum calculations, we need to set $\OO = \OO^*$ in both cases.

Let $|\Psi(m)\rangle$ be the ground state of the clean translationally invariant system on a torus in the presence of $m$ flux quanta, $m$ being an integer. Then
\begin{align}
    \tilde{C}_{M',\text{o}} |\Psi\rangle = e^{2\pi i l_{\text{o}}/M'} | \Psi \rangle,
\end{align}
where recall that $M'$ is the largest integer such that $\OO$ is invariant under $2\pi/M'$ rotations centered at $\OO$.  
We find that the angular momentum $l_{\text{o}}$ obeys the formula
\begin{align}
    l_{\text{o}}(m) = \mathscr{S}_{\text{o}} m + C \frac{m^2}{2} + K(C, L) \mod {M'} 
\end{align}
where $K(C,L)$ is a constant independent of $m$, depending on the system size $L$ and the Chern number $C$. The numerical data of $l_{\text{o}}$ is shown in 
Fig.~\ref{fig:bare_momentums}. For $|\Psi\rangle$ to be an eigenstate of $\tilde{C}_{M',\text{o}}$, appropriate global holonomies of the background gauge field and certain commensurate system sizes must be chosen, as discussed in Ref.~\cite{zhang2022fractional}. We note that one can also recover $\mathscr{S}_{\text{o}}$ by locally inserting flux and performing partial rotations \cite{zhang2022fractional}.

\bgroup
\def\arraystretch{1.8}
\begin{table*}
    \centering
\begin{tabular}{ c||p{0.5\textwidth}|p{0.3\textwidth}  }
\hline

Charge response &
$
    Q_W=C\frac{\delta\Phi_{W,\text{o}}}{2\pi}+\frac{\Omega_W \mathscr{S}_{\text{o}}}{2\pi}+\vec{\mathscr{P}}_{\text{o}}\cdot \vec{b}_{\OO}+\nu(k+n_{\text{irreg},\text{o},\Omega,\vec{b}}) \mod 1
$ & $\OO = $ origin of loop used to measure Burgers vector $\vec{b}_{\OO}$\\
\hline
Angular momentum & $
    l_{\text{o}}(m) = \frac{C m^2}{2} + \mathscr{S_\text{o}} m+ K(C,L) \mod M
$
& $\OO = $ rotation center
\\
\hline

Linear momentum & $    p_{\lambda,y}(m) = -\mathscr{P}_{\Oy} m+ K_y \mod 1$ & $\OO$ is determined by the `gauge origin' $\bar{\OO}$ through Eq.~\eqref{eq:generalobar}\\

\hline

1d polarization & $-\mathcal{P}_{\mathcal{O},x}=\frac{C k\phi}{2\pi} L_y +L_y\mathscr{P}_{\OO,y}+K \mod 1$ & $\mathcal{O} = $ origin used in Resta's formula, Eq.~\eqref{eq:resta}; $\OO,\mathcal{O}$ must satisfy Eq.~\eqref{eq:1dorelation}.\\

\hline

\end{tabular}
\caption{General equations for charge response (including dislocation and disclination charge), angular momentum and linear momentum. The last column explains the different special points that arise in each calculation.
}
\label{tab:geneqs}
\end{table*}
\egroup

\subsubsection{$\vec{\mathscr{P}}_{\text{o}}$ from linear momentum}

The topological field theory, Eq.~\eqref{eq:chargeresponse}, predicts that the polarization $\vec{\mathscr{P}}_{\text{o}}$ also specifies the linear momentum of $U(1)$ flux \cite{manjunath2021cgt}. We have found empirically 
that $\vec{\mathscr{P}}_{\text{o}}$ can be extracted by studying expectation values of an approximate translation operator, as we briefly summarize below. See Sec. \ref{sec:c4_PO} for additional details.

Suppose we wish to measure $\mathscr{P}_{\Oy}$ on the square lattice. We consider a state on a clean torus with $m = \frac{\phi}{2\pi} L_x L_y$ total flux quanta, where $L_x$, $L_y$ are the number of unit cells in the $x$ and $y$ directions. While the infinite plane possesses an infinite magnetic translation symmetry along the two directions, on the torus with magnetic flux it is not possible to fully preserve translation symmetry along $y$ unless $m/L_y = \frac{\phi}{2\pi} L_x$ is an integer. For general $m$, on the torus we  can insert the flux using a Landau-like gauge that is almost translation symmetric along $y$, except for a small strip which forms a cycle along $y$. 

We then define an approximate translation operator
\begin{equation}
    \tilde{T}_{\bf y} := \hat{T}_{\bf y} e^{i \sum_j \lambda_j c_j^{\dagger} c_j}.
\end{equation}
The expectation value of $\tilde{T}_{\bf{y}}$ defines the linear momentum $p_{\lambda,y}$ in the $y$ direction:
\begin{equation}
    \bra{\Psi(m)} \tilde{T}_{\bf{y}}\ket{\Psi(m)}=e^{ - \gamma + i 2\pi p_{\lambda,y}}.
\end{equation}

In our numerics we define $\lambda$ using Eq.~\eqref{eq:tyfullgaugeapp}. In particular, we find empirically that there exist special choices of $\lambda$ for which $p_{\lambda,y}$ determines the quantized polarization $\mathscr{P}_{\OO,y}$ throughout the Hofstadter butterfly, as follows. 

We find that for the Hofstadter model,  for appropriately chosen $\lambda$, the amplitude $e^{- \gamma(m)}$ in general oscillates as a function of $m$ and it vanishes for certain special values of $m$. Whenever the amplitude is nonzero, the linear momentum is found to obey the following relation:
\begin{equation}\label{eq:linearmomentum}
 p_{\lambda,y}(m) = 
     -\mathscr{P}_{\Oy} m+ K_y \mod 1 
\end{equation}
where $K_y$ is piecewise constant in $m$ (it can jump at the values of $m$ where the amplitude vanishes). The numerical data of $p_{\lambda,y}$ is shown in Fig.~\ref{fig:bare_momentums}.

The origin $\OO$ is determined as follows. We first define a point $\overline{\text{o}}$, referred to as the `gauge origin' for the vector potential, which has the property that the holonomy of the vector potential is trivial along the $x$ and $y$ cycles of the torus that meet at $\overline{\OO}$ (see Eq.~\eqref{eq:generalgauge}). Then the origin $\OO$ used to obtain $\PO$ is expressed in terms of $\overline{\OO}$; see Eq.~\eqref{eq:generalobar}.  

In defining $\tilde{T}_{\bf y}$ we in principle have the freedom to combine it with an arbitrary global $U(1)$ rotation:
\begin{align}
     \tilde{T}_{\bf y} \rightarrow   \tilde{T}_{\bf y} e^{i \chi \hat{N}},
\end{align}
which corresponds to a shift $\lambda_j \rightarrow \lambda_j + \chi$ for each $j$. Once $\OO$ is fixed, then $\chi$ is fixed to be an integer multiple of $\phi$ by fixing the flux through a dislocation created using $\tilde{T}_{\bf y}$, as explained in Sec.~\ref{sec:c4_PO} and Appendix~\ref{sec:dislocationConstruction}.
Thus we only need to consider the case where
\begin{align}
    \chi = \phi v_x = \frac{2 \pi m}{L_x L_y} v_x, 
\end{align}
for some integer $v_x$. Under such a shift in $\tilde{T}_{\bf y}$, 
\begin{align}
    p_{\lambda,y} &\rightarrow p_{\lambda,y} + m v_x \nu \nonumber \\
    &= p_{\lambda,y} + m v_x \left( \frac{C m}{L_x L_y} + \kappa \right)
\end{align}
If we consider only the term linear in $m$, this implies that
\begin{equation}
    \mathscr{P}_{\Oy} \rightarrow \mathscr{P}_{\Oy} - \kappa v_x
    = \mathscr{P}_{\OO, y} \mod 1.
\end{equation}

Note that one could consider the case where $v_x$ is fractional but quantized, and this would effectively correspond to a shift of the HSP $\OO$ by $(v_x,0)$ to a different HSP, as explained in Sec. \ref{sec:c4_PO}. 

Analogous equations hold for $\mathscr{P}_{\Ox}$, if we instead start with a Landau-like gauge along $x$. Furthermore, our procedure straightforwardly generalizes to rotational symmetries of order $M=2,3,4,6$; we discuss this in Sec.~\ref{sec:general_PO}.

We have also measured $\vec{\mathscr{P}}_{\text{o}}$ by studying the expectation values of a partial translation operator $\tilde{T}_{\bf y}|_D$, which is $\tilde{T}_{\bf y}$ restricted to some suitably chosen region $D$. This method also allows to extract a quantized $\vec{\mathscr{P}}_o$ consistent with dislocation charge, for a suitable choice of $\overline{\OO}$ and of the region $D$, when $M$ is even. We discuss this further in Secs.~\ref{sec:c4_PO} and~\ref{sec:general_PO}.

\subsubsection{$\vec{\mathscr{P}}_{\text{o}}$ from dimensional reduction and 1d polarization}

One can also define a 1d polarization along the $\hat{i}$ direction by treating the system as an effectively 1d system along $\hat{i}$. Let us first consider $\hat{i} = \hat{x}$. Then we can calculate the 1d polarization using Resta's formula \cite{resta1998eq}:  

\begin{equation}\label{eq:resta}
        \mathcal{P}_{\mathcal{O},x} = \frac{1}{2\pi}\text{arg } \langle \Psi | e^{i \frac{2\pi}{L_x} \sum_{j} j_x \hat{n}_{j}} |\Psi \rangle
\end{equation}
The above expression depends on a choice of origin $\mathcal{O}_x$ for $j_x$, which we have made explicit, 
i.e. $j_x\in\{-\mathcal{O}_x,1-\mathcal{O}_x,2-\mathcal{O}_x,\dots,L_x-1-\mathcal{O}_x\}$. Empirically, we find that
\begin{align}\label{eq:polarization}
 -\mathcal{P}_{\mathcal{O},x}=\frac{C k\phi}{2\pi} L_y  +L_y\mathscr{P}_{\OO,y}+K' \mod 1. 
\end{align}
where $k:=\frac{L_x}{2}-\overline{\OO}_x+\mathcal{O}_x$. Knowing the value of $k$ is not crucial in extracting $\mathscr{P}_{\Oy}$. This result agrees with a field theory prediction which is derived in Sec.~\ref{sec:c4_1dpolarization}.

Similar to the linear momentum calculation, we extract a value of $\vec{\mathscr{P}}_{\text{o}}$ that is consistent with other approaches only when $\OO,\mathcal{O}$ satisfy a certain relation, see Eq.~\eqref{eq:1dorelation}. This calculation is independent of the specific details of the gauge, as long as $\Phi_y$ is linear in $\phi$. The full discussion is contained in Sec.\ref{sec:c4_1dpolarization}. 

\section{Basic properties of lattice defects}\label{sec:useful_bgd}

\begin{figure}[t]
    \centering
    \includegraphics[width=8cm]{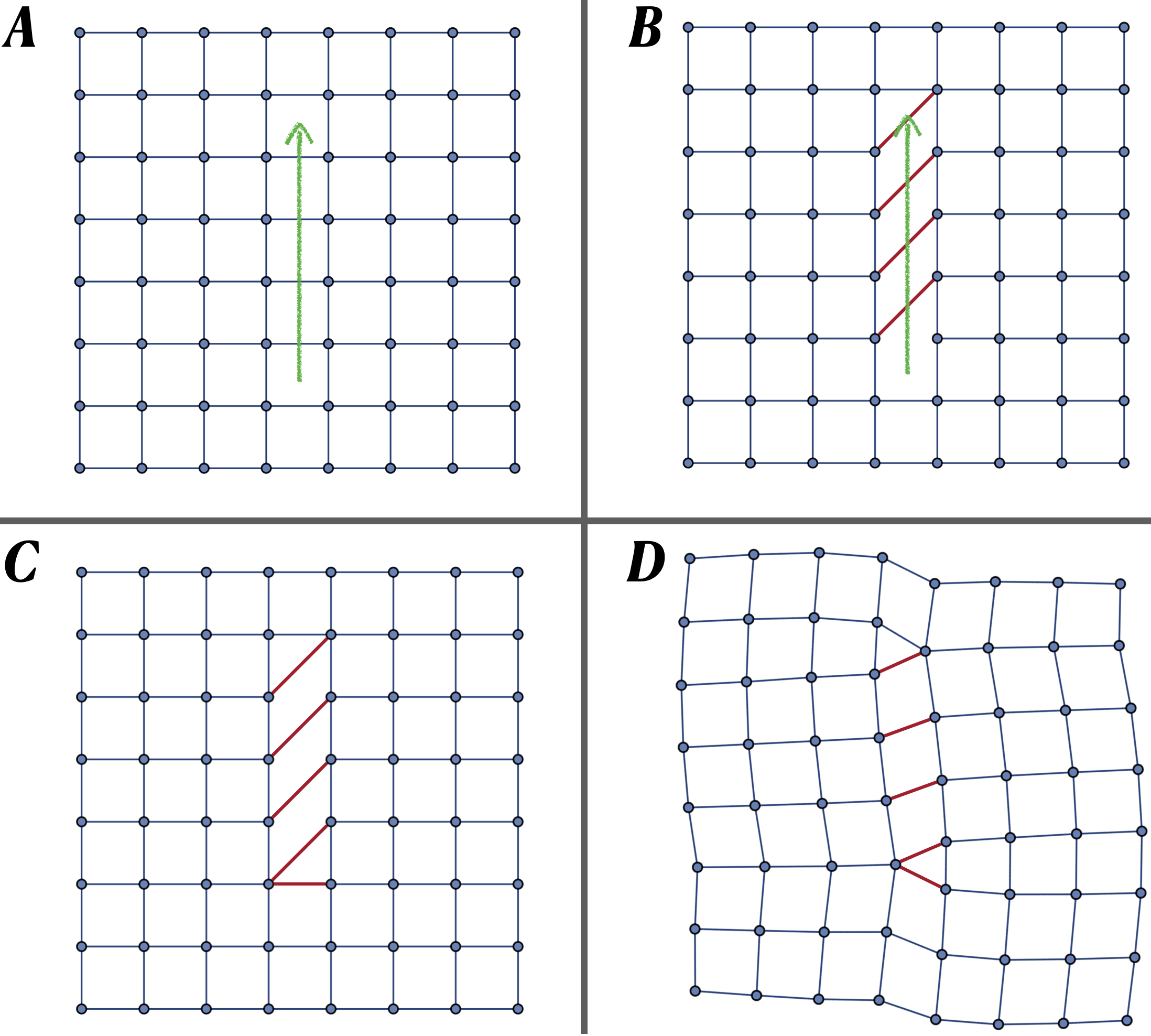}
    \caption{Dislocation construction on a square lattice. \textbf{A.} Drawing a cut in the $\hat{y}$ direction. \textbf{B.} Conjugate hoppings. \textbf{C.} Doing local moves. \textbf{D.} Reorganizing.}
    \label{fig:c4dislocation_main}
\end{figure}

\begin{figure}[t]
    \centering
    \includegraphics[width=8cm]{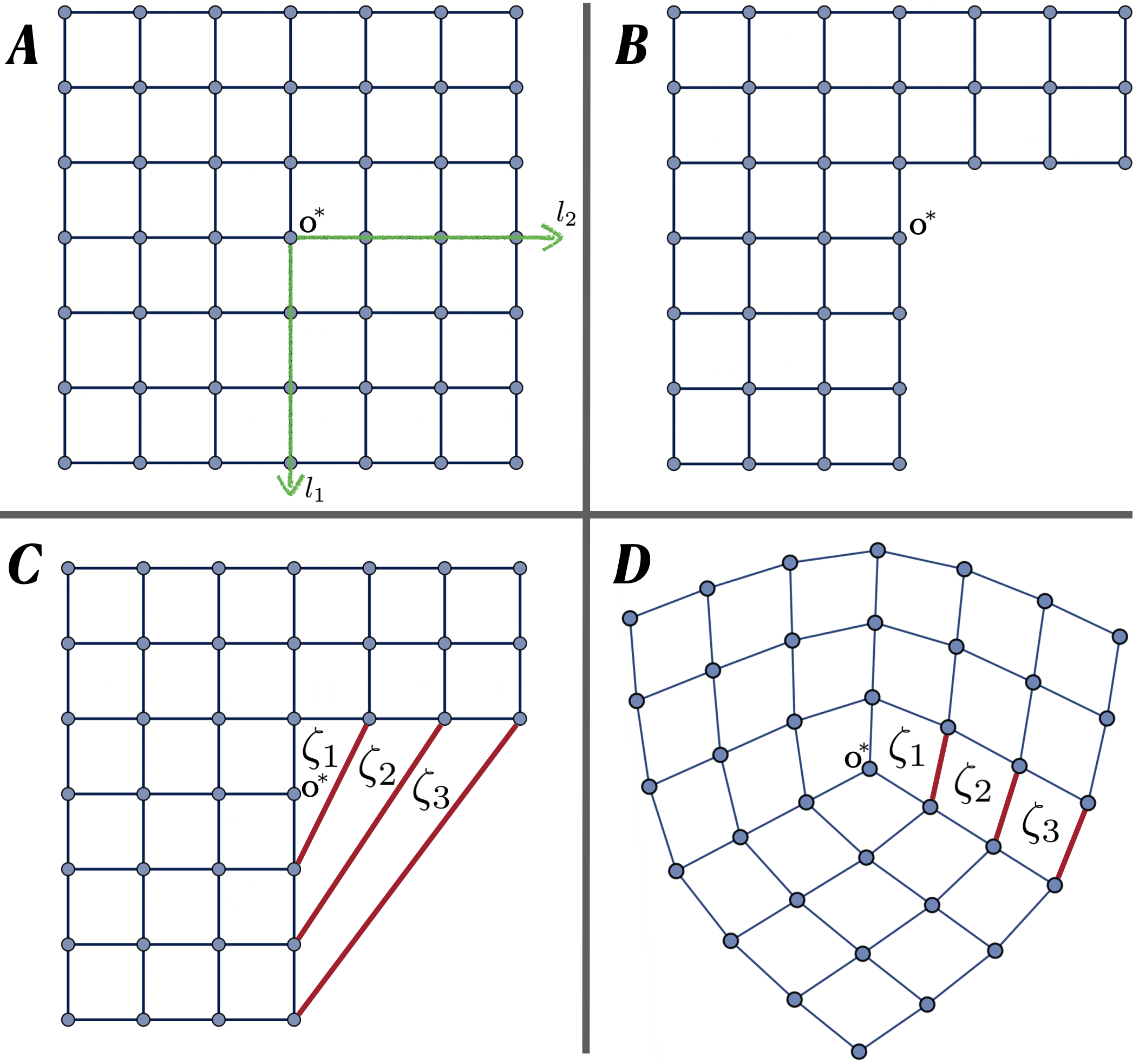}
    \caption{\label{fig:cutNglue} Cut and glue procedure of constructing a disclination. \textbf{A.} Original lattice on an open plane; \textbf{B.} Cutting; \textbf{C.} Gluing, this creates new plaquettes $\zeta_i$; \textbf{D.} Reorganizing.}
\end{figure}

Before discussing the numerical calculations in detail, we review some useful background material on lattice defects and their properties. A more extensive background review can be found in App.~\ref{sec:bgd}. The quantum mechanical details of constructing a defect Hamiltonian  $H_{\text{defect}}$ through a cut-and-clue procedure are described in detail for dislocation defects in App.~\ref{sec:dislocationConstruction}.

We illustrate the procedure for constructing a dislocation defect in Fig.~\ref{fig:c4dislocation_main}. First we make a cut on an infinite clean lattice and define the left $(L)$ and right $(R)$ sides of the cut. We replace all bonds that cross the cut so that a point $L_i$, originally connected to some point $R_i$, is now connected to $R_{i + \vec{b}^*}$. Here $\vec{b}^*$ is an integer vector related to the dislocation Burgers vector, which we define below.

We illustrate the construction of a disclination defect in Fig.~\ref{fig:cutNglue}. We draw two rays $l_1, l_2$ which meet at the point $\text{o}^*$, such that $l_2$ is obtained from $l_1$ by rotating about $\text{o}^*$ through the angle $\Omega>0$. Now we delete all points within the wedge enclosed by $l_1$ and $l_2$, except those that lie exactly on $l_1$. We then reconnect the bonds so that a link $i_1 j$ (where $i_1$ lies on $l_1$) is replaced by a link $i_1 j'$, where $j'$ is obtained from $j$ upon rotating by $\Omega$ about $\OO^*$.

The disclination angle $\Omega$ can be directly measured from the defect lattice alone. It is the angle by which a unit vector is rotated upon being parallel transported around the defect. `Pure' dislocation defects are those with zero disclination angle. They are characterized by a dislocation Burgers vector $\vec{b}$, which is defined as follows. Starting from a point $\text{o}$, consider a sequence of lattice translations which encircles the defect (and no other defects) in counterclockwise fashion, and ends at $\text{o}$. For a dislocation, the sum of these translations will not equal zero, but instead some integer vector, which we define as $\vec{b}$. The above dislocation construction gives two dislocations at the two ends of the cut, with $\vec{b} = \pm \vec{b}^*$. For a pure dislocation, the shape of the loop and the choice of $\OO$ do not affect the value of $\vec{b}$. 

\begin{figure}[t]
    \centering
    \includegraphics[width=8cm]{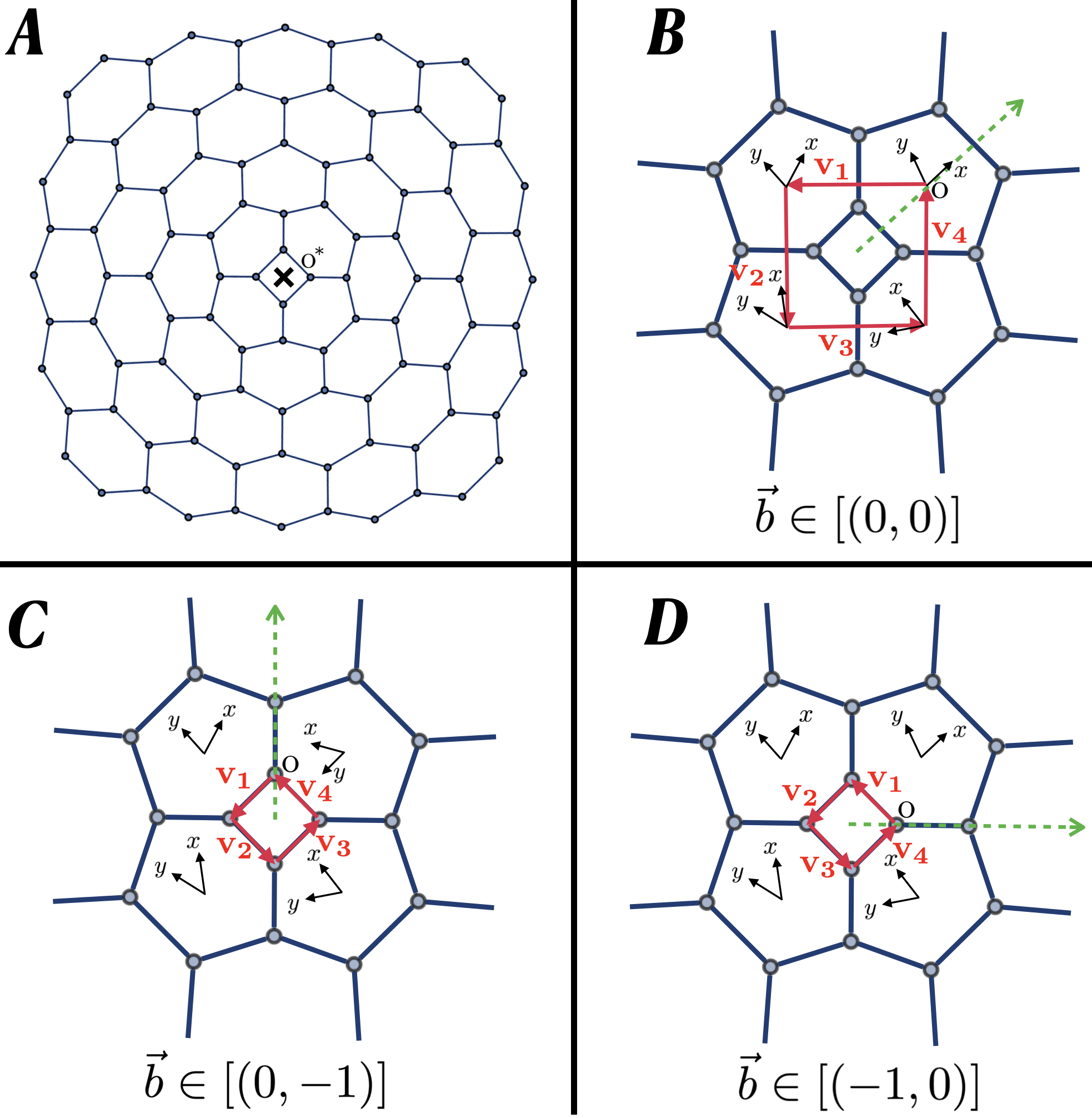}
    \caption{\textbf{A.} A pure disclination with $\Omega=\frac{2\pi}{3}$ created using $\tilde{C}_{3,\OO^*}$ via a cut-and-glue construction. The disclination core $\OO^*$ is marked as `$\times$'. We can define a `frame' i.e. an $x,y$ basis at every point on the defect lattice (they are plotted at the four plaquette centers near the disclination). The frame rotates by $-2\pi/3$ upon crossing the green dotted branch cut which passes through $\OO^*$ and $\OO$. 
    We calculate $\vec{b}_{\OO}=\vec{v}_1+\vec{v}_2+\vec{v}_3+\vec{v}_4$ for different $\OO$
    \textbf{B.} $\OO$ at a plaquette center. $\{\vec{v}_1,\vec{v}_2,\vec{v}_3,\vec{v}_4\}=\{(-1,1),(-1,0),(0,-1),(1,-1)\}$.\newline \textbf{C.} $\OO$ at one of the sites. $\{\vec{v}_1,\vec{v}_2,\vec{v}_3,\vec{v}_4\}=\{(-2/3,1/3),(-1/3,-1/3),(1/3,-2/3),(2/3,-1/3)\}$.\newline \textbf{D.} $\OO$ at the other site. $\{\vec{v}_1,\vec{v}_2,\vec{v}_3,\vec{v}_4\}=\{(-1/3,2/3),(-2/3,1/3),(-1/3,-1/3),(1/3,-2/3)\}$}
    \label{fig:exampleLattice}
\end{figure}

The Burgers vector of a disclination with $\Omega \ne 0$ can be measured similarly; note that $\text{o}^*$ and $\text{o}$ need not be related. Importantly, the Burgers vector when $\Omega \ne 0$ sensitively depends on the choice of $\OO$, assuming $\OO^*$ is fixed. If we shift $\OO \rightarrow \OO + \vec{v}$, then
\begin{equation}\label{eq:bOplusv}
    \vec{b}_{\OO+\vec{v}} = \vec{b}_{\OO} + (1-U(\Omega)) \vec{v}.
\end{equation}
This is derived in App.~\ref{sec:bgd}. As an illustration, Fig.~\ref{fig:exampleLattice} shows the same $\Omega = \frac{2\pi}{3}$ disclination but with different Burgers vectors depending on the choice of $\OO$. Since choosing $\vec{v} = \vec{\Lambda}$ to be an integer vector should give us an equivalent characterization of the defect, we have the equivalence
\begin{equation}
    \vec{b}_{\OO} \simeq \vec{b}_{\OO} + (1-U(\Omega)) \vec{\Lambda}.
\end{equation}
Thus, if $\vec{v}$ is fractional, $\vec{b}_{\OO + \vec{v}}$ will not be equivalent to $\vec{b}_{\OO}$. If $\Omega = \frac{2\pi k}{M'}$ where $k,M'$ are coprime, the distinct classes of Burgers vectors form a group $K_{M'}$ as defined in Sec.~\ref{sec:ovviewmain}.

Note that our construction ensures the following relation:
\begin{equation}
    \vec{b}_{\text{o}^*}\simeq(0,0),
\end{equation}
which can be verified by constructing the various classes of disclinations for $M=2,3,4,6$. Setting $\text{o} = \text{o}^*$ thus ensures that the defect has trivial Burgers vector. This will be a convenient choice to make in the following sections.
Note that we sometimes also use the notation $\vec{b}_{\OO}\in[(0,0)]$ to indicate that $\vec{b}_{\OO}$ is in the same equivalence class of $(0,0)$.

\section{Calculation of $\mathscr{S}_{\text{o}}$ on the square lattice}\label{sec:c4_SO}

This section and the next are devoted to numerically checking the predictions of the field theory for the square lattice. This section reviews and generalizes the main results from Ref.~\cite{zhang2022fractional}. Analogous calculations for $M=2,3,6$ are discussed in Sec.~\ref{sec:general}. 

\begin{figure*}[t]
    \centering
    \includegraphics[width=14cm]{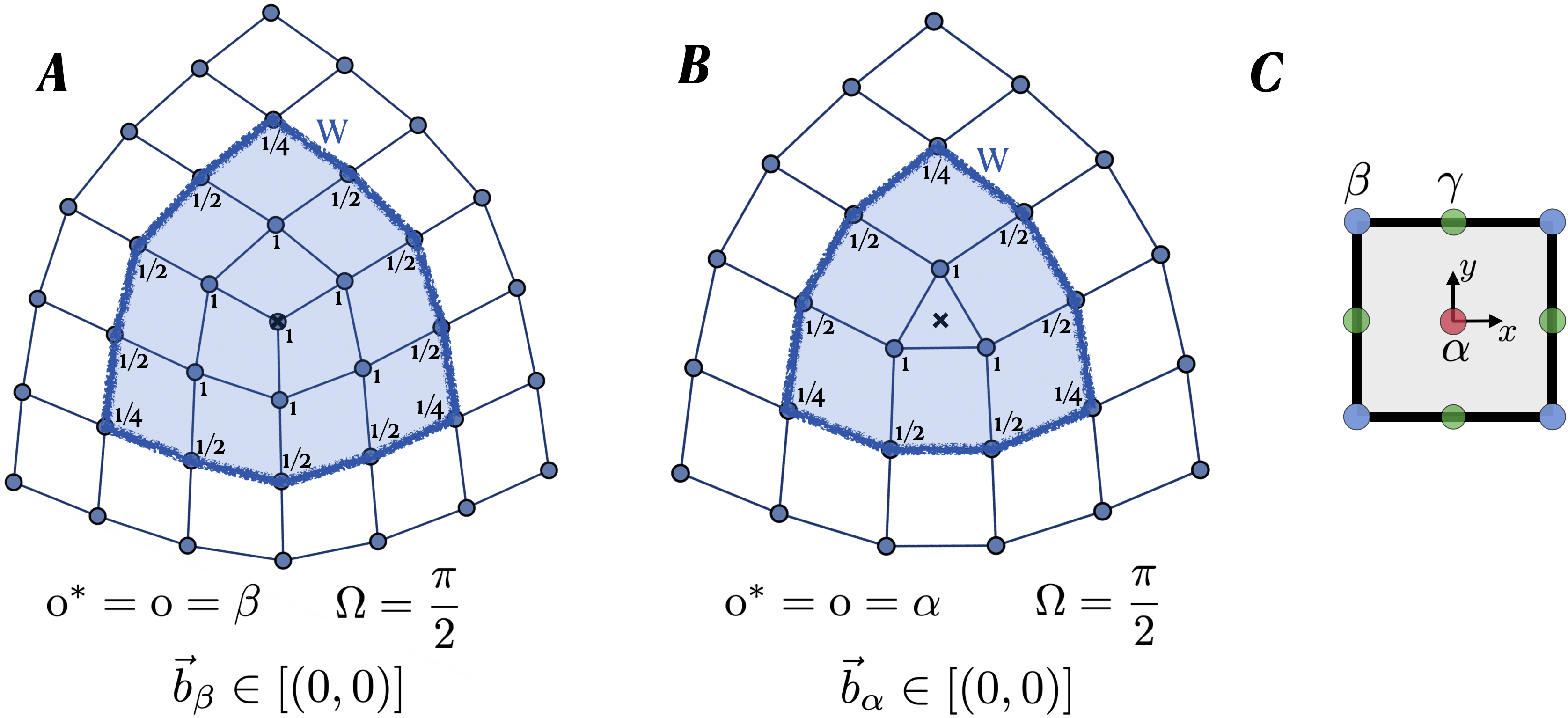}
    \caption{Pure disclinations with 2 different $\text{o}$ \textbf{A.}~$\OO=\beta$ \textbf{B.}~$\OO=\alpha$. The disclination core $\OO^*$ is marked as `$\times$'. The weightings wt($i$) within a representative choice of $W$ are marked near each site. When $\OO=\alpha$, there is a irregular unit cell at the disclination core. \textbf{C.} The unit cell choice, black solid lines represent hoppings. }
    \label{fig:disclinations}
\end{figure*}

We fix our origin at a HSP $\text{o}$ which has fourfold rotational symmetry. There are two choices, $\alpha$ and $\beta$, as shown in Fig.~\ref{fig:wyckoff}. $\alpha$ denotes the center of the unit cell, while $\beta$ denotes a corner. For calculations on the simplest square lattice, we pick the unit cell shown in Fig.~\ref{fig:disclinations}C, where $\alpha$ corresponds to a plaquette center and $\beta$ corresponds to a site. Note that the formulas for $\OO = \gamma$ are contained in the discussion for $C_2$ symmetric systems given in Sec.~\ref{sec:general}. 

For either choice of $\text{o}$, we have two topological invariants, $\mathscr{S}_{\text{o}}$ and $\vec{\mathscr{P}}_{\text{o}} = \frac{1}{2} \overline{\mathscr{P}}_{\text{o}} (1,1)$.  $\mathscr{S}_{\text{o}}$ is defined mod $4$ and satisfies $\SO = \frac{C}{2} \mod 1$. $\overline{\mathscr{P}}_{\text{o}}$ is an integer defined mod 2. 

We extract $\mathscr{S}_{\text{o}}$ 
in two physically different ways, through the disclination charge and the angular momentum of flux. The exact choice of unit cell does not affect the final result; we show this in App.~\ref{sec:trim}.

\subsection{Symmetry operators}

First we define the magnetic rotation operator $\tilde{C}_{4,\OO^*}$ which is used to create a disclination centered at $\OO^* = \alpha,\beta$:
\begin{equation}
    \tilde{C}_{4,\OO^*} \equiv \hat{C}_{4,\OO^*} e^{i \sum_j \lambda_j c_j^{\dagger}c_j}.
\end{equation}
We require that a system with a pure disclination at $\OO^*$ constructed using $\tilde{C}_{4,\OO^*}$ has flux $\phi$ in each regular unit cell. This condition forces $\tilde{C}_{4,\beta}^4 = +1$ \cite{zhang2022fractional}. When $\OO^* = \beta$, all unit cells are regular, and then this condition in fact completely fixes $\lambda_j$; this is an example where there is a unique canonical choice for the rotation operator $\tilde{C}_{4,\OO^*}$ (once the Hamiltonian is fixed). 
For consistency in the definitions of our operators, we demand that $\tilde{C}_{4,\alpha}^4 = +1$ as well.

We also define translation operators $\tilde{T}_{\bf{x}}$ and $\tilde{T}_{\bf{y}}$ which obey the magnetic translation algebra:
\begin{equation}\label{eq:MTA}
    \tilde{T}_{\bf y}^{-1} \tilde{T}_{\bf x}^{-1}\tilde{T}_{\bf y}\tilde{T}_{\bf x} =e^{i \phi \sum_j c_j^{\dagger}c_j}.
\end{equation}
The gauge transformations used to define the translation operators will be discussed in Sec.~\ref{sec:c4_PO}.

\subsection{Construction of clean Hamiltonian $H_{\text{clean}}$}

In our numerical work we consider the Hofstadter model, which has a spinless free fermion Hamiltonian of the form
\begin{align}
    H_{\text{clean}} = -\sum_{\langle ij\rangle} t_{ij} c_i^\dagger c_j + \text{h.c.} 
\end{align}
where the nearest neighbor hopping terms $t_{ij} = t e^{i A_{\text{clean}, ij}}$ depend on a background vector potential $A_{\text{clean}}$, which assigns flux $\phi$ per unit cell. The parameters in $H_{\text{clean}}$ are discussed in detail in App.~\ref{sec:bgd}. 

Although we mainly consider nearest-neighbor hopping in our numerics, our theoretical predictions as well as our numerical scheme apply much more generally. For example, we can consider arbitrary next neighbor hoppings. To illustrate this, in App.~\ref{app:nnnhopping} we give evidence showing that invariants extracted numerically in the square lattice Hofstadter model remain well-defined upon adding next neighbor hopping terms. Below we will also argue that our procedure works if $H_{\text{clean}}$ has $N$-body interaction terms for $N>2$.

An important point is that we require the magnetic field $B$ to be defined everywhere within the unit cell. This means that the total magnetic flux within any subregion of the unit cell is specified as a real number. This requirement goes beyond what is directly specified by the crystalline symmetry (which only demands flux $\phi \mod 2\pi$ per unit cell). But it is a physically natural requirement, since the most general lattice models with the given symmetry have some small amount of further neighbor hopping, between different points within a single unit cell. In fact, specifying a nearest-neighbor Hamiltonian with just $\phi \text{ mod } 2\pi$ is ill-defined in a sense, because it does not specify how to consistently perturb the model with further neighbor hopping terms. Such a specification requires a choice of $\phi$ as a real number, not just modulo $2\pi$.

\subsection{$\SO$ from disclination charge}\label{sec:C4_SO_disc}

\subsubsection{Construction of defect Hamiltonian $H_{\text{defect}}$}\label{sec:amb_Hdef}

Start with a clean lattice Hamiltonian $H_{\text{clean}}$ which depends on a background $U(1)$ vector potential $A_{\text{clean}}$ through the variables $e^{i A_{ij,\text{clean}}}$. Suppose we create a lattice disclination (a detailed construction can be found in Ref.~\cite{zhang2022fractional}), and arrive at a defect Hamiltonian $H_{\text{defect}}[e^{i A_{ij,\text{defect}}}]$ through a cut-and-glue procedure. Here $A_{\text{defect}}$ is the vector potential on the defect lattice. Note that irregular unit cells may exist at the center of each defect. For example, there could be a triangular unit cell at the center of an $\Omega = \frac{\pi}{2}$ disclination (see Fig.~\ref{fig:disclinations}B), or a triangular unit cell at the center of a square lattice dislocation (See Fig.\ref{fig:dislocationW}). Irregular unit cells  can have different shapes depending on the value of $M$.

Away from the defect, the flux in any region is fully determined by $A_{\text{clean}}$, and we can ensure that the system has flux $\phi$ per unit cell. However, the flux in the immediate vicinity of the defect depends on $A_{\text{clean}}$ as well as on the definition of the symmetry operator which creates the defect. In particular, if we require a specific value of flux in the unit cells immediately adjacent to the defect, there will be a constraint on the definition of the symmetry operators we use.

\subsubsection{Charge prediction from field theory}

Having fixed $H_{\text{defect}}$, we compute the charge $Q_W$ in a region $W$ containing the defect. We always require that the boundary of $W$ coincides with the boundary of a unit cell. This ensures that the only irregular unit cells in $W$ are near the center of the defect. 
$Q_W$ is defined by the formula
\begin{equation}
    Q_W = \sum\limits_{i \in W} \text{wt}(i) Q_i.
\end{equation}
The weights $\text{wt}(i)$ were defined in Sec.~\ref{sec:ovviewmain}; this definition ensures that $Q_W + Q_{W'} = Q_{W \sqcup W'}$ when two regions $W,W'$ overlap only on their boundaries, as required by the response theory. Note that the value of $Q_W$ depends on the definition of the unit cell.

This method of measuring $Q_W$ applies to any local gapped Hamiltonian, even if it has interaction terms. As long as the system has a correlation length much smaller than the linear size of $W$, the ambiguity in $H_{\text{defect}}$ near the defect core will not affect the value of $Q_W$ for sufficiently large $W$.

The next issue is how to assign this charge to the different terms in the response theory, which predicts that for large enough $W$,
\begin{align}\label{eq:QWfield}
Q_W = C \frac{\delta \Phi_{W, \text{o}}}{2\pi} + \mathscr{S}_{\text{o}} \frac{\Omega}{2\pi} + \vec{\mathscr{P}}_{\text{o}} \cdot \vec{b}_{\text{o}} + \nu n_{W, \text{o}} \mod 1.
\end{align}
Here $\Omega$ and $\vec{b}_{\text{o}}$ are the disclination angle and Burgers vector of the defect, respectively. We fix $\vec{b}_{\OO} = (0,0)$ throughout this section, so the term with $\PO$ does not contribute (this is equivalent to saying that $\OO = \OO^*$).

The parts of this equation which require special care are $\delta \Phi_{W,\text{o}}$ and $n_{W, \text{o}}$\footnote{This was referred to as $n_{u.c.,W}$ in Ref.~\cite{zhang2022fractional}.}. We discuss how to define these below. 

\subsubsection{Definition of $\delta \Phi_{W,\OO}$ and $n_{W,\OO}$}

$\delta \Phi_{W,\OO}$ is, intuitively,  the excess flux in the region $W$. To define the excess flux, we must compute the flux in $W$, and compare it to some background reference flux. There are two possibilities. If $\OO=\beta$, then all unit cells in $W$ are regular (see Fig.~\ref{fig:disclinations}A), and $n_{W,\beta} = 0 \mod 1$. Here $\delta \Phi_{W,\OO}=0$ if all unit cells have flux $\phi$. On the other hand, if $\OO=\alpha$, then there is an irregular unit cell at the disclination core (see Fig.~\ref{fig:disclinations}B).

In general, we can arrange to have at most one irregular unit cell in the core of the defect, with some flux $\phi_{\text{irreg}}$ which is determined by our choice of rotation operators. Its area is denoted $n_{\text{irreg},\text{o}}$, which can be fractional.  Note that 
\begin{equation}
    n_{W,\text{o}} = n_{\text{irreg},\text{o}} \mod 1.
\end{equation}

Then, if we choose our symmetry operators so that the flux through each regular unit cell in $H_{\text{defect}}$ is $\phi$, we have 
\begin{align}\label{eq:deltaphi}
    \delta \Phi_{W,\OO} = \phi_{\text{irreg}} - n_{\text{irreg},\text{o}} \phi.
\end{align}
Thus the value of $\delta \Phi_{W,\OO}$ depends on the value we assign to $n_{\text{irreg},\text{o}}$.

\subsubsection{Computation of $n_{\text{irreg},\OO}$}\label{sec:amb_lift}

If $\OO = \beta$, there are no irregular unit cells, as discussed above, so we simply define $n_{\text{irreg},\beta}=0 \mod 1$. If $\OO=\alpha$, we find $n_{\text{irreg},\alpha} = \frac{3}{4} \mod 1$. A heuristic argument is that the cut-and-glue procedure for a $\pi/2$ disclination removes 1/4 of a unit cell at the disclination center, i.e. the irregular unit cell at the disclination core consists of 3 subcells. Each subcell contributes 1/4th of a full unit cell, therefore $n_{\text{irreg},\alpha} = \frac{3}{4} \mod 1$. (A consistency check on this result is to demand that the value of $\SO$ corresponding to some physical point $\OO$ be the same whether we choose the unit cell to satisfy $\OO = \alpha$ or $\OO = \beta$. If $n_{\text{irreg}, \beta} = 0 \mod 1$, we find it necessary to have $n_{\text{irreg},\alpha} = \frac{3}{4} \mod 1$.)

We note that the previous step involved another subtlety. Even if we know $n_{\text{irreg},\text{o}}$, there can be a further ambiguity in $\delta \Phi_{W,\OO}$. If $n_{\text{irreg},\text{o}}$ is an integer, then Eq.~\eqref{eq:deltaphi} is perfectly well-defined modulo $2\pi$. However, if $n_{\text{irreg},\text{o}} = a/b$ is a fraction with $a,b$ coprime, $\delta \Phi_{W,\OO}$ is not invariant under the transformation $\phi \rightarrow \phi + 2\pi$. To keep $\delta \Phi_{W,\OO}$ invariant, we need to pick a lift of $\phi$ from $[0,2\pi)$ to $[0, 2\pi b)$. But this is why we insist on specifying the actual magnetic field everywhere in the unit cell. This fixes $\phi$ as a real number, not just mod $2\pi$, and so fractions of $\phi$ can be defined unambiguously.

\subsubsection{Computing $\SO$}
For a $\pi/2$ disclination with $\vec{b}_{\OO} = (0,0)$, Eq.~\eqref{eq:QWfield} predicts that
\begin{equation}
    \frac{\SO}{4} = Q_W - \nu n_{W,\OO} - C \frac{\delta \Phi_{W,\OO}}{2\pi} \mod 1.
\end{equation}
This determines $\SO/4 \mod 1$ in terms of well-defined quantities. Note that our procedure to define $n_{\text{irreg},\OO}$ and $\delta \Phi_W$ is independent of the details of the Hamiltonian, in particular if there are further neighbor hopping terms and interaction terms.
Therefore we expect that $\SO$ can be robustly extracted for any $H_{\text{clean}}$ with a symmetric, gapped ground state.

\subsection{Angular momentum of flux}

We can also compute $\mathscr{S}_{\text{o}}$ from the angular momentum eigenvalues of $\tilde{C}_{4,\text{o}^*}$ after inserting additional flux, if we set $\OO = \OO^*$. 
On the torus, there are two positions on the torus that are left invariant under a $\hat{C}_{4,\OO^*}$ rotation, $\OO^*$ and $\OO^*+(L/2,L/2)$. If $L$ is odd, then the two positions are not the same point in the unit cell. This is deemed unnatural since we only want a single origin. Thus,
we consider an even length system on a torus, insert $m$ total flux quanta uniformly, and define
\begin{equation}
    \tilde{C}_{4,\text{o}} \ket{\Psi(m)} \equiv e^{i \frac{\pi}{2} l_{\text{o}}} \ket{\Psi(m)}. 
\end{equation}
The field theory predicts that there will be a contribution to $l_{\text{o}}$ which equals $\mathscr{S} m$. Indeed, we numerically find that
\begin{align}\label{eq:angularMomentum}
    l_{\text{o}}(m) &= \frac{C m^2}{2} + m\mathscr{S}_{\text{o}}  + K(C,L) \mod 4.
\end{align}
The numerical data is shown in Fig.~\ref{fig:bare_momentums}. Additional technical details in these calculations, in particular a discussion of partial rotations, can be found in Ref.~\cite{zhang2022fractional}. 

\subsection{Application to Hofstadter model}

Ref.~\cite{zhang2022fractional} obtained $\SO$ for the Hofstadter model, taking $\OO=\beta$ to be at a site. Here we also study the case where $\OO=\alpha$ is at a plaquette center. The values of $\mathscr{S}_\text{o}$ obtained using both disclination charge and angular momentum are consistent. In the limit of small $\phi$ and $\nu$, our procedure also agrees with known results for continuum Landau levels, i.e. $\mathscr{S}_\text{o}= \frac{C^2}{2} \mod 4$ for either $\OO=\beta$ or $\OO=\alpha$. The Hofstadter butterflies for $\mathscr{S}_{\text{o}}$ are plotted in Fig.~\ref{fig:SandP}. 

We find that they obey the following empirical formulas (recall that $\beta$ is chosen to be a site). For $C>0$,
\begin{equation}
\label{eq:s_formula_b}
    \mathscr{S}_{\beta}(\phi)= \frac{C^2}{2}  -(C+1)\left\lfloor\frac{C\phi}{2\pi}\right\rfloor
    + 2\sum_{\substack{\frac{p}{q}<\frac{\phi}{2\pi}\\\text{odd }q
    }} \left\lfloor \frac{C+q}{2q} \right\rfloor  \mod 4,
\end{equation}
where the third term of Eq.\eqref{eq:s_formula_b} we sum over all $\frac{p}{q}$ in the Farey sequence of order $C$ that satisfy $\frac{p}{q}<\frac{\phi}{2\pi}$ and $q$ odd. 
The value of $\mathscr{S}_{\beta}$ for $C < 0$ can be obtained from the transformation $\mathscr{S}_{\beta}(\mu,\phi) = 1-\mathscr{S}_{\beta}(-\mu,\phi)$ which flips the sign of $C$. $\mathscr{S}_{\beta}=1$ for the fully filled state with $\nu=1,C=0$.

Similarly, we can set $\OO=\alpha$ which is the plaquette center. For $C>0$,
\begin{equation}
\label{eq:s_formula_a}
    \mathscr{S}_{\alpha}(\phi)= \frac{C^2}{2}  +C\left\lfloor\frac{C\phi}{2\pi}\right\rfloor
    + 2\sum_{\substack{\frac{p}{q}<\frac{\phi}{2\pi}\\\text{odd }q
    }} \left\lfloor \frac{C+q}{2q} \right\rfloor  \mod 4.
\end{equation}
The value of $\mathscr{S}_{\alpha}$ for $C < 0$ lobes can be obtained from the transformation $\mathscr{S}_{\alpha}(\mu,\phi) = 2C-\mathscr{S}_{\alpha}(-\mu,\phi)$ which flips the sign of $C$. $\mathscr{S}_{\alpha}=0$ when $C=0$.

One can verify that $\mathscr{S}_{\beta}$ has period $2\pi$ in $\phi$, but $\mathscr{S}_{\alpha}$ has period $8\pi$ in $\phi$.
The physical reason is that the defect lattice with $\text{o}=\alpha$ has a triangular plaquette at the disclination core. Under $\phi\rightarrow\phi+2\pi$, the background flux through this triangular plaquette transforms as $3\phi/4\rightarrow 3\phi/4+\frac{3\pi}{4}$. Therefore $H_{\text{defect}}$ is invariant under $\phi\rightarrow\phi+8\pi$ up to gauge transformations.

\section{Calculation of $\vec{\mathscr{P}}_{\text{o}}$ on the square lattice}\label{sec:c4_PO}

We next discuss three different ways to calculate $\vec{\mathscr{P}}_{\text{o}}$  for $\OO = \alpha,\beta$ directly on the square lattice, using dislocation charge, linear momentum, and a 1d polarization response. (The case $\OO = \gamma$ will be handled in Sec.~\ref{sec:general}.) In fact, we show that $\vec{\mathscr{P}}_{\alpha}, \vec{\mathscr{P}}_{\beta}$ can also be computed if we know $\mathscr{S}_{\alpha}, \mathscr{S}_{\beta}$ alone: see the results in Table \ref{table:shiftingO}, in particular Eq.~\eqref{eq:SaSbPa}. These results are derived in Sec.\ref{app:ftrelabel} and App.~\ref{app:ftcomps}. So in principle we can also measure the polarization indirectly from the disclination charge and angular momentum responses. All the calculations yield the same numerical values of $\vec{\mathscr{P}}_{\alpha(\beta)}$, and are consistent with field theory predictions. 

In this section we will not require the rotation operator $\tilde{C}_{4,\OO^*}$, but will still use the translation operators satisfying Eq.~\eqref{eq:MTA}. A notational remark: in Secs~\ref{sec:c4_PO},~\ref{sec:general} and the related Apps.~\ref{sec:dislocationConstruction},~\ref{app:linmom_details} we will use $A$ to denote the entire vector potential on a lattice without dislocation or disclination defects, so $A$ has the same meaning as $A_{\text{clean}}$ from the previous section. The vector potential in a system with lattice defects will be denoted as $A_{\text{defect}}$.

\subsection{$\PO$ from dislocation charge}\label{sec:c4_PO_DC}
First we consider the charge bound to a single dislocation (for an explicit construction of such a dislocation defect, see App.~\ref{sec:dislocationConstruction}). We follow the procedure outlined in Sec.~\ref{sec:C4_SO_disc}; similar arguments can be applied in this case.

For a defect with zero disclination angle and dislocation Burgers vector $\vec{b} = (0,\pm 1)$ or $(\pm 1,0)$, there is a triangular plaquette at the dislocation, and all other plaquettes within a large radius are ensured to be regular through our construction. Suppose the triangular plaquette has flux $\phi_{\text{irreg}}$. If the translation operator used to construct the dislocation is
$$\tilde{T}_{\bf y} = \hat{T}_{\bf y} e^{i \sum_j \lambda_j \hat{n}_j},$$ 
we find that when $\vec{b} = (0,1)$,
\begin{equation}\label{eq:phi-irreg}
    \phi_{\text{irreg}} = A_{j_0,j_0+\hat{y}} - \lambda_{j_0}
\end{equation}
where $j_0$ is a specific point in the irregular unit cell at the dislocation (see App.~\ref{app:phi-irreg} for a proof). Thus, the value of $\phi_{\text{irreg}}$ is set by the combined choice of $A,\lambda$ and $j_0$. In particular, we show in App.~\ref{app:phi-irreg} that for $v_x \in \Z$, the following transformations all take $\phi_{\text{irreg}} \rightarrow \phi_{\text{irreg}} + \phi v_x$: (1) taking $j_0 \rightarrow j_0 + (v_x,0)$ keeping $A,\lambda$ fixed; (2) taking $\lambda_j \rightarrow \lambda_j -\phi v_x$ for each $j$, keeping $A$ and $j_0$ fixed.
(Here we can even take $v_x \in \mathbb{R}$.) In this section it will turn out that fixing a precise value of $\phi_{\text{irreg}}$ is not essential, but the above discussion will also be useful in Sec.~\ref{sec:linearmomentum}.

The field theory predicts that the total charge in a region $W$ surrounding the dislocation is
\begin{equation}\label{eq:QWdisloc}
    Q_W = C\frac{\delta\Phi_{W,\text{o}}}{2\pi}+\frac{\overline{\mathscr{P}}_{\text{o}}}{2} + (k+n_{\text{irreg},\text{o}})\nu \mod 1,
\end{equation}
where we have written $n_{W,\text{o}} = k + n_{\text{irreg},\text{o}}$ for some integer $k$. Here $n_{\text{irreg},\OO}$ is the effective number of unit cells we assign to the irregular unit cell at the dislocation. We use the subscript $\text{o}$ in $n_{\text{irreg},\text{o}}$ because the value assigned to it will turn out to be different for different high symmetry points $\OO$.

Recall from the general discussion in Sec.~\ref{sec:C4_SO_disc} that if all regular unit cells have flux $\phi$, 
$$\delta \Phi_{W,\text{o}} = \phi_{\text{irreg}} - \phi n_{\text{irreg},\text{o}}$$ 
where $\phi n_{\text{irreg},\text{o}}$ is the background flux assigned to the dislocation plaquette. 

\begin{figure}[t]
    \centering
    \includegraphics[width=6cm]{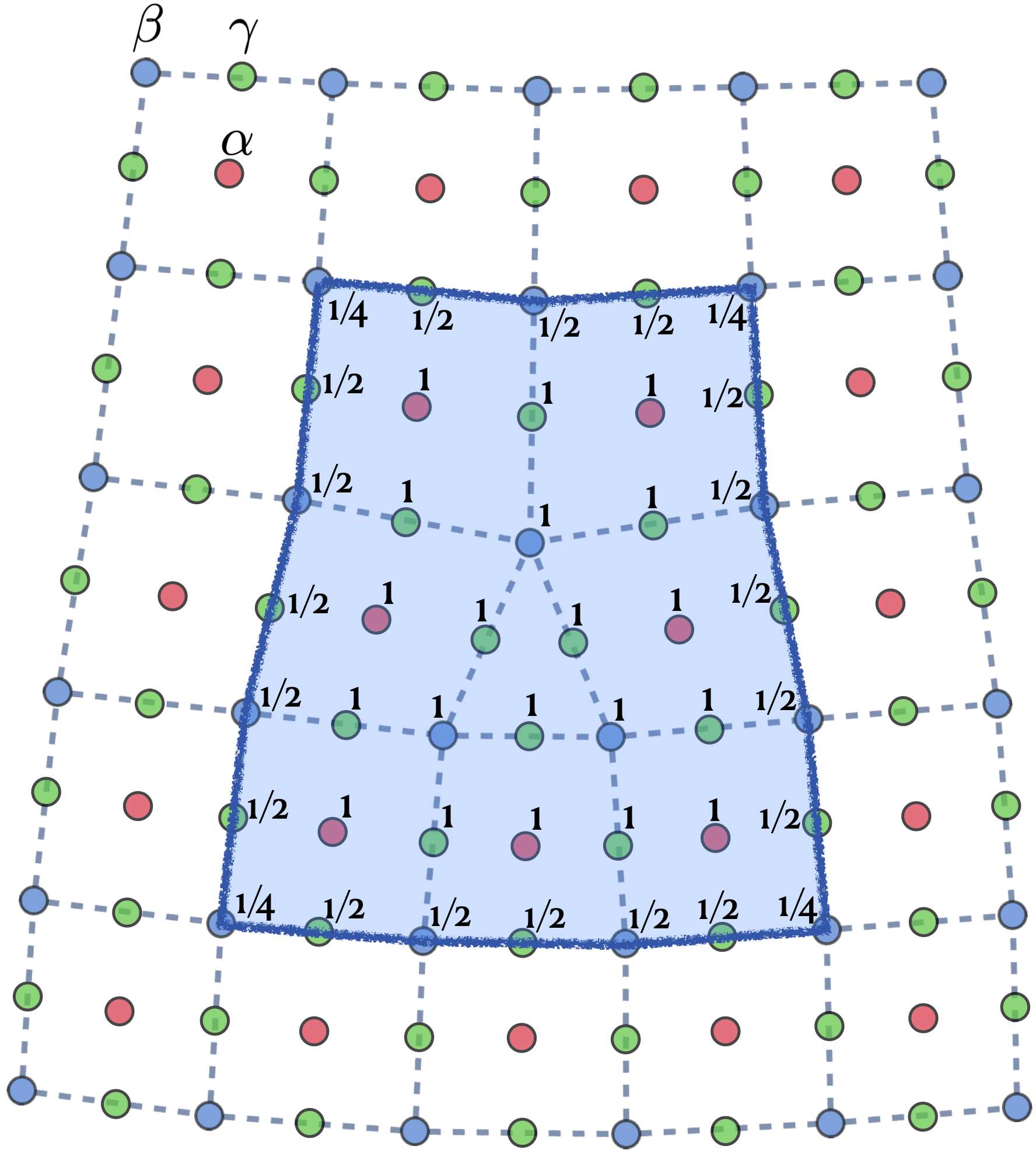}
    \caption{Region $W$ for a square lattice dislocation with $\vec{b}=(1,0)$. The weightings $\text{wt}(i)$ are labeled on each relevant MWPs. Dashed lines represent unit cell boundaries, and colored circles represent the MWPs $\alpha$ (red), $\beta$ (blue) and $\gamma$ (green).}
    \label{fig:dislocationW}
\end{figure}

Next we determine $n_{\text{irreg},\OO}$. First we consider $\OO = \beta$ (see Fig.~\ref{fig:dislocationW}).
The choice of $n_{\text{irreg},\beta}$ should satisfy a few sanity checks. First, it should lead to an integer-valued $\overline{\mathscr{P}}_{\beta}$. Numerically, only $n_{\text{irreg},\beta} = 0$ or $1/2$ give integer values of $\overline{\mathscr{P}}_{\beta}$; this can also be seen directly from Eq.~\eqref{eq:QWdisloc} by looking at the limit of zero Chern number and $\nu \in \Z$. 

Second, the system with the same Hamiltonian but with all orbitals filled has zero Chern number. It can be adiabatically connected to one in which the points in the $\alpha, \beta, \gamma$ maximal Wyckoff positions have integer charge $N_{\alpha}, N_{\beta}, N_{\gamma}$  respectively. In this situation, the polarization can also be defined using the dipole moment within each subcell, and it is possible to show, independent of any dislocation charge calculation, that
\begin{align}
   \overline{\mathscr{P}}_{\alpha} &= N_{\beta} + N_{\gamma} \mod 2; \label{eq:PAC4}\\
   \overline{\mathscr{P}}_{\beta} &= N_{\alpha} + N_{\gamma} \mod 2.\label{eq:PBC4}
\end{align}
See App.~\ref{app:C=0calcs} for the details. Now we show that Eq.~\eqref{eq:QWdisloc} only agrees with this result if we set $n_{\text{irreg},\beta} = \frac{1}{2}$.

To derive this explicitly, consider Fig.~\ref{fig:dislocationW}. We choose $W$ so that $\partial W$ overlaps with $\beta$ sites. Using our weighting procedure, we get
\begin{equation}
    Q_W = 7 N_{\alpha} + 7.5 N_{\beta}+ 15.5 N_{\gamma},
\end{equation}
while from the field theory, if we set $\OO=\beta$,
\begin{equation}
    Q_W = \kappa(7 + n_{\text{irreg},\beta}) + \frac{\overline{\mathscr{P}}_{\beta}}{2} \mod 1,
\end{equation}
where $\kappa = N_{\alpha} + N_{\beta} + 2N_{\gamma}$. Using Eq.~\eqref{eq:PBC4} in the above equations, and simplifying, we obtain $n_{\text{irreg},\beta} = 1/2 \mod 1$.

The above calculation was done in a limit with Chern number zero. We now assume that $n_{\text{irreg},\beta} = 1/2$ even when $C \ne 0$. This is reasonable because in the actual model under study, changing the Chern number should not change $n_{\text{irreg},\beta}$. After this step, we get the following prediction:
\begin{equation}\label{eq:dislocationCharge}
    Q_W = \frac{\overline{\mathscr{P}}_{\beta}}{2} + (k+\frac{1}{2})\nu +C\frac{\delta\Phi_{W,\beta}}{2\pi}\mod 1
\end{equation}
where $W$ encloses only one dislocation, with one triangular plaquette. Note that 
$$\delta \Phi_{W,\beta} = \phi_{\text{irreg}}-\frac{\phi}{2}$$
is completely well-defined mod $2\pi$ since $\phi$ is specified as a real number (see Sec.~\ref{sec:amb_lift}). This allows us to find $\overline{\mathscr{P}}_{\beta} \mod 2$ unambiguously.

Now let us discuss $\overline{\mathscr{P}}_{\alpha}$. Again consider Fig.~\ref{fig:dislocationW}. Let us use the exact same $W$ (i.e. the same unit cell choice), so 
\begin{equation}
    Q_W = 7 N_{\alpha} + 7.5 N_{\beta} + 15.5 N_{\gamma}.
\end{equation}
However, if we choose $\text{o}=\alpha$, the field theory now predicts that
\begin{equation}
    Q_W = \kappa(7 + n_{\text{irreg},\alpha}) + \frac{\overline{\mathscr{P}}_{\alpha}}{2}\mod 1.
\end{equation}
In the Chern number 0 limit, demanding consistency with Eq.~\eqref{eq:PAC4} implies that $n_{\text{irreg},\alpha} = 0 \mod 1$. 

We have thus found that $n_{\text{irreg},\text{o}}$ depends on whether $\text{o}$ lies at the corner or the center of a unit cell. The values of $n_{\text{irreg},\text{o}}$ are tabulated in Table \ref{table:nuc}. When we consider other values of $M$ in Sec.~\ref{sec:general}, we will again find that $n_{\text{irreg},\text{o}}$ must be fixed for each $\text{o}$ by requiring consistency with analytical results in the limit of full filling. 

\subsection{Application to Hofstadter model}

We apply Eq.~\eqref{eq:dislocationCharge} to the usual square lattice Hofstadter model, choosing $\alpha$ at a plaquette center and $\beta$ at a site as before. The Hofstadter butterfly for $\overline{\mathscr{P}}_{\beta}$ is plotted in Fig.~\ref{fig:SandP}. $\overline{\mathscr{P}}_{\beta}$ follows the empirical formula
\begin{equation}
\label{eq:t_formula}
\overline{\mathscr{P}}_{\beta}(\phi)=C\kappa\mod 2.
\end{equation}
Note that $\overline{\mathscr{P}}_{\beta}$ has period $4\pi$ in $\phi$ and not $2\pi$, because its definition involves the quantity $\phi/2$. (Also, if $\phi \rightarrow \phi + 2\pi$ for fixed $\nu$, Eq.~\eqref{eq:t_formula} changes by $C \mod 2$ so a shift of $4\pi$ is needed to leave it invariant.) We use Eq. \eqref{eq:t_formula}, along with an eigenvalue database \cite{osadchy2001db}, to generate the Hofstadter butterfly for $\overline{\mathscr{P}}_{\beta}$ in Fig. \ref{fig:SandP}.
We have only plotted it for $0<\phi\le 2\pi$. The values for $2\pi<\phi\le 4\pi$ can be obtained either using Eq. \ref{eq:t_formula} or by reflecting the butterfly about $\phi=2\pi$. 

Similarly, with $\alpha$ at a plaquette center, and $n_{\text{irreg},\alpha} = 0$, we find that
\begin{equation}\label{eq:t_formula2}
    \overline{\mathscr{P}}_{\alpha} = (C+1) \kappa \mod 2.
\end{equation}
$\overline{\mathscr{P}}_{\alpha}$ has period $2\pi$ in $\phi$, and is also plotted in Fig.~\ref{fig:SandP}. Shifting $\phi \rightarrow \phi+2\pi$ for fixed $\nu$ changes Eq.~\eqref{eq:t_formula2} by $C(C+1) \mod 2$, but as $C(C+1)$ is even, this change is trivial.

Finally, note that there is a general relation between $\overline{\mathscr{P}}_{\alpha}$ and $\overline{\mathscr{P}}_{\beta}$ which goes beyond just the Hofstadter model:
\begin{equation}\label{eq:pApB}
    \overline{\mathscr{P}}_{\alpha} - \overline{\mathscr{P}}_{\beta} = \kappa \mod 2.
\end{equation}
Eq.~\eqref{eq:pApB} can be proven from field theory; we do this in Sec.~\ref{app:ftrelabel}.

\subsection{$\PO$ from linear
momentum}\label{sec:linearmomentum}

On a closed manifold, the field theory term $\frac{\PO}{2\pi} \cdot A \wedge \vec{T}$ can be rewritten as $\frac{\PO}{2\pi} \cdot \vec{R} \wedge F$, plus some additional terms, where $F$ is the total magnetic flux, and $\vec{R} = (X,Y)$ is the translation gauge field mentioned below Eq.~\eqref{eq:chargeresponse}.
Consider the linear momentum in, say, the $y$ direction associated to a state with total flux $2\pi m = \phi L_x L_y$. The linear momentum of the inserted flux, which is the charge under the translation gauge field, is predicted to receive a contribution 
\begin{equation}
    \frac{\delta \mathcal{L}}{\delta Y_0} =  \frac{\mathscr{P}_{\Oy}}{2\pi}\int F = \mathscr{P}_{\Oy} m = \frac{1}{2}\overline{\mathscr{P}}_{\text{o}} m \mod 1
\end{equation}
from this term. The mod 1 normalization is due to the convention chosen to define $Y$. We now verify this prediction from numerical calculations on the Hofstadter model for $\OO = \alpha,\beta$ by studying the system on a torus.

On a torus with magnetic flux $2\pi m = \phi L_x L_y$, a translation by ${\bf y}$ cannot be an exact symmetry of the Hamiltonian in general, even after applying a gauge transformation. This is because, for a fixed $x$, the holonomy $\oint A_x dx$ along a noncontractible cycle of the torus changes by $\frac{2\pi m}{L_y}$ under $\tilde{T}_{\bf y}$. If $\frac{m}{L_y} \in \Z$, $\tilde{T}_{\bf y}$ is an exact symmetry because the change in $\oint A_x dx$ can be undone by a large gauge transformation. For other values of $m$, $\tilde{T}_{\bf y}$ needs to be accompanied by an operator $\mathcal{F}_{\frac{2\pi m}{L_y}}$ which adiabatically inserts flux $\frac{2\pi m}{L_y}$ through the cycle running along $x$, to make it an exact symmetry. In our work we do not use the exact translation operator because it is difficult to numerically implement $\mathcal{F}_{\frac{2\pi m}{L_y}}$.

If the value of $|m| \mod L_y$ is of order 1, the inserted flux from $\mathcal{F}_{\frac{2\pi m}{L_y}}$ is $O(1/L_y)$, which vanishes in the thermodynamic limit; therefore
$\tilde{T}_{\bf y} \simeq   \tilde{T}_{\bf y} \mathcal{F}_{\frac{2\pi m}{L_y}}$ is a good approximation \cite{Song2021polarization}.
For general values of $m$, we find that the expectation value of $\tilde{T}_{\bf y}$ oscillates between 0 and 1 in amplitude as a function of $m$ (indicating the closeness of the approximation).

Apart from the flux in each unit cell, the vector potential has another gauge invariant quantity called the `gauge origin' $\overline{\OO}$, which we define below. The operator $\tilde{T}_{\bf y}$ also requires us to fix a gauge transformation $\{\lambda_j\}$. The main result of this section is that for a fixed system size and a fixed choice of $\OO$, we find only one choice of $\overline{\OO}$ and $\lambda$ for which $\PO$ is quantized throughout the Hofstadter model. And remarkably, this value of $\PO$ agrees with results from the dislocation charge calculation. 

We find that we can similarly measure $\PO$ using the expectation value of a partial translation operator $\tilde{T}_{\bf{y}}|_D$ (again not an exact symmetry), which is $\tilde{T}_{\bf{y}}$ restricted to an appropriate region $D$. Further details of the linear momentum calculations can be found in App.~\ref{app:linmom_details}.

\subsubsection{Definition of vector potential}

We assume, as in the charge calculation, that the holonomies of the vector potential $A$ can be specified along any loop in the continuum in which the lattice is embedded. We define a gauge-invariant point, the `gauge origin' $\overline{\OO} = (\overline{\OO}_x,\overline{\OO}_y)$, such that the holonomy of $A$ is zero on the $x$ and $y$ cycles of the torus that intersect $\overline{\OO}$. Note that $\overline{\OO}$ need not correspond to a lattice site in general. Vector potentials with different values of $\overline{\OO}$ are not gauge equivalent, and will be distinguished in the treatment below.

\begin{figure}[t]
    \centering
    \includegraphics[width=7.5cm]{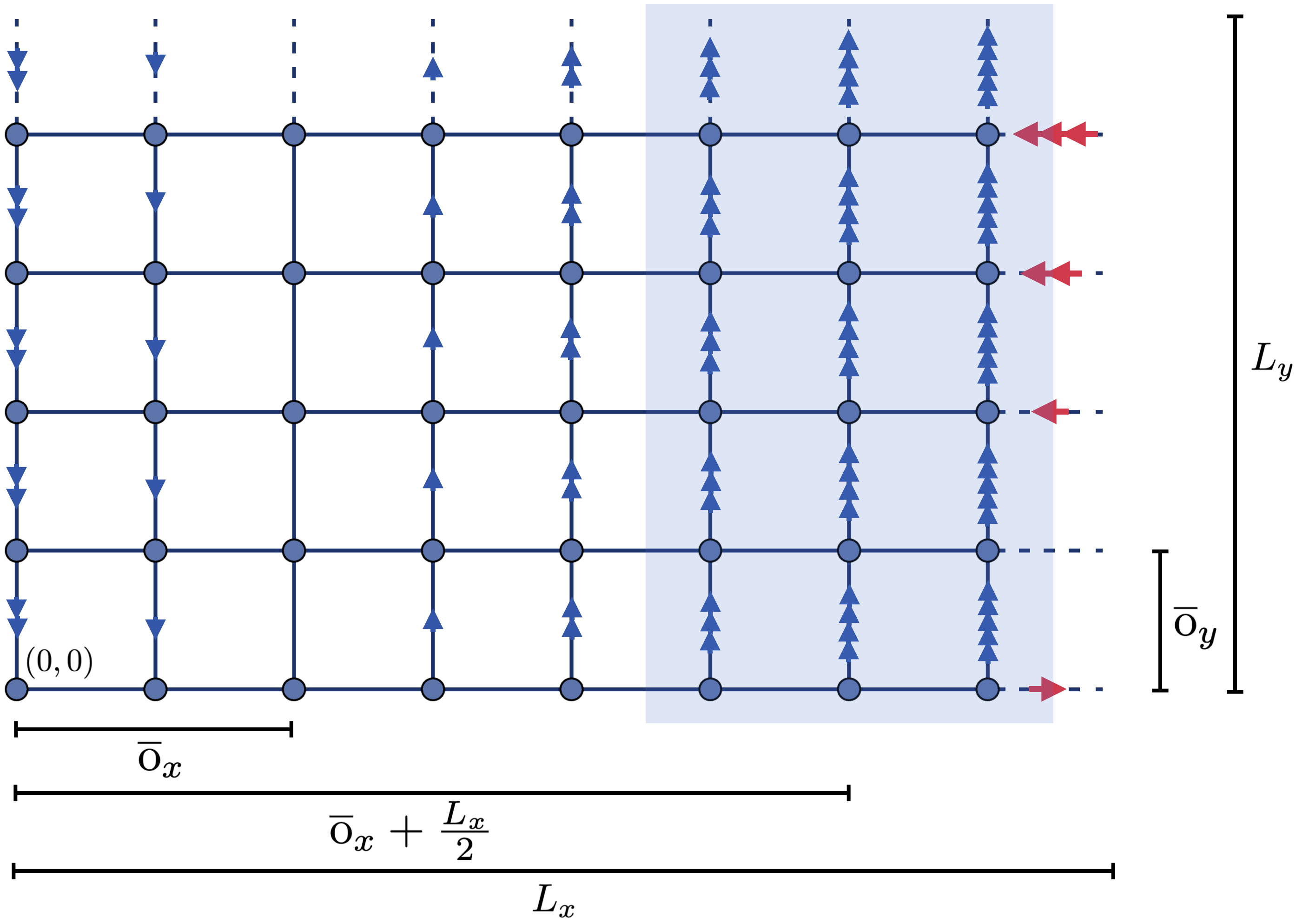}
    \caption{Gauge choices defined in Eq.~\eqref{eq:generalgauge} on an $L_x \times L_y$ torus. Each blue arrow represents a vector potential $A_{ij}=\frac{2\pi}{L_xL_y}$. Each red arrow represents a vector potential $A_{ij}=\frac{2\pi}{L_y}$. $\overline{\text{o}}_x$ and $\overline{\text{o}}_y$ mark the distance between $(0,0)$ and the cycle with trivial holonomy in the $x$ and $y$ directions respectively. The blue region is the partial translation region $D$ which is centered around $\overline{\OO}_x + \frac{L_x}{2}$.}
    \label{fig:gauge}
\end{figure}

To measure $\mathscr{P}_{\Oy}$ in our numerics, we insert a total of $m$ flux quanta on the torus using a Landau-like gauge along $y$. One form of this gauge, shown in Fig. \ref{fig:gauge}, is
\begin{equation}\label{eq:generalgauge}
    \begin{aligned}
        A_{j,j+\hat{x}}&=-\frac{2\pi (y-\overline{\text{o}}_y)m}{L_y}\delta_{j_x,L_x-1}\\
        A_{j,j+\hat{y}}&=\frac{2\pi (x-\overline{\text{o}}_x )m}{L_x L_y}\\
        j=(j_x,j_y),~ &j_x\in \{0,\dots L_x-1\}, j_y\in\{0,\dots L_y-1\}.
    \end{aligned}
\end{equation}
Here, $m\equiv\frac{\phi}{2\pi} L_xL_y$ is the total number of flux quanta through the torus. $\OO$ can be determined from $\overline{\text{o}}$, as we explain in the subsequent section. 

To measure $\mathscr{P}_{\Ox}$, we define a similar gauge, but instead along $x$.

\subsubsection{Approximate translation operator}

Next we define the operator
\begin{equation}
    \tilde{T}_{\bf y} := \hat{T}_{\bf y}e^{i \sum_j \lambda_j c_j^{\dagger} c_j}.
\end{equation}
$\lambda_j$ is a function of $m$. If $\tilde{T}_{\bf y}$ were an exact symmetry, there would be a set $\{\lambda_j\}$ satisfying
\begin{equation}\label{eq:Ty-lambda}
    A_{i+\hat{y}, j + \hat{y}} = A_{ij} - \lambda_i + \lambda_j
\end{equation}
everywhere on the torus. But as mentioned above it is not possible to have exact symmetry when $m/L_y \notin \Z$. Thus we can at best ensure that this relation holds everywhere except for a small strip that we require to be centered at a fixed $x$. Then the freedom in choosing $\lambda$ is completely fixed up to (i) an overall shift $\lambda_j \rightarrow \lambda_j + \chi$ for each $j$, (ii) the geometry of the strip, including the position of its center and its thickness, and (iii) the choice of $\lambda$ for points that lie within the strip (if any). We will discuss these further below. 

In our numerics we use the gauge
\begin{align}
\label{eq:tyfullgaugeapp}
    \tilde{T}_{\bf{y}}=\begin{cases}
      \hat{T}_{\bf{y}}e^{i\sum_j -\frac{\pi m}{L_y} c_j^{\dagger}c_j}\qquad   &j_x< \overline{\OO}_x\\
      \hat{T}_{\bf{y}}e^{i\sum_j \frac{\pi m}{L_y} c_j^{\dagger}c_j}\qquad   &j_x> \overline{\OO}_x\\
      \hat{T}_{\bf{y}}\qquad &j_x=\overline{\OO}_x.
    \end{cases}
\end{align}
Here $j_x$ is the $x$ coordinate of the site $j$. For this $\tilde{T}_{\bf y}$, Eq.~\eqref{eq:Ty-lambda} is violated only on the horizontal links that touch the line $j_x = \overline{\OO}_x$.

To measure $\mathscr{P}_{\Oy}$, we need to express $\OO$ in terms of gauge-invariant properties of $A$ or $\lambda$. If we fix $A,\lambda$ as above, we can define $\OO$ as follows. First we find the flux $\phi_{\text{irreg}}$ in a dislocation plaquette created using $\tilde{T}_{\bf y}$. Recall from Sec.~\ref{sec:c4_PO_DC} that $\phi_{\text{irreg}}$ depends on $A$ and $\lambda$ as well as the position of the dislocation, which is fixed by a point $j_0$; and if we shift $j_0$ by an integer vector, $\phi_{\text{irreg}}$ can change by multiples of $\phi$. For the above choices of $A$ and $\lambda$, we thus define $\OO$ so as to satisfy the following relation, with $j_{0,x} \in \Z$:
\begin{equation}\label{eq:lm_odef}
    \phi_{\text{irreg}} = \phi n_{\text{irreg},\OO} \mod \phi,
\end{equation}
where $n_{\text{irreg},\OO}$ was computed in Sec.~\ref{sec:c4_PO_DC}. In the language of that section, our choice of $\OO$ ensures that the excess flux $\delta \Phi_{W,\OO}$ in the dislocation plaquette constructed using $\tilde{T}_{\bf y}$ is zero (mod $\phi$). Note that for an arbitrary choice of $\phi_{\text{irreg}}$, there may be no solution corresponding to any high symmetry point $\OO$. But in App.~\ref{sec:dislocationConstruction}  we prove that for our choice of $A,\lambda$, Eq. \ref{eq:lm_odef} can be solved and implies
\begin{equation}\label{eq:generalobar}
\OO_x=\overline{\OO}_x+\frac{L_x}{2}+\frac{1}{2} \mod 1.
\end{equation}

Having fixed $\lambda_j$, we define the linear momentum $p_{\lambda,y}$ as the expectation value of $\tilde{T}_{\bf{y}}$ in the ground state $\ket{\Psi}$ with $m$ total flux quanta:
\begin{equation}\label{eq:pyD}
    \bra{\Psi} \tilde{T}_{\bf{y}}\ket{\Psi} :=e^{ - \gamma + i2\pi p_{\lambda,y}}.
\end{equation}
$e^{-\gamma}$ is the amplitude of the expectation value. Empirically we find that
\begin{equation}\label{eq:pOdef}
p_{\lambda,y} = -\mathscr{P}_{\Oy} m+ K_y \mod 1
\end{equation}
The term linear in $m$ is predicted by the response theory; $K_y$ is piecewise constant in $m$ in our numerics, and can change only when the amplitude vanishes. The numerical details are shown in Fig.~\ref{fig:bare_momentums} 

Using the above choice of $\lambda$ in the Hofstadter model, we find that the amplitude $e^{-\gamma}$ oscillates with $m$ whenever $C \ne 0$, and for a discrete set of $m$ values it vanishes.\footnote{For example, when $L_y$ is even, these are given by $m=\frac{L_y}{2}+k L_y$, where $k\in \Z$. This verifies our earlier comment that $\tilde{T}_{\bf y}$ is a worse approximate symmetry when $|m| \mod L_y$ is large.} For other values of $m$, the magnitude is non-vanishing and the linear momentum is well-defined. As a result, for a fixed $L_x$, we can obtain the expected value of $\mathscr{P}_{\Oy}$ throughout the Hofstadter butterfly except for a finite set of $\phi$ values where $\mathscr{P}_{\Oy}$ is not determined. But for these $\phi$ values, we can pick some other $L_x$ and then extract $\mathscr{P}_{\Oy}$. The values of $\mathscr{P}_{\Oy}$ are quantized and agree with results from the dislocation charge measurement.

Now we discuss alternative choices of $\lambda$. First there is the freedom in shifting $\lambda_j \rightarrow \lambda_j + \chi$ for each $j$. As stated in Sec.~\ref{sec:c4_PO_DC}, taking $\chi = \phi v_x = \frac{2\pi m v_x}{L_x L_y}$ changes the flux $\phi_{\text{irreg}}$ assigned to an irregular unit cell at a dislocation by $-\phi v_x$ (where $v_x \in \mathbb{R}$). If we continue to define $\OO$ using Eq.~\eqref{eq:lm_odef}, then when $v_x$ is an arbitrary real number there is no solution for $\OO$. But if $v_x$ is quantized to 0 or 1/2 mod 1, we see that $\OO$ must change (in fact, we show in App.~\ref{sec:dislocationConstruction} that $\OO_x \rightarrow \OO_x - v_x$). Hence, only discrete sets of global $U(1)$ transformations are allowed, and the effect of such transformations is simply to change the value of $\OO$. 

There is additional freedom in choosing the strip that violates translation symmetry. The location of the center of the strip can in principle be shifted by $\delta \in \mathbb{R}$ if we define $\lambda_j$ as in Eq.~\eqref{eq:tyfullgaugeapp_delta} (see Appendix). In the Hofstadter model, we find empirically that in order to get any quantized result for $\mathscr{P}_{\Oy}$ throughout the butterfly, $\delta$ must be fixed so that the center of the strip coincides with $\overline{\OO}_x$, as in Eq.~\eqref{eq:tyfullgaugeapp}, i.e. $\delta=0$. Remarkably, the quantized $\PO$ extracted from the above linear momentum calculation agrees fully with the result from dislocation charge calculations.  

When $C=0$ in this model, taking $\delta \ne 0$ is equivalent to taking $\lambda_j \rightarrow \lambda_j + \chi$ for some $\chi$.\footnote{A quick argument is that in this limit, $\ket{\Psi(m)}$ is an almost exact eigenstate of $\hat{n}_j$ with eigenvalue $\kappa$, for each $j$. Thus $\tilde{T}_{\bf y} = \hat{T}_{\bf y} e^{i \kappa \sum_j \lambda_j}$. Therefore any change in $\lambda_j$ (e.g. by redefining $\delta$) is equivalent to shifting $\lambda_j \rightarrow \lambda_j + \chi$ for each $j$ with some suitable $\chi$.} In this case any choice of $\delta$ will give a quantized result $\mathscr{P}_{\OO',y}$, but for a high symmetry point $\OO'$ that depends on $\delta$. In App.~\ref{app:linmom_details} we discuss an alternative way to understand the choice of $\delta$ using gauge-invariant quantities.

Empirically we find that the thickness of the strip does not matter, as long as $\lambda_j$ varies linearly in $j_x$ within the strip (otherwise we do not obtain quantized results throughout the butterfly for $\mathscr{P}_{\OO,y}$). We can pick the parameters $\{L_y, \overline{\OO}_y\}$ arbitrarily, and this does not affect the result for $\mathscr{P}_{\Oy}$. In the limit of zero strip thickness, Eq.~\eqref{eq:tyfullgaugeapp} is the only choice of $\lambda$ that we have found that gives quantized results throughout the butterfly.

\subsubsection{Partial translations}

We can also use a partial translation operator $\tilde{T}_{\bf y}|_D := \hat{T}_{\bf{y}}|_D e^{i \sum_{j \in D} \lambda_j c_j^{\dagger} c_j}$ to extract the same linear momentum. To get quantized results for $\PO$, we empirically find that $D$ needs to be mirror symmetric about some cycle $\ell$ along which the holonomy $e^{i \sum_{j \in \ell} A_{j,j+\hat{y}}}$ is real valued.   

Here we define the linear momentum of a state with $m$ flux quanta as
\begin{equation}
    \bra{\Psi} \tilde{T}_{\bf{y}}|_D\ket{\Psi} :=e^{ - \gamma_D + i 2\pi p_{\lambda,y}}.
\end{equation}
Numerically, we obtain $\overline{\mathscr{P}}_{\text{o}}$ through the equation
\begin{equation}
    p_{\lambda,y} =-\mathscr{P}_{\Oy}m+K_{y} \mod 2.
\end{equation}
Here $K_{y}$ is some constant which only depends on the Chern number and the system size.

We find that there are only two choices of the parameters $\{D,\overline{\OO},\lambda\}$ which give a quantized $\mathscr{P}_{\Oy}$ throughout the butterfly. In one case,
\begin{enumerate}
    \item $D$ is centered at the cycle with $x=\overline{\text{o}}_x$ (trivial holonomy)
     \item $\overline{\text{o}}$ is arbitrary
    \item $\tilde{T}_{\bf{y}}|_D:= \hat{T}_{\bf{y}}|_D$, i.e. $\lambda_j \equiv 0$ for $j \in D$.
\end{enumerate}
However, empirically these choices give $\mathscr{P}_{\Oy} = 0$ everywhere in the butterfly,  which does not agree with the results obtained from other methods. 

In the second case, we choose
\begin{enumerate}
    \item $D$ centered at the cycle with $x=\overline{\text{o}}_x-\frac{L_x}{2}$ (holonomy $(-1)^m$), for each $m$ 
    \item $\overline{\text{o}}=\text{o}+\frac{L_x}{2}+(\frac{1}{2},\frac{1}{2})$
    \item $\tilde{T}_{\bf{y}}|_D$ is defined as
    \begin{align}\label{eq:tygauge}
    \tilde{T}_{\bf{y}}|_D :=\begin{cases}
      \hat{T}_{\bf{y}}|_De^{i\sum_j -\frac{\pi m}{L_y} c_j^{\dagger}c_j}\qquad   &j_x\le \overline{\OO}_x\\
      \hat{T}_{\bf{y}}|_De^{i\sum_j \frac{\pi m}{L_y} c_j^{\dagger}c_j}\qquad   &j_x> \overline{\OO}_x.
    \end{cases}
\end{align}
\end{enumerate}
This choice of $\lambda,\overline{\OO},D$ is the only one we have found that gives a quantized but non-trivial result for $\mathscr{P}_{\Oy}$ throughout the butterfly. Furthermore, remarkably this agrees with the other methods used in this section (dislocation charge, linear momentum from full translations, and the 1d polarization discussed below). We motivate this choice further in App.~\ref{app:linmom_details}.

\subsection{$\vec{\mathscr{P}}_{\text{o}}$ from 1d polarization}\label{sec:c4_1dpolarization}

It is natural to ask what the invariant $\vec{\mathscr{P}}_{\text{o}}$ as defined above has to do with other traditional many-body definitions of the polarization. In this section we provide one concrete answer: we consider a torus and compute the 1d polarization along $x$ as defined by Resta \cite{resta1994}, treating $L_y$ as a parameter. We show that in addition to the usual term proportional to $C$, there is a term $L_y \mathscr{P}_{\Oy}$ which fixes the dependence of the 1d polarization on the dimensionally reduced coordinate.

Consider the 1d polarization in the $x$ direction, denoted $\mathcal{P}_x$. It is defined by the following action: 
\begin{equation}\label{eq:L1d}
    \mathcal{L}_{1d} = -\mathcal{P}_x \int\limits dx dt E_x,
\end{equation}
where $E_x = \partial_t A_x - \partial_x A_t$ is the $x$ component of the electric field. In this section we assume that $A$ denotes the entire vector potential and not just its deviation from some background. The sign in Eq.~\eqref{eq:L1d} is chosen so that the 1d current $j$ satisfies $j = \frac{\delta \mathcal{L}_{1d}}{\delta A_t} = \partial_t \mathcal{P}_x$. 

We now use the (2+1)D field theory to make a prediction for $\mathcal{P}_x$ if the original 2d system is dimensionally reduced to an effective 1d system in the $x$ direction. We set the rotation gauge field $\omega$ to zero for this calculation. First consider the term with $C$:
\begin{equation}
    \begin{aligned}
           &\frac{C}{4\pi}\int\limits dx dy dt  A \wedge dA\\
   &= \frac{C}{4\pi}\int\limits dx dy dt (A_t B - A_x E_y + A_y E_x). 
    \end{aligned}
\end{equation}
Here $E_x,E_y,B$ are the full electric and magnetic fields respectively. Note that the Chern-Simons term is usually written in terms of the deviation of the vector potential from some background, but we have instead used the full vector potential. This rewriting is motivated by the topological field theory, derived in App.~\ref{sec:field}, and will be justified by the empirical results we show below. 

If $A_y \ne 0$ and $E_x$ is independent of $y$, then we can rewrite this term as
\begin{align}
     \frac{C}{2\pi} \int\limits dx dt (\oint A_y(x) dy) E_x + \dots 
\end{align}
This step involves an integration by parts, which contributes a factor of 2.
Since the system on the torus is not exactly translationally symmetric, the holonomy $\Phi_y(x) = \oint A_y(x) dy$ has an $x$-dependence. In order to extract a spatially averaged polarization, which is what we calculate microscopically using Resta's formula below, we assume $E_x$ is a constant. Then the above term becomes
\begin{align}
     \frac{C}{2\pi} \Phi_{y,\text{avg}}\int\limits dx dt E_x + \dots
\end{align}
Here, $\Phi_{y,\text{avg}} := \frac{1}{L_x}\int dxdy A_y$ is the average of the global holonomies in the $y$ direction.  Thus the Chern number contributes to the 1d polarization. The value of $\Phi_{y,\text{avg}}$ depends on the range of integration for $x$. If we assume $\mathcal{O}_x \le x \le L_x + \mathcal{O}_x$ for some $\mathcal{O}_x$ and $\Phi_y(x)$ is continuous in this range, we can show that 
\begin{equation}
    \Phi_{y,\text{avg}}=\phi L_y (\frac{L_x}{2}-\overline{\OO}_x+\mathcal{O}_x) \mod 2\pi.
\end{equation}

Now notice that there is also a contribution from the field theory term $Y \wedge dA$:
\begin{equation}
    \begin{aligned}
            &\frac{\mathscr{P}_{\OO,y}}{2\pi} \int\limits dx dy dt  Y \wedge dA \\
            &= \frac{\mathscr{P}_{\OO,y}}{2\pi}\int\limits dx dy dt (Y_t B - Y_x E_y + Y_y E_x) \\
    &= \frac{\mathscr{P}_{\OO,y}}{2\pi}\oint\limits dy Y_y \int\limits dx dt E_x + \dots \\
   &=  \mathscr{P}_{\OO,y}L_y \int\limits dx dt E_x + \dots
    \end{aligned}
\end{equation}
Note that due to a normalization convention, $\oint\limits dy Y_y = 2\pi L_y$. The remaining terms in the field theory do not contribute. Thus we naively expect that the 1d polarization should be equal to
\begin{equation}\label{eq:pol1Dold}
-\mathcal{P}_{\mathcal{O},x}=\frac{C \phi}{2\pi} L_y (\frac{L_x}{2}-\overline{\OO}_x+\mathcal{O}_x) +L_y\mathscr{P}_{\OO,y}+K' \mod 1,
\end{equation}
where we assume $K$ is a constant that is independent of $L_x$ and $L_y$. We have used the subscript $\mathcal{O}$ in $\mathcal{P}_{\mathcal{O},x}$ to specify the dependence of this quantity on $\mathcal{O}_x$. As we now discuss, to match Eq.~\eqref{eq:pol1Dold} to a numerical calculation, we demand that $\mathcal{O}_x$ be the origin of coordinates in that calculation.

Numerically we compute the quantity $\mathcal{P}_{\mathcal{O},x}$ using Resta's formula:
\begin{equation}
        \mathcal{P}_{\mathcal{O},x} = \frac{1}{2\pi}\text{arg } \langle \Psi | e^{i \frac{2\pi}{L_x} \sum_{j} j_x c_{j}^{\dagger}c_{j}} |\Psi \rangle
\end{equation}
where $j_x$ is defined w.r.t. the origin $\mathcal{O}\equiv(\mathcal{O}_x,\mathcal{O}_y)$, i.e. $j_x\in\{-\mathcal{O}_x,1-\mathcal{O}_x ,2-\mathcal{O}_x,\dots, L_x-1-\mathcal{O}_x\}$. Here $\mathcal{O}$ refers to the point on the torus where the coordinate is chosen to be $(0,0)$.

Empirically, we find that $\mathcal{P}_{\mathcal{O},x}$ exactly obeys Eq.~\eqref{eq:pol1Dold}.
Let us take a square lattice with $L_x=20$, $L_y=12,13$ and $\mathcal{O}_x=0,\frac{1}{2}$ as an example. The constant in $\phi$ contribution $L_y\mathscr{P}_{\OO,y}+K’$ for the two different $L_y$ is plotted in Fig.~\ref{fig:1dpol}. Empirically we find that the difference of the two gives the desired answer
$\mathscr{P}_{\text{o},y}$ when 

\begin{equation}\label{eq:1dorelation}
    \text{o}_x=\mathcal{O}_x+\frac{L_x}{2}+\frac{1}{2} \mod 1.
\end{equation}
The presence of the $\frac{L_x}{2}+\frac{1}{2}$ term may seem mysterious, but we show why it should be present in the $C=0$ case below. 

Consider $\mathcal{P}_{\mathcal{O},x}$ for a square lattice with one orbital per site and at full filling $\nu=1$. We analytically calculate $\mathcal{P}_{\mathcal{O},x}|_{\nu=1}$ using Resta's formula:

\begin{figure}[t]
    \centering
    \includegraphics[width=8cm]{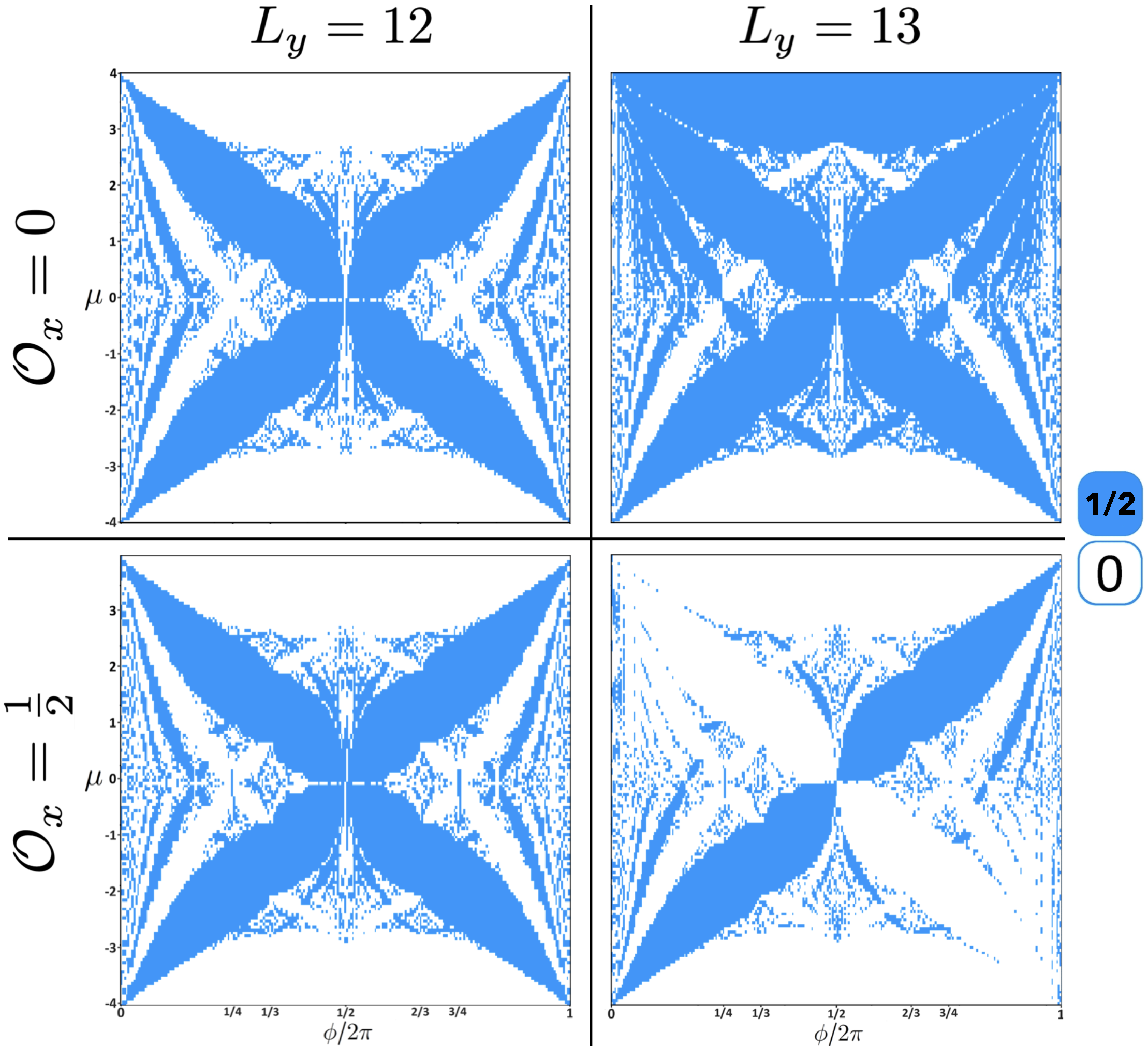}
    \caption{Butterflies showing the constant in $\phi$ contribution $L_y\mathscr{P}_{\OO,y}+K’$ of the 1d polarization $\mathcal{P}_{\mathcal{O},x}$, for different $\mathcal{O}_x$ and $L_y$. Fixing $L_x=20$, the result depends only on the parity of $L_y$. For a fixed $\mathcal{O}$, $\mathscr{P}_{\OO,y}$ is the difference between the $\phi=0$ intercept for $L_y=13$ and $L_y=12$. $\mathcal{O}$ and $\OO$ satisfy the relation Eq.~\eqref{eq:1dorelation}. For $\OO_x=\frac{1}{2}$, only half of the period $0<\phi<2\pi$ is plotted.}
    \label{fig:1dpol}
\end{figure}

\begin{equation}
\begin{aligned}
        \mathcal{P}_{\mathcal{O},x}|_{\nu=1}
        &=\sum_{r}\frac{j_x}{L_x} \mod 1\\
        &=\frac{L_x(L_x-1-2\mathcal{O}_x)}{2}\frac{L_y}{L_x} \mod 1\\
        &=L_y(\frac{L_x}{2}-\frac{1}{2}-\mathcal{O}_x) \mod 1.
\end{aligned}
\end{equation}
The coefficient of $L_y$ equals $-\mathscr{P}_{\Oy}$. In this special case $\mathscr{P}_{\Oy}$ just equals $\OO_x \mod 1$ (zero at a site, and 1/2 at a plaquette center). By equating these two expressions for $\mathscr{P}_{\Oy}$, we get Eq.~\eqref{eq:1dorelation}.
We expect that when $C\neq 0$ Eq.~\eqref{eq:1dorelation} will continue to hold, and we have verified this expectation numerically.

The relationship between $\mathcal{O}$ and $\OO$ is reminiscent of the linear momentum calculation in which one has to shift the gauge origin $\overline{\OO}$ by $(\frac{L_x}{2},\frac{L_y}{2})+(\frac{1}{2},\frac{1}{2})$ relative to $\text{o}$ in order to obtain $\PO$. We have numerically checked that Eq.~\eqref{eq:pol1Dold} gives consistent answers for different gauge choices and for different system sizes.

\section{Calculations for general $M$}\label{sec:general}

We now generalize our charge response, linear momentum, and 1d polarization calculations. We discuss $M=2,3,4,6$ together, and assume that $\text{o}$ can be at any HSP within the unit cell. We keep the choice of unit cell $\Theta$ arbitrary. We will focus on the details that differ from those outlined previously when $M=4$.

Recall that $M'$ is the order of the rotation symmetry which preserves $\OO$, while $M$ is the largest possible value of $M'$ considering all possible HSPs $\OO$. For the computations below we will need the $M'$-fold rotation operator $\tilde{C}_{M',\OO^*}$ defined so that $(\tilde{C}_{M',\OO^*})^{M'}=1$.

\subsection{ $\mathscr{S}_{\text{o}}$ from pure disclination charge}

We can use the operator $\tilde{C}_{M',\OO^*}$ to create a pure disclination at the high symmetry point $\OO^*$ and calculate its charge response. As previously discussed, the origin $\OO$ used to measure the Burgers vector of the disclination need not be equal to $\OO^*$. And by pure disclination, we mean that $\vec{b}_{\text{o}} \simeq (0,0)$. In this case the field theory predicts that 
\begin{equation}\label{eq:generalpuredisclinationcharge}
\begin{aligned}
        Q_W&=C\frac{\delta\Phi_{W,\text{o}}}{2\pi}+\frac{\Omega_W \mathscr{S}_{\text{o}}}{2\pi}
        + \nu(k+n_{\text{irreg},\text{o},\Omega}) \mod 1,
\end{aligned}
\end{equation}
where $k$ is an integer. Note that the field theory is only sensitive to $\OO$ and not $\OO^*$; it cares only about the measured value of the Burgers vector and not about how the defect was created microscopically.

 The computation of $\delta\Phi_{W,\OO}$ and then $\SO$ is done according to the procedure discussed in Sec.~\ref{sec:C4_SO_disc}. As we discussed there, a crucial detail is the area of the irregular unit cell at the center of the disclination, which we here call $n_{\text{irreg},\text{o},\Omega}$. We have included a subscript $\Omega$ because in general, this number also depends on the disclination angle $\Omega$. $n_{\text{irreg},\text{o},\Omega}$ is defined as

\begin{equation}\label{eq:uco}
    n_{\text{irreg},\text{o},\Omega}\equiv\begin{cases}
    (1-\frac{\Omega}{2\pi}) \quad\text{if } \text{o} \text{ is at the center}\\
    \hspace{1.7cm}\text{of the unit cell } \\
    0 \quad\hspace{1.1cm}\text{otherwise}.
    \end{cases}
\end{equation}
This can be intuitively understood as follows. If the origin is at the center of the unit cell, the disclination construction process would remove a fraction $\frac{\Omega}{2\pi}$ of the central unit cell (i.e. $\frac{M\Omega}{2\pi}$ subcells). If $\OO$ is not at the center of the unit cell, then it must be at the boundary of the unit cell. In this case, there is no irregular unit cell, thus $n_{\text{irreg},\OO,\Omega}=0$. Note that when $M =2,3,4$, we can always choose $\text{o}$ to be on the boundary of the unit cell, so that there are no irregular unit cells. On the other hand, this definition of $n_{\text{irreg},\text{o},\Omega}$ is needed especially for $2\pi/6$ disclinations when $M=6$, because the only $C_6$ symmetric points are at the unit cell centers, so in any $2\pi/6$ disclination there will necessarily be an irregular unit cell. 

\subsubsection{Comments on definition of $n_{\text{irreg},\OO}$}

In the discussion above we emphasize that $n_{\text{irreg},\text{o},\Omega}$ depends on the relative position between $\text{o}$ and the unit cell $\Theta$, and not the absolute position of either. We can pick a different unit cell $\Theta$ which changes $Q_W$ and $n_{W,\text{o}}$ simultaneously, leaving $\mathscr{S}_{\text{o}}$ invariant. We discuss this in detail in App.~\ref{sec:trim}. 

An important general point is that our definitions of $n_{\text{irreg},\text{o},\Omega}$ (and a similar quantity $n_{\text{irreg},\text{o},\vec{b}}$ defined in the next section) are only sensitive to the value of $M$ and not to the actual fine structure of the microscopic lattice. Conceptually, we can imagine tiling the plane with unit cells (square for $M=4$, hexagonal for $M=3,6$, and so on), and the only constraint is that the centers and corners of these unit cells need to be at appropriate high symmetry points of the microscopic lattice. In particular, the tiling does not have to match the structure of hopping or interaction terms in the microscopic Hamiltonian. Then, we note that the number of unit cells in any region $W$ is a property of the tiling alone. Similarly, we can apply a cut and glue procedure on the infinite plane tiling to get a tiling for a surface with a dislocation or disclination defect. Then the quantities $n_{\text{irreg},\text{o},\Omega}$ and $n_{\text{irreg},\text{o},\vec{b}}$ are properties of this defect tiling alone. As such, these are independent of microscopic details of the Hamiltonian such as the distribution of sites and the hoppings between them. The fact that our prescription depends on the tiling rather than on microscopic details such as hopping and interaction terms makes it readily generalizable beyond the nearest neighbor Hofstadter models that we have mainly studied in this work.    

\subsection{$\vec{\mathscr{P}}_{\text{o}}$ from dislocation charge}

We next calculate $\vec{\mathscr{P}}_{\text{o}}$ from the charge response of a dislocation (assume $\Omega = 0$).
If a defect has a nontrivial Burgers vector $\vec{b}$, it will generally also have an irregular unit cell. This irregular unit cell is triangular if $M=2,4$ and quadrilateral if $M=3,6$. In this case, we find that the area which should be assigned to the irregular unit cell depends on $\vec{b}$, so we use the notation $n_{\text{irreg},\text{o},\vec{b}}$. 

The results are shown in Table \ref{table:nuc}. They are derived by matching the dislocation charge result with known values of $\vec{\mathscr{P}}_{\text{o}}$ in the $C=0$ limit (See App. \ref{app:C=0calcs}) from the same real space picture discussed in the previous section, when $M=4$. The origin dependence of $n_{\text{irreg},\text{o},\vec{b}}$ can also be understood from field theory. In Sec.~\ref{app:ftrelabel} we argue that the following general formula holds:
\begin{equation}
n_{\text{irreg},\text{o}+\vec{v},\vec{b}} = n_{\text{irreg},\text{o},\vec{b}} - v_x b_y + v_y b_x,
\end{equation}
where $(b_x, b_y)$ is the dislocation Burgers vector, and $(v_x,v_y)$ is a fractionally quantized vector that shifts $\text{o}$ to another HSP. Additionally, if $\vec{b}=(0,0)$ then $n_{\text{irreg},\text{o},\vec{b}}=0$; and if $\vec{b}\rightarrow -\vec{b}$, then $n_{\text{irreg},\text{o},\vec{b}}\rightarrow -n_{\text{irreg},\text{o},\vec{b}} \mod 1$. This condition means that a dislocation-antidislocation pair has a total unit cell number which is an integer.

\bgroup
\def\arraystretch{1.3}
\begin{table}
    \centering
\begin{tabular}{ |c|c|c||c|c|c|  }
 \hline
 \multicolumn{5}{|c|}{$n_{\text{irreg},\text{o},\vec{b}}$ } \\
 \hline\hline

$M$ & $\text{o}$ & $M'$ & $n_{\text{irreg},\OO,\vec{b}=(1,0)}$ & $n_{\text{irreg},\OO,\vec{b}=(0,1)}$ \\
 \hline
\hline
2&$\alpha$&2&0&0\\
\hline
2&$\beta$&2&$\frac{1}{2}$&$\frac{1}{2}$\\
\hline
2&$\gamma$&2&$\frac{1}{2}$&0\\
\hline
2&$\delta$&2&0&$\frac{1}{2}$\\
\hline

4&$\alpha$&4&0&0\\
\hline
4&$\beta$&4&$\frac{1}{2}$&$\frac{1}{2}$\\
\hline
4&$\gamma_2$&2&$\frac{1}{2}$&0\\
\hline
4&$\gamma_1$&2&0&$\frac{1}{2}$\\
\hline

3&$\alpha$&3&0&0\\
\hline
3&$\beta$&3&$\frac{1}{3}$&$\frac{2}{3}$\\
\hline
3&$\gamma$&3&$\frac{2}{3}$&$\frac{1}{3}$\\

\hline

6&$\alpha$&6&0&0\\
\hline
6&$\beta_1$&3&$\frac{1}{3}$&$\frac{2}{3}$\\
\hline
6&$\beta_2$&3&$\frac{2}{3}$&$\frac{1}{3}$\\
\hline
6&$\gamma_1$&2&0&$\frac{1}{2}$\\
\hline
6&$\gamma_2$&2&$\frac{1}{2}$&0\\
\hline
6&$\gamma_3$&2&$\frac{1}{2}$&$\frac{1}{2}$\\
\hline

\end{tabular}
\caption{$n_{\text{irreg},\text{o},\vec{b}}$ for all possible high symmetry points $\OO$, for $M=2,3,4,6$. Points belonging to the same MWP that rotate into each other are given the same $\{\beta,\gamma\}$ symbol but are distinguished by their coordinates, measured with respect to $\alpha$. $M'$ is the site symmetry group at $\text{o}$.}\label{table:nuc}

\end{table}
\egroup

\begin{figure*}[t]
    \centering
    \includegraphics[width=16.5cm]{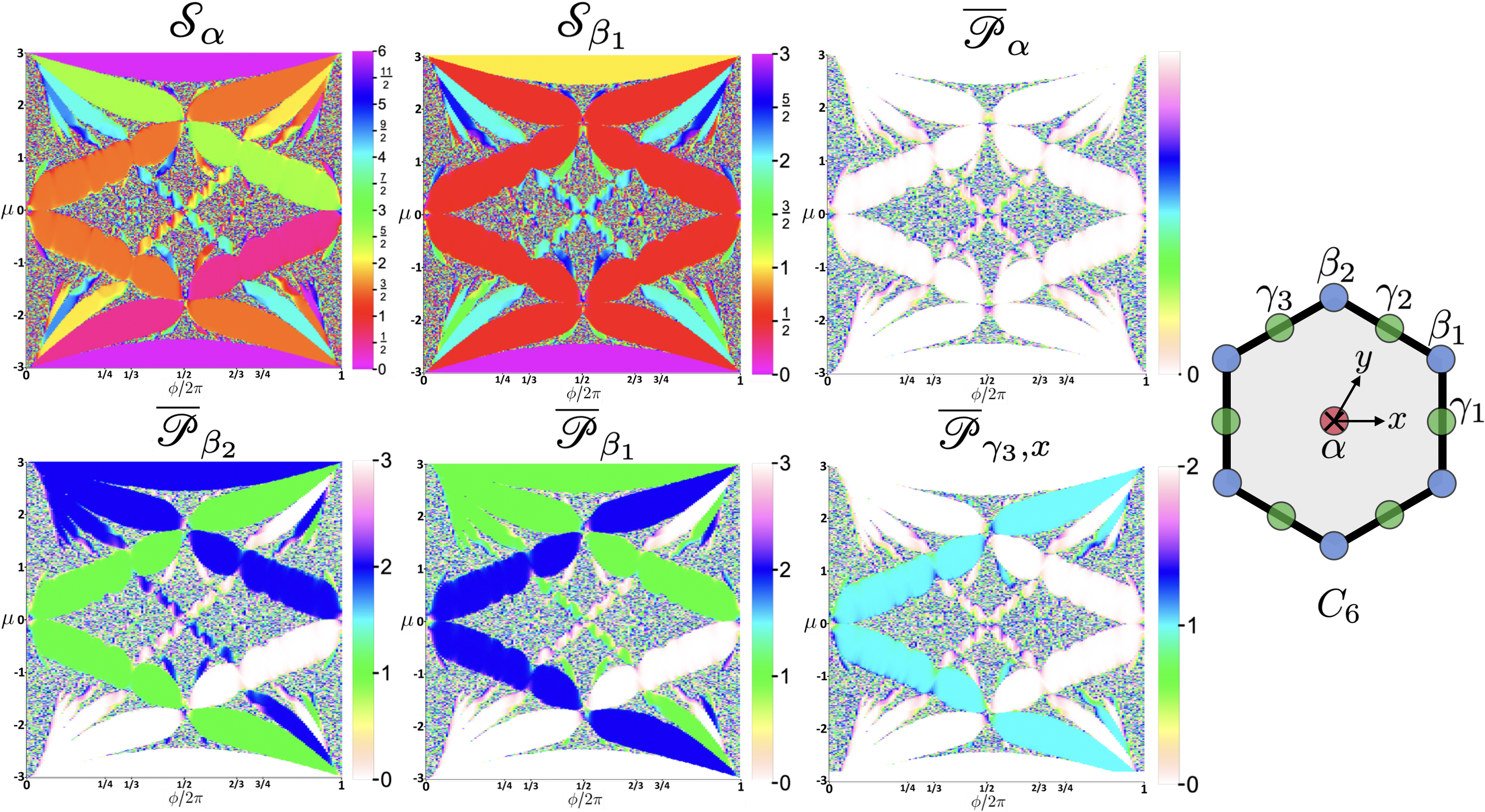}
    \caption{$\SO$ and $\PO$ in the honeycomb Hofstadter model.
    The site symmetry group is $\Z_{M'}$, with $M’=6,3,2$ for the MWPs $\{\alpha, \beta, \gamma\}$ respectively.
    $\mathscr{S}_{\alpha}, \mathscr{S}_{\beta_1}, \overline{\mathscr{P}}_{\alpha}, \overline{\mathscr{P}}_{\beta_2}, \overline{\mathscr{P}}_{\beta_1},\overline{\mathscr{P}}_{\gamma_3,x}$ 
    are all calculated from the charge response (Eq.~\eqref{eq:generalcharge}) of suitable lattice defects (here we have not shown $\SO$ and $\PO$ for all possible $\OO$ but only some representative ones). 
    The numerical calculations are done on an open disk with a radius $R=20$. The defects are located at the center of the disk. The noisy features appear since the butterfly is numerically calculated on a finite size system rather than analytically derived with an empirical formula as in Fig.~\ref{fig:SandP}. In the first 3 main Landau level, $\mathscr{S}_\alpha$ quantize to $\{\frac{1}{2}\pm 0.0002,2\pm 0.004, \frac{9}{2}\pm 0.04 \}$, other invariants in the first 3 main Landau levels have similar standard deviations.}
    \label{fig:hexnum}
\end{figure*}

With this information, we can write down the charge response for a dislocation:
\begin{equation}\label{eq:generaldislocationcharge}
\begin{aligned}
        Q_W=C\frac{\delta\Phi_{W,\text{o}}}{2\pi}+\vec{\mathscr{P}}_{\text{o}}\cdot \vec{b}
        +\nu(k+n_{\text{irreg},\text{o},\vec{b}}) \mod 1.
\end{aligned}
\end{equation}
This is the main equation that we use for calculations involving $\vec{\mathscr{P}}_{\text{o}}$ for $M=2,3,4,6$. As an example,
we numerically calculate $\vec{\mathscr{P}}_{\text{o}}$ in a Hofstadter model on the honeycomb lattice with different choices of $\OO$, whose site symmetry groups are $\Z_2, \Z_3$ and $\Z_6$. The raw data for $\vec{\mathscr{P}}_{\text{o}}$ is shown in Fig. \ref{fig:hexnum}. Again, we note that the exact choice of unit cell does not affect our calculation of $\vec{\mathscr{P}}_\text{o}$ and $\SO$. We show this in Appendix \ref{sec:trim}.

Note that in order to find $\PO$, it is enough to compute dislocation charge with $\vec{b} = (1,0),(0,1)$; this is why Table~\ref{table:nuc} explicitly contains these values of $\vec{b}$. Nevertheless, our procedure to compute dislocation charge applies for general choices of $\vec{b}$. The main observation is that $n_{\text{irreg},\OO,\vec{b}}$ only depends on the equivalence class of $\vec{b}$ under the equivalence
\begin{equation}
    \vec{b} \simeq \vec{b} + (1-U(2\pi/M')) \vec{\Lambda},
\end{equation}
where $\vec{\Lambda}$ is an integer vector and $2\pi/M'$ is the minimal rotation that preserves $\OO$. For example, $n_{\text{irreg},\OO,\vec{b}}$ is invariant if $\vec{b}$ is rotated by the angle $2\pi/M'$. Since any $\vec{b}$ is in the same equivalence class as either $(1,0)$ or $(0,1)$, Table~\ref{table:nuc} is sufficient for computations with general $\vec{b}$.

Combining Eqs.~\eqref{eq:generalpuredisclinationcharge} and~\eqref{eq:generaldislocationcharge}, the complete charge response of disclination and dislocation defects with arbitrary $\text{o}$, $\Omega_W$ and $\vec{b}_{\OO}$ is 

\begin{equation}\label{eq:generalcharge}
\begin{aligned}
        Q_W&=C\frac{\delta\Phi_{W,\text{o}}}{2\pi}+\frac{\Omega_W \mathscr{S}_{\text{o}}}{2\pi}+\vec{\mathscr{P}}_{\text{o}}\cdot \vec{b}\\
        &+\nu(k+ n_{\text{irreg},\text{o},\Omega,\vec{b}})
\end{aligned}
\end{equation}
where the effective irregular unit cell number now depends on both $\Omega$ and $\vec{b}$. For all the examples that we have studied, we have observed the relation
\begin{equation}
    n_{\text{irreg},\text{o},\Omega,\vec{b}} = n_{\text{irreg},\text{o},\Omega,(0,0)} + n_{\text{irreg},\text{o},0,\vec{b}} \mod 1.
\end{equation}
Here $n_{\text{irreg},\text{o},\Omega,(0,0)},n_{\text{irreg},\text{o},0,\vec{b}}$ are the quantities we previously referred to as $n_{\text{irreg},\text{o},\Omega},n_{\text{irreg},\text{o},\vec{b}}$ respectively. This relation matches our expectation from the field theory, in which we can imagine separating a defect with parameters $(\vec{b},\Omega)$ into two defects with parameters $((0,0),\Omega),(\vec{b},0)$ without changing the effective number of unit cells mod 1.

\subsection{$\mathscr{S}_{\text{o}}$ from angular momentum}

We denote as $l_{\text{o}}$ the angular momentum with $\text{o}$ as the rotation center. The many body ground state on a clean torus with $m$ total flux quanta is $\ket{\Psi}$. $l_{\text{o}}$ is defined by
\begin{equation}
\tilde{C}_{M,\OO}\ket{\Psi}=e^{il_{\OO} \frac{2\pi}{M}}\ket{\Psi}.
\end{equation}
Exactly as for $M=4$, we find that
\begin{equation}
    l_{\text{o}} = \frac{C m^2}{2} + \mathscr{S_\text{o}} m+ K(C,L) \mod M.
\end{equation}
The above equation works for any $M=2,3,4,6$. In this work we have also numerically calculated $l_{\alpha}$ in the honeycomb lattice, choosing $\OO=\alpha$. We extract the same $\mathscr{S}_{\alpha}$ from the disclination charge response (See Fig.\ref{fig:hexnum}). This verifies the prediction that $\SO$ can be measured using an angular momentum dual response.

One can also perform a partial rotation to extract $\mathscr{S}_{\text{o}}$. The details (specialized to $M=4$) are in Ref.~\cite{zhang2022fractional}. 

\subsection{$\vec{\mathscr{P}}_{\text{o}}$ from linear momentum}\label{sec:general_PO}

The procedure to calculate a linear momentum $(p_{\lambda,x},p_{\lambda,y})$ and hence $\PO$ on the square lattice generalizes straightforwardly to the case where $M=2,3,6$. For $M$ even we can define $\tilde{T}_{\bf y}$ precisely as for $M=4$. But for $M=3$, we find that a slight modification is required in the definition of $\lambda$:
\begin{align}\label{eq:tyfullgaugeappm3}
    \tilde{T}_{\bf{y}}=\begin{cases}
      \hat{T}_{\bf{y}}e^{i\sum_j -\frac{\pi m}{L_y} c_j^{\dagger}c_j}\qquad   &j_x< \overline{\OO}_x\\
      \hat{T}_{\bf{y}}e^{i\sum_j \frac{\pi m}{L_y} c_j^{\dagger}c_j}\qquad   &j_x> \overline{\OO}_x\\
      \hat{T}_{\bf{y}}e^{i\sum_j (\overline{\OO}_x-j_x)\frac{\pi m}{L_y} c_j^{\dagger}c_j} \qquad &\overline{\OO}_x-1<j_x<\overline{\OO}_x+1.\end{cases}
\end{align}
This can be seen as a generalization of Eq.~\eqref{eq:tyfullgaugeapp} which smooths out the gauge transformation around the gauge origin $\overline{\OO}$.
The relevant equations for  $M=3$ are identical to the ones for $M=2,4$:
\begin{align}\label{eq:py-M=3}
    \bra{\Psi} \tilde{T}_{\bf{y}}\ket{\Psi} &:=e^{ - \gamma + i 2\pi p_{\lambda,y}} \\
p_{\lambda,y} &= -\mathscr{P}_{\Oy} m+ K_y \mod 1.
\end{align}
We have numerically checked that on the honeycomb lattice we extract the same $\PO$ as with the other methods (see Fig.~\ref{fig:hexnum}), assuming Eq. \eqref{eq:generalobar}. Since $\mathscr{P}_{\Oy}$ is quantized mod 3, the minus sign is crucial in this case, unlike in the case $M=2,4,6$. 

On the other hand, the partial translation calculation does not generalize to the case where $\OO$ is a $C_3$ symmetric HSP. This is because we appear to need the region $D$ for partial translations to be mirror symmetric along the cycle with holonomy $e^{i\pi m}$. But this is not possible for $C_3$ symmetric points. It is not clear whether a mirror symmetric region $D$ is essential for our calculation to work, or whether an alternative method might work in this case; we leave this question for future study.

\subsection{$\vec{\mathscr{P}}_{\text{o}}$ from 1d polarization}
Finally, we can dimensionally reduce our $2d$ space along $y$, and compute a 1d polarization $\mathcal{P}_x$ in the $x$ direction. For $M=4$ we obtained an empirical formula, Eq.~\eqref{eq:pol1Dold}, which matches the field theory prediction. 
We rewrite it here:
\begin{align}
-\mathcal{P}_{\mathcal{O},x}=\frac{C \phi}{2\pi} L_y (\frac{L_x}{2}-\overline{\OO}_x+\mathcal{O}_x) +L_y\mathscr{P}_{\OO,y}+K' \mod 1,
\end{align}
where $\overline{\OO}$ is the gauge origin, and $\mathcal{O}$ is the origin in Resta's formula, Eq.~\eqref{eq:resta}.

This equation turns out to also work for $M=2,3,6$. 
As an example, in the honeycomb lattice, the position of the sites are:
\begin{equation}
\begin{aligned}
     &i=(i_x-\frac{1}{3}-\mathcal{O}_x,i_y-\frac{1}{3}-\mathcal{O}_y),\\ &j=(j_x+\frac{1}{3}-\mathcal{O}_x,j_y+\frac{1}{3}-\mathcal{O}_y).
    \end{aligned}
\end{equation}
$i$ and $j$ are at the $\beta_1$ and $\beta_2$ MWPs of the $C_6$ unit cell respectively (see Fig.~\ref{fig:wyckoff} for the unit cell convention), and $i_x,j_x\in \{0,1,\dots L_x-1\}, i_y,j_y\in\{0,1,\dots L_y-1\}$.
Similar to the linear momentum calculation, in order to get a result for $\vec{\mathscr{P}}_{\text{o}}$ which is consistent with the known $C=0$ result, we need to choose $\mathcal{O}_x=\text{o}_x+\frac{L_x}{2}+\frac{1}{2}\mod 1$.

\section{Origin dependence of $\mathscr{S}_{\text{o}},\vec{\mathscr{P}}_{\text{o}}$}\label{app:ftrelabel}

In this section we sketch how to derive the formulas in Table~\ref{table:shiftingO} that express $\mathscr{S}_{\text{o}+\vec{v}}, \vec{\mathscr{P}}_{\text{o}+\vec{v}}$ in terms of $\mathscr{S}_{\text{o}}, \vec{\mathscr{P}}_{\text{o}}$, and $\kappa$. $\vec{v}$ is a quantized vector (perhaps fractional) such that $\OO,\OO+\vec{v}$ are both symmetric under $M$-fold rotations. These relationships were commented on previously in Secs.~\ref{sec:c4_SO},~\ref{sec:c4_PO} and~\ref{sec:general}. We leave the technical computations to App.~\ref{app:ftcomps}. Background details and definitions regarding the response theory which we use in this section can be found in App.~\ref{sec:field}.

Let $A$ be the full vector potential for the system (meaning that $dA = F$ measures the total magnetic field). The full Lagrangian which involves $A$ is 
\begin{align}
    \mathcal{L} &= \frac{C}{4\pi} A \wedge dA + \mathcal{L}_{A,\OO}, \\
    \mathcal{L}_{A,\OO} &:= \frac{1}{2\pi}A\wedge\left( \mathscr{S}_{\text{o}} d\omega + \vec{\mathscr{P}}_{\text{o}} \cdot \vec{T} + \kappa A_{XY}\right).
\end{align}
Here $dA$ is the total $U(1)$ flux. The total $U(1)$ charge and flux in any region should physically be invariant under a shift of origin, and therefore the contribution from the term with $C$ to any response property remains invariant upon shifting the origin. Hence we only consider the transformation of the remaining terms as given by $\mathcal{L}_{A,\OO}$. Crucially, in $\mathcal{L}_{A,\OO}$ the coefficient of the term with $\frac{1}{2\pi} A \wedge A_{XY}$ is $\kappa$ instead of $\nu$, because the difference, given by $ \frac{C\phi}{4\pi^2} A \wedge A_{XY}$, is now contained in the term $\frac{C}{4\pi} A \wedge dA$.

To understand the transformation of $\mathcal{L}_{A,\OO}$, it will be most convenient to use the discrete simplicial formulation of the response theory, as developed in \cite{manjunath2021cgt}; in that case, the above wedge product should be interpreted as a cup product, but we will stick to wedge product notation below. First note that a space group operation which performs a rotation about $\text{o}$ and then a translation takes the form ${\bf g} = ({\bf r}, h)$, where ${\bf r} \in \Z^2, h \in \Z_M$. The same group element with respect to the shifted origin $\OO'=\OO+\vec{v}$ takes the form 
$${\bf g}' = (\vec{v},0)({\bf r}, h)(-\vec{v},0).$$ 
Since the crystalline gauge fields encode some configuration of group elements on a manifold, a shift of origin by a fractional lattice vector $\vec{v}$ can be viewed as a `fractional' gauge transformation of the crystalline gauge fields. In particular, because this is not a true gauge transformation, the crystalline gauge fluxes also get redefined, leading to a transformation of the response coefficients.

In the simplicial formulation, if the gauge field is flat, the crystalline gauge fluxes are defined by the terms $d\omega, A_{XY}, (1-U(\frac{2\pi}{M}))^{-1} d\vec{R}$. (Note that in the discrete case we can equivalently write $\vec{T}$ as $d\vec{R}$; we will do this for the rest of the section.)\footnote{These fluxes correspond to a set of 2-cocycle representatives which generate the group $\H^2(G_{\text{space}},\Z)$ when $G_{\text{space}} = \Z^2 \rtimes \Z_M$ \cite{manjunath2021cgt}.  We have $\H^2(G_{\text{space}},\Z) \cong \Z_M \times \Z\times K_M $. The 2-cocycles corresponding to $d\omega, A_{XY}, (1-U(\frac{2\pi}{M}))^{-1} d\vec{R}$ generate these three factors respectively.}  Under a shift of origin, these fluxes transform into new quantities $d\omega', (1-U(\frac{2\pi}{M}))^{-1}d\vec{R}',A'_{XY}$. In App.~\ref{app:ftcomps} we show that
\begin{align}
    d\omega' &= d\omega \label{eq:k1}\\
    d\vec{R}' &= d\vec{R} + \vec{\tau}_M d\omega \label{eq:k2}\\
    A'_{XY} &= A_{XY} + \rho_M d\omega + \vec{\mu}_M \cdot d\vec{R} \label{eq:k3},
\end{align}
where $\vec{\tau}_M,\rho_M, \vec{\mu}_M$ depend on $\vec{v}$. This transforms the Lagrangian as follows:
\begin{equation}\label{eq:F1}
    \mathcal{L}_{A,\OO'}= \frac{1}{2\pi}A\wedge\left( \mathscr{S}_{\text{o}'} d\omega' + \vec{\mathscr{P}}_{\text{o}'} \cdot d\vec{R}' + \kappa' A'_{XY}\right).
\end{equation}
Assuming a fixed choice of unit cell, the total charge measured in any given region $W$ should be the same for either choice of origin. Thus
\begin{align}
    & \int_W (\SO d\omega + \PO \cdot d\vec{R} + \kappa A_{XY}) \nonumber \\
    = & \int_W (\mathscr{S}_{\OO'} d\omega' + \P_{\OO'} \cdot d\vec{R}' + \kappa' A'_{XY}).
\end{align}
We now use \eqref{eq:k1},\eqref{eq:k2},\eqref{eq:k3}, and compare coefficients. This gives us
\begin{align}
    \kappa &= \kappa' \\
     \PO &= \vec{\mathscr{P}}_{\text{o}'-\vec{v}} = \vec{\mathscr{P}}_{\text{o}'} + \kappa' \vec{\mu}_M \\
   \SO  &= \mathscr{S}_{\text{o}'-\vec{v}} = \mathscr{S}_{\text{o}'} +  \vec{\mathscr{P}}_{\text{o}'} \cdot \vec{\tau}_M + \kappa' \rho_M.
\end{align}
These equations finally give the results in Table \ref{table:shiftingO}: the calculation reduces to showing \eqref{eq:k1},\eqref{eq:k2},\eqref{eq:k3}, and finding $\rho_M, \vec{\mu}_M, \vec{\tau}_M$ in terms of $\vec{v}$. We have performed these calculations in Appendix \ref{app:ftcomps}; to obtain values for $\rho_{M=2,4}$ which are consistent with results for Chern number 0, we need to make an assumption on the functional form of $A_{XY}$, which we explain there. The results are contained in Eqs.~\eqref{eq:tau} ($\vec{\tau}_M$),~\eqref{eq:rho} ($\rho_M$) and in Sec.~\ref{app:change_PO} ($\vec{\mu}_M$). 

Note that Eq.~\eqref{eq:tau} implies
\begin{equation}\label{eq:dRprime}
    d\vec{R}' = d\vec{R} + M(1-U(2\pi/M))d\omega.
\end{equation}
This is in fact equivalent to Eq.~\eqref{eq:bOplusv}, which reads
\begin{equation}
    \vec{b}_{\OO+\vec{v}} = \vec{b}_{\OO} + (1-U(\Omega))\vec{v}.
\end{equation}
The equivalence can be shown as follows. First let us identify $d\vec{R}' = 2\pi \vec{b}_{\OO+\vec{v}},d\vec{R} = 2\pi \vec{b}_{\OO},d\omega = \Omega = \frac{2\pi k}{M}$. Then Eq.~\eqref{eq:dRprime} becomes
\begin{equation}
    \vec{b}_{\OO+\vec{v}} = \vec{b}_{\OO} + k(1-U(2\pi/M)) \vec{v}.
\end{equation}
To prove the equivalence, we need to show that the two expressions for $\vec{b}_{\OO+\vec{v}}$ are in the same equivalence class:
\begin{equation}
    (1-U(2\pi k/M))\vec{v} = k(1-U(2\pi/M)) \vec{v} + (1-U(2\pi/M))\vec{\Lambda}
\end{equation}
for some $\vec{\Lambda} \in \Z^2$. But this follows from the fact that $(1-U(2\pi/M)) \vec{v}$ is always an integer vector.

As a check on Table~\ref{table:shiftingO}, we can see that the values of $\SO$ and $\PO$ for zero Chern number given in App.~\ref{app:C=0calcs} follow the equations in the Table. To check Table~\ref{table:shiftingO} when $C \ne 0$, we consider an example with $C_6$ rotational symmetry in App.~\ref{app:SOExample}.

In this derivation we made the crucial assumption that $F$ is invariant under a shift of origin. Note that $F = d \delta A + \frac{\phi}{2\pi} A_{XY}$ where $\delta A$ is the deviation of the $U(1)$ gauge field from its background value. $F$ can indeed be made invariant under the transformation of Eq.~\eqref{eq:k3}, if we take
\begin{equation}
    \delta A \rightarrow \delta A - \frac{\phi}{2\pi} (\rho_M \omega +\vec{\mu}_M \cdot \vec{R}).
\end{equation}
Note that $d\delta A$ only changes around a defect, where $d\omega$ and $d\vec{R}$ can be nonzero. The transformations of $d\delta A, A_{XY}$ indicate that our conventions for `background flux' and `excess flux' in each plaquette (measured by $\phi A_{XY}$ and $d\delta A$ respectively) change by equal and opposite amounts. But the total flux in each plaquette is invariant. 

\subsection{Transformation of $\PO$}\label{sec:trans_PO}
Although $\rho_M, \vec{\tau}_M$ depend sensitively on $M$, our calculations show that 
$$\vec{\mu}_M = (v_y, -v_x),$$ 
irrespective of the choice of $M$. This implies that
\begin{equation}\label{eq:Pshift}
    \vec{\mathscr{P}}_{\OO + \vec{v}} = \PO + \kappa (-v_y,v_x)
\end{equation}
for $M \in2,3,4,6$. 

The fact that $\PO$ transforms proportionally to $\kappa$ has an important consequence. If we consider two Hamiltonians 1 and 2, which can for example be the end points of some path in parameter space, we may naively imagine differences of the form $\PO^{(2)} - \PO^{(1)}$ to be completely independent of $\OO$. But for $\PO$ as defined in this work, this is true only if $\kappa_2 = \kappa_1$. Indeed, Eq.~\eqref{eq:Pshift} implies that
\begin{equation}
   \vec{\mathscr{P}}_{\OO + \vec{v}}^{(2)}-\vec{\mathscr{P}}_{\OO + \vec{v}}^{(1)} = \PO^{(2)} - \PO^{(1)} + (\kappa_2-\kappa_1) (-v_y,v_x).
\end{equation}
Thus, in order to measure an origin-independent quantity through differences of $\PO$, we must ensure that the initial and final values of $\kappa$ are equal.

\subsection{Formula for $n_{\text{irreg},\OO+\vec{v},\vec{b}}$}

An interesting corollary of these results is that they allow us to determine the correct assignments of $n_{\text{irreg},\OO,\vec{b}}$ for a fixed $\vec{b}$, as we vary $\text{o}$. On an infinite plane lattice with only dislocations, we can set $\omega=0$, and so Eq.~\eqref{eq:k3} becomes
\begin{equation}
    A'_{XY} = A_{XY} + \vec{\mu}_M \cdot d\vec{R}.
\end{equation}
With $\vec{\mu}_M$ as above, we integrate over a region $W$ surrounding the dislocation, and reduce mod 1 to obtain
\begin{equation}\label{eq:nirreg}
    n_{\text{irreg},\OO+\vec{v},\vec{b}} = n_{\text{irreg},\OO,\vec{b}} -v_x b_y + v_y b_x \mod 1.
\end{equation}
Once $n_{\text{irreg},\OO,\vec{b}}$ is known for a single origin $\text{o}$, the values of $n_{\text{irreg},\OO+\vec{v},\vec{b}}$ can be determined using Eq.~\eqref{eq:nirreg}, as we have listed in Table \ref{table:nuc}. This non-trivial transformation rule is confirmed by our numerical results and the analytical results at $C = 0$. 

\section{Discussion}

In this paper we have described several complementary many-body approaches to measure the quantized charge polarization $\PO$ and the discrete shift $\SO$ in gapped topological phases, including those with a nonzero Chern number and magnetic field. We extract $\PO$ by studying the fractional charge bound to dislocations, the linear momentum bound to flux, and from an effective 1d polarization response. We have obtained explicit numerical results for the spinless Hofstadter model with $C_M$ rotational symmetry, for $M=2,3,4,6$, by matching our microscopic calculations to field theory predictions. We have also obtained a theoretical understanding of the origin dependence of $\PO$ and the discrete shift $\SO$. Together with the Chern number $C$ and $\kappa\equiv \nu-C\phi/2\pi$, the quadruple $\{C,\SO,\PO,\kappa\}$ completely specifies the quantized charge response in systems with charge conservation, magnetic translation, and point group rotation symmetry.

An important issue we wish to emphasize is how to understand $\SO,\PO$ as invariants describing a topological phase. In the case of an origin-independent invariant such as $C$ or $\kappa$, knowing that the invariant differs between two systems is enough distinguish them as topological phases. However, in order to distinguish two systems based on their respective values of $\SO$ or $\PO$, it is essential to first fix a common origin $\OO$ for both systems, and then compare the various numbers. This is because two systems are in the same phase only if they can be adiabatically connected to each other without closing the gap or breaking the symmetry, \textit{and while keeping their common origin $\OO$ fixed}. Without fixing this common origin, we cannot meaningfully define the notion of adiabatic equivalence between two systems.

In this paper we have focused on the quantization of $\vec{\mathscr{P}}_{\OO}$ due to non-trivial point group symmetry, $M > 1$. In the case where we do not have point-group symmetry, our methods allow us to define an intrinsically two-dimensional many-body polarization $\vec{\mathscr{P}}_{\OO}$ for any real-space origin $\OO$, even when $C \neq 0$. It is an interesting question to understand the relationship between our definition of polarization for $C \neq 0$ and the one based on free fermion band theory proposed by Coh and Vanderbilt \cite{coh2009}. 


Since $\SO$ and $\PO$ are topological invariants that depend on crystalline symmetry, their values will cease to be quantized if we break the crystalline symmetry by adding disorder. In particular, if we introduce onsite or bond disorder which breaks the rotational symmetry, $\SO$ is no longer quantized. If the disorder breaks either the translational or the rotational symmetry, $\PO$ is no longer quantized. Nevertheless, the disorder averages of $\SO,\PO$ remain quantized to their respective clean values, although their standard deviations increase with an increase in disorder strength. This was shown numerically for $\SO$ in Ref.~\cite{zhang2022fractional} and is expected to hold for $\PO$ as well. 

Next we comment on some unresolved issues in this work. In both the linear momentum and 1d polarization calculations, we need to pick a distinguished point and relate it to $\OO$. For the linear momentum calculation $\overline{\OO}$ is the gauge origin, and in the 1d polarization calculation $\mathcal{O}$ determines the coordinates for each site in Resta's formula, Eq.~\eqref{eq:resta}. In both calculations, we extract $\PO$ correctly only if $\OO$ satisfies a certain relation with $\overline{\OO}$ or $\mathcal{O}$. In the linear momentum calculation, we also found that if $\tilde{T}_{\bf y} = \hat{T}_{\bf y} e^{i \sum_j \lambda_j c_j^{\dagger} c_j}$, for a fixed $\overline{\OO}$ there is only one choice of $\lambda$ that gives a quantized $\mathscr{P}_{\Oy}$ throughout the Hofstadter butterfly. These observations are completely empirical; we leave a full explanation for future work.

For our linear momentum calculations we used an approximate translation operator and thus had to work with its expectation values rather than exact eigenvalues. It would be useful to compute the linear momentum exactly by incorporating a flux insertion operator that makes the translation symmetry exact. 

Additionally, under a shift of origin, the field theory does not fix a unique transformation rule for $\SO,\PO$ when $M=2,4$; instead it gives a few different possibilities as explained in detail in App.~\ref{app:ftcomps}. To fix the transformation consistent with numerics and physical expectations, we need to make some additional choices in the field theory that are allowed but do not have an obvious physical interpretation. 

We close by pointing out some related open questions. Previously, Ref.~\cite{coh2009} proposed a way to define the charge polarization as a single-particle Berry phase in momentum space when $C \ne 0$, by picking a suitable origin for the Brillouin zone. It would be useful to understand whether this choice can be related to the ones we have made in defining $\PO$ using the dislocation charge and the linear momentum. 

We have not commented on how the invariants $\SO, \PO$ manifest at corners and edges of the system. The relation between disclination charge and fractional corner charge when $C=0$ has been discussed in several places, see e.g. Ref.~\cite{Benalcazar2019HOTI}. Refs.~\cite{coh2009,Song2021polarization} specifically discussed an edge charge interpretation of the polarization; Ref.~\cite{coh2009} did so in the context of Chern insulators, and Ref.~\cite{Song2021polarization} did so from the perspective of a boundary Luttinger theorem when $C=0$. It would be interesting to understand the corner and edge charges in the context of our results, which apply for general $C$ and in the presence of a magnetic field. 

Another interesting direction is to study the charge polarization in \textit{fractional} Chern insulators. In fact, Refs.~\cite{manjunath2021cgt,Manjunath2020fqh} use field theory to predict that this can indeed be defined systematically and that its quantized fractional values are sensitive to $M$ as well as to the anyon content of the theory. In such topologically ordered phases, the charge polarization can encode a novel, nontrivial form of symmetry fractionalization which was called the `discrete torsion vector' in Refs.~\cite{manjunath2021cgt,Manjunath2020fqh}.

\section{Acknowledgements}

GN thanks D. Bulmash and V. Galitski for discussions during the initial stages of this project. This work is supported by the Laboratory for Physical Sciences through the Condensed Matter Theory Center, and by NSF CAREER (DMR- 1753240) (MB, NM), ARO W911NF-20-1-0232 (GN), QLCI grant OMA-2120757.

\appendix 

\section{Review}\label{sec:bgd}

This section has three parts. First we introduce some standard definitions of quantities on a clean lattice. Next we introduce dislocation and disclination defects, and define the disclination angle and dislocation Burgers vector. After that we discuss the basic properties of $\SO$ and $\PO$ from the perspective of disclination and dislocation charge respectively.

\subsection{Definitions and background}
\subsubsection{Maximal Wyckoff positions}
The definitions in this section are taken from Ref.~\cite{Cano_2021}. Fix an origin $\text{o}$. Suppose we are given a set of points in the infinite plane which form a lattice with spatial symmetry group $G_{\text{space}}$. On this lattice (assumed to be without any defects), we define a Hamiltonian $H_{\text{clean}}$ with translation operators  $\hat{T}_{\bf x},\hat{T}_{\bf y}$ corresponding to translations by the elementary lattice vectors, and additional point group symmetry operators, defined w.r.t. $\OO$. The full symmetry group $G$ of $H_{\text{clean}}$ can contain operations in $G_{\text{space}}$ as well as internal symmetry operations.  

For any point $p$ in the plane, the \textit{site symmetry group} $G_p$ is the subgroup of operations in $G_{\text{space}}$ that leaves $p$ invariant. Two points $p,p'$ are said to have site symmetry groups that are conjugate to each other if $G_{p'} = {\bf g} G_p {\bf g}^{-1}$ for some ${\bf g} \in G_{\text{space}}$. The \textit{Wyckoff position} containing $p$ is the set of all points whose site symmetry groups are conjugate to $G_p$. For example, every point $p'$ obtained from $p$ by a lattice translation or rotation is in the same Wyckoff position as $p$. Furthermore, every point with a trivial site symmetry group belongs to a single Wyckoff position. 

$p$ is in a \textit{maximal} Wyckoff position (MWP) if $G_p$ is not a proper subgroup of $G_{p'}$ for any other site $p'$. In Fig. \ref{fig:wyckoff}, we show the MWPs for the wallpaper groups p2, p3, p4, p6. Note that when $G_{\text{space}} = \text{p4}$, the high symmetry point $\gamma_i$ has site symmetry group $G_{\gamma} \cong \Z_2$ but is still in a MWP, even though there are other points with site symmetry group $\Z_4$. This is because there is no point with site symmetry group $\Z_4$ that contains $G_{\gamma}$. We will always choose $\text{o}$ to belong to a MWP. 

It is important to distinguish a MWP, which is a collection of points, from a single high symmetry point (HSP) of a unit cell. Our notation for HSPs is $\beta_i, \gamma_i$ where $\beta, \gamma$ denotes the MWP and $i$ runs over the corresponding HSPs which are inequivalent under lattice translations. 

\subsubsection{Unit cells and subcells}

A \textit{unit cell} $\Theta$ for the given lattice corresponds to a division of lattice points into elementary repeating units.
Starting with a clean lattice on the infinite plane, pick a HSP $\alpha$ for which the site symmetry group contains the full point group. (This choice may not be unique.) Then define $\alpha$ and its lattice translates as the centers of each unit cell. This is the convention used in Fig.~\ref{fig:wyckoff}. Now consider any other point $q$. If $q$ is equidistant from $n>1$ unit cell centers, $q$ is assigned to be on the common boundary of $n$ unit cells centered around $\alpha$. If $q$ is closest to one particular $\alpha$ point, we say that $q$ lies in the interior of the unit cell. By convention, the corners of the unit cell are labelled as $\beta$ (or $\beta_i$, if there are multiple HSPs in the same MWP); any other points on the unit cell boundary are denoted $\gamma, \delta, \dots$ as required.

Note that this definition does not fully determine the shape of the unit cell; we only require the sub-cells to rotate into each other, and there is still a lot of freedom in drawing the exact boundaries. 

\begin{figure}[t]
    \centering
    \includegraphics[width=7cm]{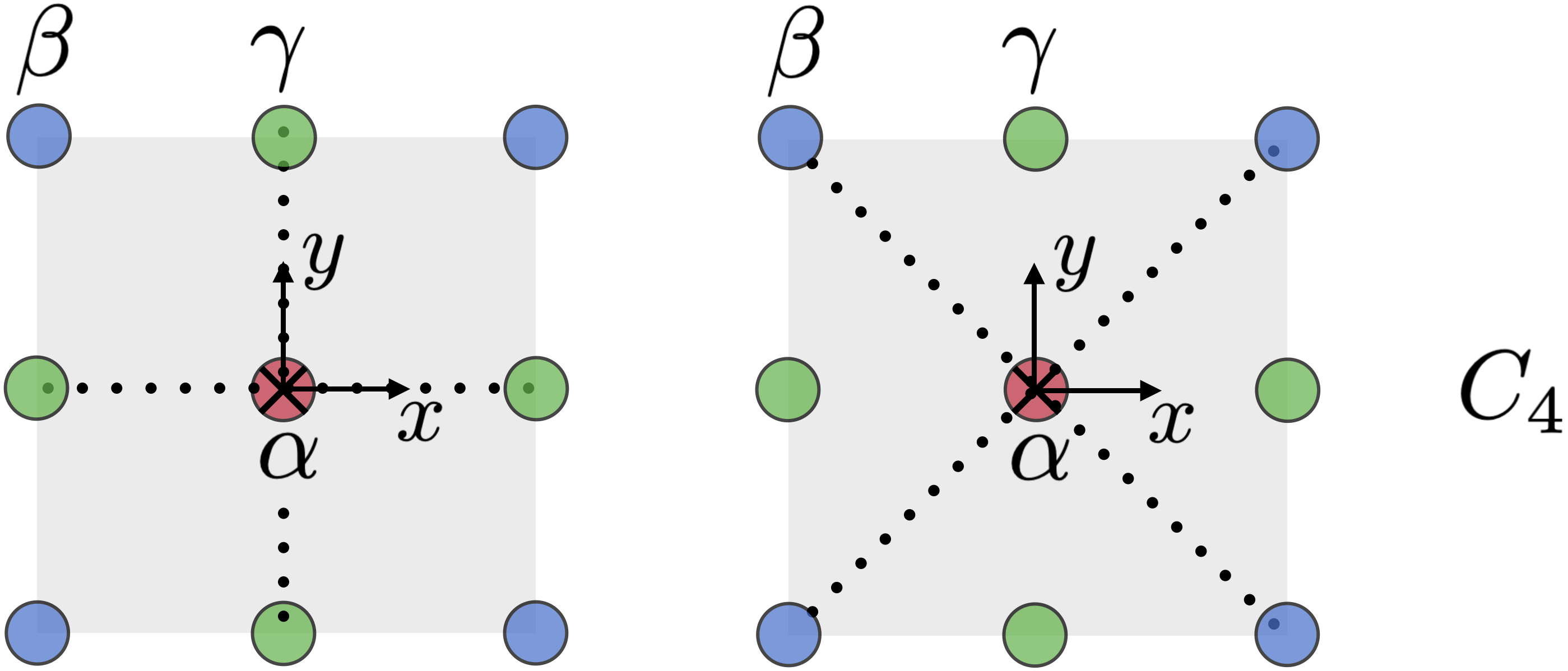}
    \caption{Two representative ways of dividing a $C_4$ unit cell into subcells.}
    \label{fig:unitcell_division}
\end{figure}

Now let us specialize to the case where $H_{\text{clean}}$ has an $M$ fold rotational symmetry. We can subdivide an $M$-fold rotationally symmetric unit cell into $M$ \textit{subcells}. Fig.~\ref{fig:wyckoff} illustrates such a division. We only require that subcells rotate into each other under rotations about $\alpha$. We have no constraint on the shape of the subcells: in Fig.~\ref{fig:unitcell_division} we have shown two equally valid shapes. Therefore, although the vertices of each subcell are fixed by our definition of the unit cell, the boundaries are otherwise arbitrary. 

\subsubsection{Parameters in $H_{\text{clean}}$}\label{app:Hclean}
In our numerics, we consider Hofstadter models, i.e. free fermion Hamiltonians of the form
\begin{align}
    H_{\text{clean}} = -\sum_{ij} t_{ij} c_i^\dagger c_j + \text{h.c.}
\end{align}
where $i,j$ are site indices and the hopping terms $t_{ij} = t e^{-i A_{\text{clean}, ij}}$ depend on a background vector potential $A_{\text{clean}}$. We mainly take $t_{ij} = 1$ if $i,j$ are nearest neighbors, and $t_{ij}=0$ otherwise. However, in App.~\ref{app:nnnhopping} we also consider the Hofstadter model with next neighbor hopping terms, as an illustration. 

For any loop $\ell$, the enclosed flux equals $\arg \left(\prod\limits_{ \langle ij \rangle \in \ell} t_{ij}\right) = \sum\limits_{\langle ij \rangle \in \ell} A_{\text{clean}, ij} \mod 2\pi$, where the sum is taken counterclockwise. We assume that $H_{\text{clean}}$ has the symmetry $G_f = U(1)^f \times_{\phi} G_{\text{space}}$, where $U(1)^f$ denotes the group $U(1)$ whose order 2 element is the fermion parity operation. This means that the total flux in each unit cell equals $\phi \mod 2\pi$. 

If $H_{\text{clean}}$ is defined on a torus with side lengths $L_1,L_2$ instead of the infinite plane, the $\Z^2$ translation symmetry is broken down to $\Z_{L_1} \times \Z_{L_2}$ if there is no flux; if there is flux, the symmetry is still broken down to a subgroup but the details are more complicated. 

As explained in the main text, we demand that the magnetic field is specified everywhere within a unit cell. This requirement goes beyond the specification of $H_{\text{clean}}$ and of the symmetry data. We also note (see in particular App.~\ref{app:nnnhopping}) that our numerical methods are expected to generalize to arbitrary gapped systems with next neighbor hopping and interaction terms.

\subsubsection{Definition of $\nu$ and $\kappa$}\label{app:defkappa}

A lattice system with $U(1)$ charge conservation symmetry has a filling per unit cell given by $ \frac{N_e\eta_{u.c.}}{N_{\text{orb}}}$ where $N_e$ is the number of electrons, $N_{\text{orb}}$ is the number of orbitals, and $\eta_{u.c.}$ is the number of orbitals per unit cell. On a clean torus, this reduces to $\nu := \frac{N_e}{N_{u.c.}}$, where $N_{u.c.}$ is the total number of unit cells. If the system has a unique, gapped ground state and has Chern number $C$, $\nu$ and $C$ are related by Eq.~\eqref{eq:k0def}, which also defines the integer
$\kappa$. Eq.~\eqref{eq:k0def} can be derived for example using the following argument \cite{Lu2017fillingenforced,Manjunath2020fqh}.

The quantity $e^{2\pi i \nu}$ is by definition the Aharonov-Bohm phase corresponding to the adiabatic transport of a $2\pi$ flux around a unit cell, which has flux $\phi$. Now the braiding phase of a flux $\phi_1$ around another flux $\phi_2$ can alternatively be expressed in terms of the Hall conductance $\sigma_H$; it equals $e^{i \phi_1 \phi_2 \sigma_H} = e^{i \phi_1 \phi_2 \frac{C}{2\pi}}$. Taking a $2\pi$ flux around a unit cell corresponds to setting $\phi_1 = 2\pi, \phi_2 = \phi$. Thus the same Aharonov-Bohm phase can be expressed in two ways:
\begin{align}
    e^{i 2\pi \nu} &= e^{i \phi C} \\
    \implies \quad \nu &= C \frac{\phi}{2\pi} \mod 1 \\
    \implies \quad \nu &= C \frac{\phi}{2\pi} + \kappa,
\end{align}
where $\kappa \in \Z$. If $C=0, \nu = \kappa$.

In the limit of small flux per unit cell, $0 \le C \phi/2\pi \le 1$, $\kappa$ has a simple interpretation. When $\kappa = 0$, a state with Chern number $C$ corresponds to a system of $C$ filled Landau levels. Therefore $\kappa$ is simply the deviation of the total filling from the Landau level limit. In a system with $\phi = 2\pi p/q$ where $p,q$ are coprime, the filling on a torus takes the form $\nu = r/q$ for some integer $r$. Then Eq.~\eqref{eq:k0def} can be written written as a Diophantine equation $r = q \kappa + p C$ for the integers $C,\kappa$. In the Hofstadter model it can further be shown that $|C| \le q/2$. This Diophantine equation is often used in the study of the Hofstadter model.

Finally, suppose we place the system on a closed manifold with lattice defects. This could for example be the surface of a cube, which has 8 corners, each being a disclination with disclination angle $\pi/2$. In this case, the actual filling per unit cell differs from $\nu$. In fact the difference is proportional to the discrete shift, as discussed in Ref.~\cite{zhang2022fractional}. 

\subsection{Origin dependence of $\vec{b}_{\OO}$}

Here we prove Eq.~\eqref{eq:bOplusv} in the main text. We define a $\Z^2$ translation gauge field $\vec{R}$ and a $\Z_M$ rotation gauge field $\omega$ (see Sec.~\ref{sec:field} for a detailed definition). The holonomy of $(\vec{R},\omega)$ around a defect equals some space group element which we denote as $(\vec{b}_{\text{o}}, \Omega)$, with $\vec{b}_{\text{o}} \in \Z^2,M\Omega/2\pi \in \Z_M$. $\vec{b}_{\text{o}}$ and $\Omega$ encode the dislocation Burgers vector and the disclination angle of each defect. We assume that the starting point of the loop is an integer vector away from $\OO$. Now suppose we choose a different starting point for the loop $\text{o}' = \text{o}+\vec{v}$, where $\vec{v}$ may be fractional. We wish to relate $\vec{b}_{\text{o}'}$ to $\vec{b}_{\text{o}}$.

The group element which describes the same defect but w.r.t. the new origin $\text{o}'$ is obtained from $(b_{\text{o}},\Omega)$ by conjugating with the translation $(\vec{v},0)$:
\begin{equation}
    (\vec{v},0) (\vec{b}_{\text{o}},\Omega) (-\vec{v},0) = (\vec{b}_{\text{o}} + \vec{v} - U(\Omega) \vec{v},\Omega).
\end{equation}
Here $U(\Omega)$ is the rotation matrix with angle $\Omega$. The above result follows from the multiplication law for the space group. This group element must give the holonomy of $\vec{R},\omega$ w.r.t a loop starting at $\OO'$. Thus we find that
\begin{equation}
    \vec{b}_{\text{o}'} = \vec{b}_{\text{o}} + \vec{v} - U(\Omega) \vec{v},
\end{equation}
as claimed.

\subsection{Properties of $\mathscr{S}_{\text{o}}, \vec{\mathscr{P}}_{\text{o}}$}

The results below were first obtained in Refs.~\cite{manjunath2021cgt,zhang2022fractional}, and are compiled here for completeness. Suppose the rotation point group of the system is $\Z_M$, and consider an origin $\text{o}$ whose site symmetry group is $\Z_{M'}$. The topological response theory which defines $\SO, \PO$ is reviewed in App. \ref{sec:field}. Although the quantization conditions are derived using group cohomology arguments that are mathematically involved, the following intuition is enough to understand them. If a lattice defect has disclination angle $\Omega$ and Burgers vector $\vec{b}_{\text{o}}$, the total charge at the defect is a sum of various independent contributions from the field theory, Eq.~\eqref{eq:chargeresponse}. In particular, the term $\frac{\SO}{2\pi} A \wedge d\omega$ assigns this defect a fractional charge $\SO \frac{\Omega}{2\pi}$, while the term $\frac{\PO}{2\pi} \cdot A \wedge \vec{T}$ assigns the defect a fractional charge $\PO \cdot \vec{b}_{\text{o}}$. 

First we argue that $\mathscr{S}_{\text{o}}$ is either an integer or a half-integer, and it is defined modulo $M'$. Note that the term $\frac{\mathscr{S}_{\text{o}}}{2\pi} A \wedge d\omega$ assigns $U(1)$ charge $\frac{\mathscr{S}_{\text{o}}}{M}$ to a defect with disclination angle $\frac{2\pi}{M}$ and zero Burgers vector. On a closed spatial manifold of genus $g$, the total charge assigned by this term equals $\frac{\mathscr{S}_{\text{o}}}{2\pi} \int d\omega = 2\mathscr{S}_{\text{o}} (1-g)$. This must be an integer, for each $g$, irrespective of the other terms in the response theory. Therefore $2\mathscr{S}_{\text{o}}$ must be an integer. 

Now, the charge at any defect can be changed by an integer through local operators which add a fermion at the defect. Therefore, the topological terms $\frac{\SO}{2\pi} A \wedge d\omega$ and $\frac{\SO+M'}{2\pi} A \wedge d\omega$, which assign charge $\frac{\mathscr{S}_{\text{o}}}{M'}$ and $\frac{\mathscr{S}_{\text{o}}}{M'}+1$ to an elementary $2\pi/M'$ disclination, are topologically equivalent (in fact, we can show that their difference does not contribute to the partition function on a closed manifold). This leads to the relation $\mathscr{S}_{\text{o}} \simeq \mathscr{S}_{\text{o}} + M'$.

The arguments given in Ref.~\cite{zhang2022fractional} further show that if the generator of rotations at $\OO$ satisfies $\tilde{C}_{M',\OO}^{M'} = +1$, there is a relation 
\begin{equation}
    \mathscr{S}_{\text{o}} = \frac{C}{2} \mod 1.
\end{equation}
Thus $\mathscr{S}$ is an integer or a half integer according as $C$ is even or odd. 
In particular, once we fix $\text{o}$ and the Chern number, $\SO$ can only take one of $M'$ distinct values. 

Next, we explain the quantization of $\vec{\mathscr{P}}_{\text{o}}$, which turns out to be very sensitive to $M'$. Consider a lattice defect with parameters $(\vec{b}_{\text{o}},\Omega)$. 
The response theory predicts that the fractional charge at such a defect receives a contribution
$$ \vec{\mathscr{P}}_{\text{o}} \cdot \vec{b}_{\text{o}} \mod 1 $$
from the term with $\PO$. Let us shift the origin by a lattice vector $\vec{\Lambda}$. This takes $\vec{b}_{\text{o}} \rightarrow \vec{b}_{\text{o}} + \vec{\Lambda} - U(\Omega) \vec{\Lambda}$. The charge at the defect with respect to the new origin is
$$ \vec{\mathscr{P}}_{\text{o}} \cdot (\vec{b}_{\text{o}} +  \vec{\Lambda} - U(\Omega) \vec{\Lambda}) \mod 1.$$
For the theory to be consistent under a shift of origin, the above quantities must agree mod 1, i.e. 
\begin{equation}
\vec{\mathscr{P}}_{\text{o}} \cdot (\vec{\Lambda} - U(\Omega) \vec{\Lambda}) = 0 \mod 1.
\end{equation}
We can now take the minimal allowed disclination angle $\Omega = \frac{2\pi}{M'}$, and obtain the desired quantization:
\begin{equation}\label{eq:Pcdotb}
\vec{\mathscr{P}}_{\text{o}} \cdot (\vec{\Lambda} - U(\frac{2\pi}{M'}) \vec{\Lambda}) = 0 \mod 1.
\end{equation}

The distinct choices of $\vec{\mathscr{P}}_{\text{o}}$ compatible with this condition form a group which we denote $K_{M'}$. Using the rotation matrices given in Table~\ref{tab:U_mats}, we show below that $\vec{\mathscr{P}}_{\OO}$ is a nontrivial topological invariant only when $M' = 2,3,4$. 
The general parametrization for $\vec{\mathscr{P}}_{\text{o}}$ is given in Eq.~\eqref{eq:parametrization}.

\subsubsection{Deriving the quantization of $\vec{\mathscr{P}}_{\OO}$} \label{app:Pcquant}
The matrix representation of $U(2\pi/M')$ for different $M'$ (chosen to be consistent with the coordinate axis definitions in Fig.~\ref{fig:wyckoff}) is given in Table~\ref{tab:U_mats}.

When $M' = 2$, Eq.~\eqref{eq:Pcdotb} gives
\begin{equation}
    2\mathscr{P}_{\OO,x} \Lambda_x + 2\mathscr{P}_{\OO,y} \Lambda_y = 0 \mod 1 
\end{equation}
for any $\Lambda_x,\Lambda_y$. Therefore we set $\mathscr{P}_{\OO,x} = \frac{\overline{\mathscr{P}}_{\OO,x}}{2},\mathscr{P}_{\OO,y} = \frac{\overline{\mathscr{P}}_{\OO,y}}{2}$ where $\overline{\mathscr{P}}_{\OO,x},\overline{\mathscr{P}}_{\OO,y}$ are integers.

When $M' = 3$, Eq.~\eqref{eq:Pcdotb} gives after simplification
\begin{align}
    \mathscr{P}_{\OO,x}  = \mathscr{P}_{\OO,y} = -2\mathscr{P}_{\OO,x}\mod 1
\end{align}
implying that we can set $\vec{\mathscr{P}}_{\OO} = \frac{\overline{\mathscr{P}}_{\OO}}{3}(1,2)$ for some integer $\overline{\mathscr{P}}_{\OO}$, with $\overline{\mathscr{P}}_{\OO} \simeq \overline{\mathscr{P}}_{\OO} + 3$.

When $M' = 4$, Eq.~\eqref{eq:Pcdotb} gives after simplification
\begin{align}
    \mathscr{P}_{\OO,x}  = \mathscr{P}_{\OO,y} = -\mathscr{P}_{\OO,y}\mod 1
\end{align}
implying that we can set $\vec{\mathscr{P}} = \frac{\overline{\mathscr{P}}_{\OO}}{2}(1,1)$ for some integer $\overline{\mathscr{P}}_{\OO}$, with $\overline{\mathscr{P}}_{\OO} \simeq \overline{\mathscr{P}}_{\OO} + 2$.

When $M' = 6$, Eq.~\eqref{eq:Pcdotb} gives after simplification
\begin{align}
    \mathscr{P}_{\OO,x}  = \mathscr{P}_{\OO,y} = 0\mod 1
\end{align}
implying that $\PO$ must be an integer vector and is therefore trivial.

\section{Calculation of $\vec{\mathscr{P}}_{\text{o}}$ and $\SO$ when $C=0$} \label{app:C=0calcs}
\subsection{$\PO$}
Here we show how to calculate $\PO$ in the limit of zero Chern number, where the system can be adiabatically connected to an insulator in which the charge density is a sum of delta functions at the maximal Wyckoff positions $i=\alpha, \beta, \gamma, \delta$. The charge at these points is always an integer, of the form $N_i$. 
In this limit, the polarization $\vec{\mathscr{P}_{\text{o}}}$ is related to the dipole moment $\vec{P}_{\text{o}}:=\sum_{j\in\Theta}Q_j\vec{r}_j\mod \Z^2$, where $\Theta$ is the unit cell, and $\vec{r}_j$ is a representative position vector for the point $j$.


Next we relate $\vec{P}_{\OO}$ to the desired quantity $\PO$: we show that
\begin{equation}
    (\mathscr{P}_{\Ox}, \mathscr{P}_{\Oy}) = (P_{\Oy}, -P_{\Ox}) \mod \Z^2.
\end{equation}
This is true irrespective of any rotational symmetries. The argument is as follows (it is adapted from statements in Ref.~\cite{Song2021polarization}, although our result differs from theirs by a sign). We consider a clean lattice in which the vector potential satisfies $A_t = 0$, and the translation gauge fields have components satisfying the `natural' choice
$$X_x = Y_y = 1; X_y = Y_x = 0$$ 
(implying that $\int_{\vec{r}_i}^{\vec{r}_f} (X_x,Y_y) = \vec{r}_f - \vec{r}_i$). We set $\omega = 0$ for convenience and assume that 
$$\mathcal{L} = \frac{C}{4\pi} A \wedge dA + \frac{\PO}{2\pi} \cdot A \wedge d\vec{R} + \nu A \wedge A_{XY},$$
neglecting all other terms. With the above assumptions, $\mathcal{L}$ can be expanded as follows:
\begin{align}
    \mathcal{L} &= \frac{C}{4\pi} (- A_x \partial_t A_y + A_y \partial_t A_x) \nonumber \\
    & -\frac{\mathscr{P}_{\Ox}}{2\pi} \partial_t A_y + \frac{\mathscr{P}_{\Oy}}{2\pi} \partial_t A_x.
\end{align}
Therefore the expression for the current ${\bf j}$ is
\begin{align}
    {\bf j} &= \left(\frac{\delta \mathcal{L}}{\delta A_x},\frac{\delta \mathcal{L}}{\delta A_y}\right) \nonumber \\
    &= \frac{1}{2\pi}\partial_t (-\mathscr{P}_{\Oy},\mathscr{P}_{\Ox}) + \frac{C}{2\pi} (-E_x,E_y)
\end{align}
where $E_x,E_y$ are the electric field components. But when $C=0$, we can also write (after picking a suitable normalization)
\begin{equation}
    {\bf j} = \frac{1}{2\pi}\partial_t \vec{P}_{\OO}.
\end{equation}
Comparing the equations for ${\bf j}$ then gives Eq.~\eqref{eq:P_vs_scrP}. The above argument holds on the infinite plane or the torus, with and without rotational symmetry.

Below we consider $M=2, 3, 4$ separately; for $M=6$, $\PO \simeq (0,0)$. We first determine $\vec{P}_{\OO}$ as a dipole moment, and then find the corresponding $\PO$.

\underline{M=2}: Using Fig.~\ref{fig:wyckoff} as reference, we find that, modulo 1,
\begin{align}
    \vec{P}_{\alpha} &= \frac{1}{2} (N_{\beta} + N_{\gamma}, N_{\beta} + N_{\delta}) \nonumber \\
    \vec{P}_{\beta} &= \frac{1}{2} (N_{\alpha} + N_{\delta}, N_{\alpha} + N_{\gamma}) \nonumber \\
    \vec{P}_{\gamma} &= \frac{1}{2} (N_{\alpha} + N_{\delta}, N_{\beta} + N_{\delta}) \nonumber \\
    \vec{P}_{\delta} &= \frac{1}{2} (N_{\beta} + N_{\gamma}, N_{\alpha} + N_{\gamma}). 
\end{align}
This implies that, modulo 1, 
\begin{align}
    \vec{\mathscr{P}}_{\alpha} &= \frac{1}{2} (N_{\beta} + N_{\delta}, N_{\beta} + N_{\gamma}) \nonumber \\
    \vec{\mathscr{P}}_{\beta} &= \frac{1}{2} (N_{\alpha} + N_{\gamma}, N_{\alpha} + N_{\delta}) \nonumber \\
    \vec{\mathscr{P}}_{\gamma} &= \frac{1}{2} (N_{\beta} + N_{\delta}, N_{\alpha} + N_{\delta}) \nonumber \\
    \vec{\mathscr{P}}_{\delta} &= \frac{1}{2} (N_{\alpha} + N_{\gamma}, N_{\beta} + N_{\gamma}). 
\end{align}

\underline{M=3}: Fig.\ref{fig:wyckoff} shows that, modulo 1,
\begin{align}
    \vec{P}_{\alpha} &= \frac{1}{3} (N_{\beta} + 2N_{\gamma}, N_{\beta} +2N_{\gamma}) \nonumber \\
    \vec{P}_{\beta} &= \frac{1}{3} (N_{\gamma} + 2N_{\alpha}, N_{\gamma} + 2N_{\alpha}) \nonumber \\
    \vec{P}_{\gamma} &= \frac{1}{3} (N_{\alpha} + 2N_{\beta}, N_{\alpha} + 2N_{\beta}).
\end{align}
This implies that, modulo 1,
\begin{align}
    \vec{\mathscr{P}}_{\alpha} &= \frac{1}{3} (N_{\beta} + 2N_{\gamma}, 2N_{\beta} + N_{\gamma}) \nonumber \\
    \vec{\mathscr{P}}_{\beta} &= \frac{1}{3} (N_{\gamma} + 2N_{\alpha}, 2N_{\gamma} + N_{\alpha}) \nonumber \\
    \vec{\mathscr{P}}_{\gamma} &= \frac{1}{3} (N_{\alpha} + 2N_{\beta}, 2N_{\alpha} + N_{\beta}). 
\end{align}
Since by convention $\vec{\mathscr{P}}_{\text{o}} = \frac{\overline{\mathscr{P}}_{\text{o}}}{3}(1,2) \mod 1$, the value of $\overline{\mathscr{P}}_{\text{o}}$ can be read off from the $x$ component of the above equations.

\underline{M=4}: Fig.\ref{fig:wyckoff} shows that, modulo 1,
\begin{align}
    \vec{P}_{\alpha} &= \frac{N_{\beta} + N_{\gamma}}{2} (1,1) \nonumber \\
    \vec{P}_{\beta} &= \frac{N_{\alpha} + N_{\gamma}}{2} (1,1) 
\end{align}
In this case, $\vec{P}_{\text{o}} = \vec{\mathscr{P}}_{\text{o}} \mod 1$, and $\overline{\mathscr{P}}_{\alpha(\beta)} = N_{\beta(\alpha)} +N_{\gamma} \mod 2$. The polarization at the points $\gamma_1,\gamma_2$ with $C_2$ symmetry can be obtained from the above $M=2$ results by taking $\gamma_2 = \gamma, \gamma_1 = \delta$, and $N_{\gamma} = N_{\delta}$.

\begin{figure*}[t]
    \centering
    \includegraphics[width=15cm]{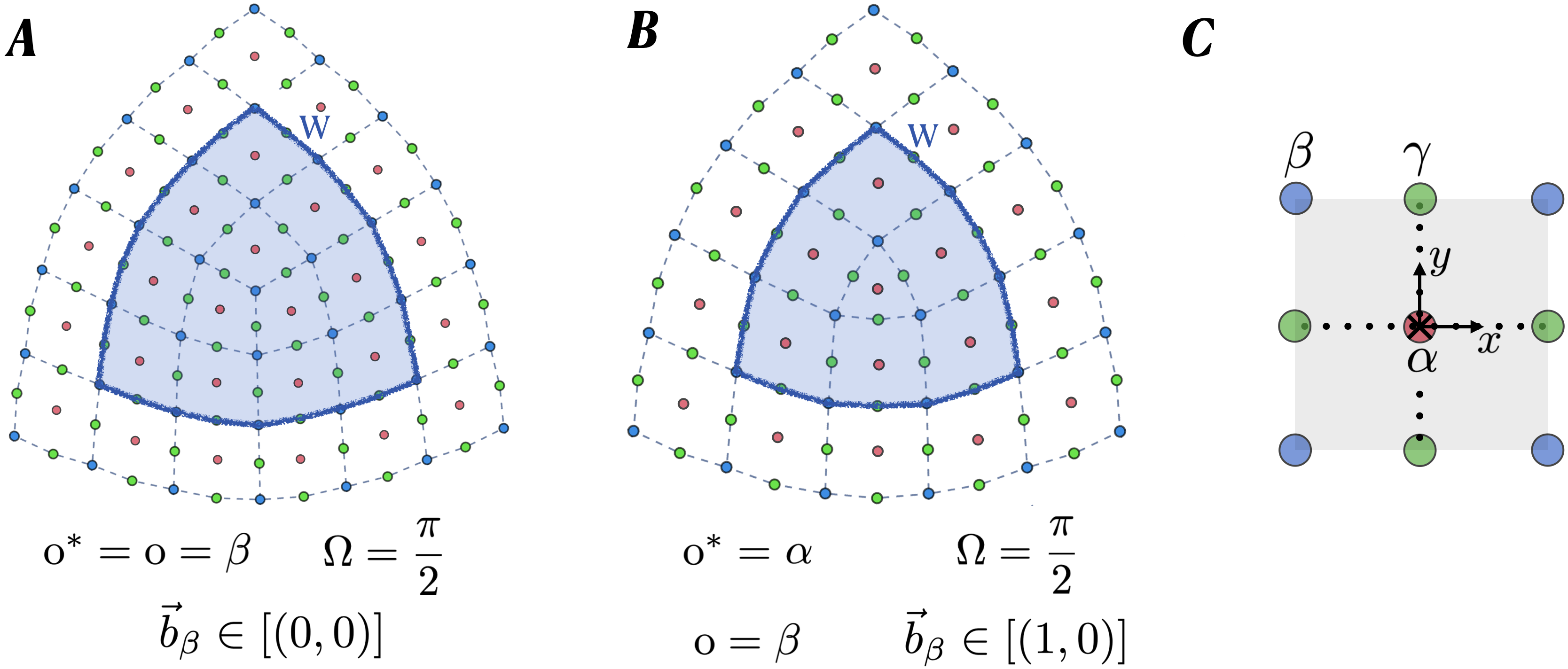}
    \caption{Pure disclinations with two different choices of $\OO^*$: \textbf{A.}~$\OO^*=\beta$, \textbf{B.}~$\OO^*=\alpha$. Dashed lines represent unitcell boundary \textbf{C.} The $M=4$ unit cell}
    \label{fig:ostarc4}
\end{figure*}

\subsection{$\SO$}
The calculation of $\SO$ in a system with $M$-fold rotational symmetry proceeds by constructing a disclination with disclination angle $2\pi/M$ at $\OO$ and measuring the fractional charge $Q_W$ in a region $W$ around the disclination. When $C=0$ we can consider the fully filled limit in which the charge at any point in the MWPs $i = \alpha, \beta, \gamma, \delta$ is equal to $N_{i}$. Then, by working out the different cases, we find that
\begin{equation}
    \SO = N_{\OO} \mod M
\end{equation}
for $\OO = \alpha, \beta, \gamma, \delta$.

\subsection{Argument that $\SO$ is independent of $\OO^*$}\label{sec:ostarindependence}

In this section we use a square lattice calculation with $C=0$ as an example to argue that $\mathscr{S}_{\OO}$ does not depend on the origin $\text{o}^*$ used to construct the disclination, but only on the point $\text{o}$ used to measure the dislocation Burgers vector. Say we want to prove that $\mathscr{S}_{\beta}$ is independent of the origin. We consider $\text{o}^*=\beta$ and $\text{o}^*=\alpha$ separately and construct two different $\Omega=\frac{\pi}{2}$ disclinations using $\tilde{C}_{4,\text{o}^*}$ (see Fig.~\ref{fig:ostarc4}). We fix $\text{o}=\beta$ in both cases and extract $\mathscr{S}_{\beta}$ from the charge response. We have numerically checked that $\mathscr{S}_{\beta}$ in for these two choices of $\OO^*$ are the same (see Fig.~\ref{fig:SandP}). Now we analytically show that $\SO$ is independent of $\OO^*$ when $C=0$.

We pick the region $W$ shown in Fig.~\ref{fig:ostarc4}. We first consider Fig.~\ref{fig:ostarc4}A where $\OO^*=\beta$. In the $C=0$ limit, the charge response is
\begin{equation}
    Q_W=\frac{\mathscr{S}_{\beta}}{4}+12\nu \mod 1 
\end{equation}
A direct counting of Wannier orbitals gives
\begin{equation}
    Q_W= 12 N_{\alpha}+ (12+\frac{1}{4}) N_{\beta}+24N_{\gamma}.   
\end{equation}
Since $\nu=N_{\alpha}+N_{\beta}+2N_{\gamma}$, we obtain $\mathscr{S}_{\beta}=\frac{1}{4}N_\beta$.

Now we switch to $\OO=\alpha$ (see Fig.~\ref{fig:ostarc4}B). The Burgers vector is non-trivial and there is a irregular unit cell in the defect core contributing $n_{\text{irreg},\vec{b},\OO}=\frac{1}{2}\nu$. Taking $C=0$, the impure disclination charge is

\begin{equation}
    Q_W=\frac{\mathscr{S}_{\beta}}{4}+\frac{\overline{\mathscr{P}}_{\beta}}{2}+(6+\frac{1}{2})\nu \mod 1.
\end{equation}
A direct counting of Wannier orbitals gives 

\begin{equation}
    Q_W= 7 N_{\alpha}+ (6+\frac{3}{4}) N_{\beta}+(12+\frac{3}{2})N_{\gamma}.   
\end{equation}
We now use the $C=0$ result $\overline{\mathscr{P}}_{\beta}=N_{\alpha}+N_{\gamma}$ which was obtained above. Combined with the above equations, we again obtain $\mathscr{S}_{\beta}=\frac{1}{4}N_\beta$. This completes our verification that $\mathscr{S}_{\beta}$ is independent of $\OO^*$. We can apply similar arguments to verify that $\mathscr{S}_{\OO}$ is independent of $\OO^*$ for each $\OO$ and each $M=2,3,4,6$.

\section{Microscopic construction of a dislocation Hamiltonian }\label{sec:dislocationConstruction}

In this section, we demonstrate a systematic method to construct a dislocation Hamiltonian $H_{\text{defect}}$ in terms of the Hamiltonian $H_{\text{clean}}$ on a clean rotationally symmetric lattice with $M=2,3,4,6$. The procedure is similar in spirit to the construction of disclination defects given in \cite{zhang2022fractional}.

\begin{figure}[t]
    \centering
    \includegraphics[width=6cm]{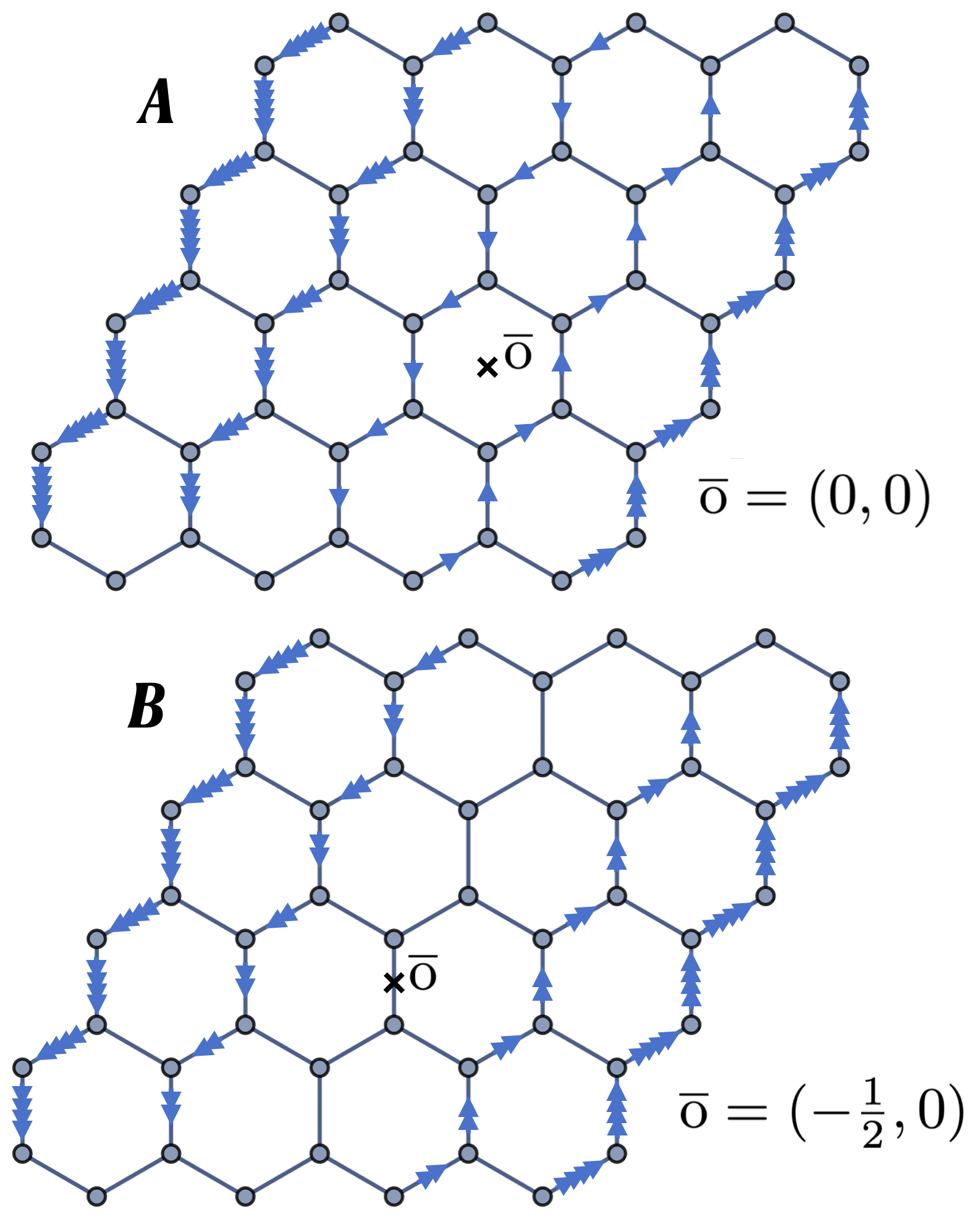}
\caption{Landau gauge on a honeycomb lattice with different gauge origin $\overline{\text{o}}$, each blue arrow represents a hopping phase $A_{ij}=\frac{\phi}{4}$. }
    \label{fig:gaugeorigin}
\end{figure}

Since the dislocation is a defect of the translation symmetry, the construction necessarily involves the magnetic translation operator $\tilde{T}\equiv e^{i\sum_j\lambda_jc_j^{\dagger}c_j}\hat{T}$, which is a translation followed by a gauge transformation which ensures that $\tilde{T}$ is a symmetry of $H_{\text{clean}}$.
For concreteness, we write a general form of the Landau gauge on the square and honeycomb lattices below (this is analogous to the previously defined gauge on the torus, Eq.\eqref{eq:generalgauge}). For the square lattice,
\begin{equation}\label{eq:generalgaugeinfinitesquare}
    \begin{aligned}
        &A_{j,j+\hat{y}}=(j_x-\overline{\text{o}}_x )\phi\\
        &j=(j_x,j_y),~ j_x\in \Z, j_y\in\Z.
    \end{aligned}
\end{equation}
This vector potential is defined on an infinite plane. In this appendix we use $A \equiv A_{\text{clean}}$ to denote the entire vector potential on the clean lattice. $\overline{\text{o}}=(\overline{\text{o}}_x,\overline{\text{o}}_y)$ is called the gauge origin.
On the honeycomb lattice, consider Fig.~\ref{fig:gaugeorigin}. Assume that the boundaries of the unit cell align with the hoppings. Then define
\begin{align}\label{eq:generalgaugeinfinitehex}
        &A_{i,i+(-\frac{1}{3},\frac{2}{3})}=A_{j,j+(\frac{1}{3},\frac{1}{3})}= (x-\overline{\text{o}}_x )\frac{\phi}{2}\\
        &i=(i_x-\frac{1}{3},i_y-\frac{1}{3}),i_x\in \Z, i_y\in\Z\nonumber\\ &j=(j_x+\frac{1}{3},j_y+\frac{1}{3}),j_x\in \Z, j_y\in\Z.\nonumber
\end{align}
$i$ and $j$ are site indices at MWP $\beta$ and MWP $\gamma$ of the $C_3$ unit cell respectively (refer to Fig.\ref{fig:wyckoff} for the unit cell convention).

Using the aforementioned gauge choices, we now explain how to construct the dislocation Hamiltonian $H_{\text{defect}}$ for each $M$, starting with $H_{\text{clean}}$. In our numerics we use this procedure to construct dislocations with $\vec{b}=(\pm1,0),(0,\pm1)$; all other dislocation Burgers vectors lie in the same equivalence class as one of these choices of $\vec{b}$.

\subsection{$C_2$, $C_4$ dislocation}

Let us consider a dislocation on the simplest square lattice as an example. The construction below generalizes naturally to $C_2$ lattice dislocations since both of them have quadrilateral unit cells. Given an origin $\text{o}$ and the desired dislocation Burgers vector $\vec{b}$, the procedure for constructing $H_{\text{defect}}$ is summarized in three steps: 

\begin{enumerate}
    \item Draw a cut parallel to $\vec{b}$
    \item Define conjugate hopping terms
    \item Perform local moves
\end{enumerate}

\begin{figure}[t]
    \centering
    \includegraphics[width=8cm]{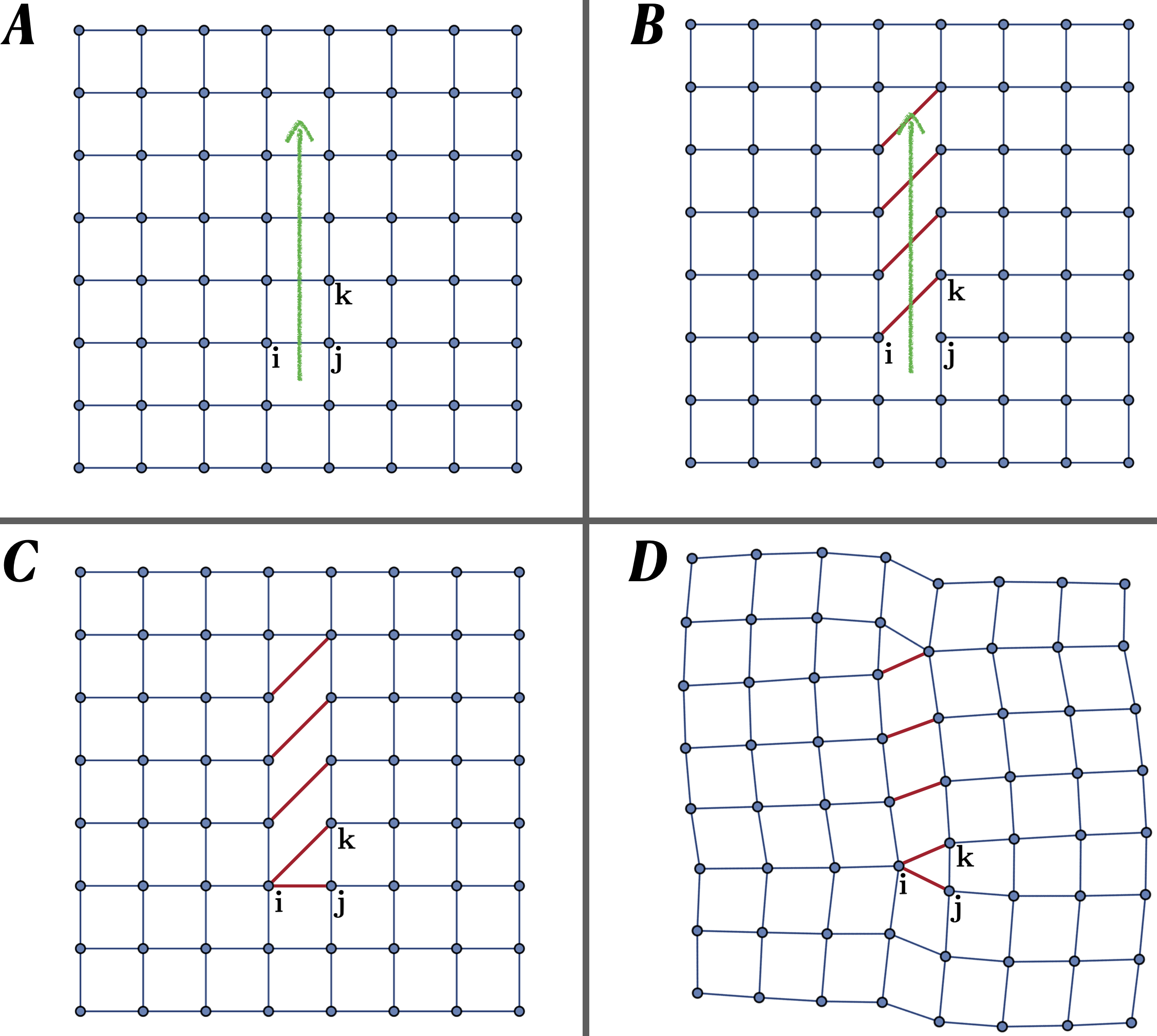}
    \caption{Dislocation construction procedure on a square lattice \textbf{A.} Drawing a cut in the $\hat{y}$ direction. \textbf{B.} Conjugate hoppings. \textbf{C.} Doing local moves. \textbf{D.} Reorganizing.}
    \label{fig:c4dislocation}
\end{figure}

The first step is to draw a linear cut in the direction of $\vec{b}$ ($\vec{b}=(0,1)$ in Fig.~\ref{fig:c4dislocation}). This cut starts and ends on two plaquette centers. The cut intersects a set of links in the lattice corresponding to hopping terms in $H_{\text{clean}}$. We will only consider 2-fermion hopping terms below.

Consider a particular term which intersects the cut, and is of the form $c_{L_j}^{\dagger}c_{R_j}e^{-iA_{L_j R_j}}$ in $H_{\text{clean}}$. $L_j$ and $R_j$ are sites on the left and right of the cut respectively.
We conjugate \textit{only} the operator $c_{R_j}$ with the translation operator $\tilde{T}_{{\bf y}}$: 
\begin{equation}\label{eq:translateHop}
    \begin{aligned}
        e^{-iA_{L_j R_j}}c_{L_j}^{\dagger}c_{R_j} &\rightarrow  e^{-iA_{L_j R_j}}c_{L_j}^{\dagger} \pqty{\tilde{T}_{{\bf y}} c_{R_j} \tilde{T}_{{\bf y}}^\dagger}\\&= e^{-iA_{L_j R_j}} \times e^{-i\lambda_{R_j}}c_{L_j}^\dagger c_{R_{j+1}}.
    \end{aligned}
\end{equation}
Starting with $H_{\text{clean}}$, we delete the original hoppings $e^{-iA_{L_j R_j}}c_{L_j}^{\dagger}c_{R_j}$, and add the new hoppings $e^{-i(A_{L_j R_j} + \lambda_{R_j})}c_{L_j}^\dagger c_{R_{j+1}}$ to obtain $H_{\text{defect}}$. In the example shown in Fig. \ref{fig:c4dislocation},
this procedure creates a dislocation with $\vec{b}=(0,1)$ on the $-y$ side of the cut and an anti-dislocation with $\vec{b}=(0,-1)$ on $+y$ side of the cut.

We want to consistently ensure that the irregular unit cells in this construction are all triangular, and not pentagonal or in some other irregular shape. Therefore, in the third step, we make one local change of the hopping terms such that both the dislocation and anti-dislocation have a triangular plaquette (see Fig. \ref{fig:c4dislocation} C). This fully determines $H_{\text{defect}}$.

\subsection{Properties of $\phi_{\text{irreg}}$ on square lattice}\label{app:phi-irreg}

Here we derive several statements that were made in Sec.~\ref{sec:c4_PO} regarding the flux $\phi_{\text{irreg}}$ in a dislocation plaquette on a square lattice.

\subsubsection{Proof of Eq.~\eqref{eq:phi-irreg}}

Consider Fig.\ref{fig:c4dislocation}. We refer to the point $j$ in this figure as $j_0$ in the main text. For a $\vec{b}=(0,1)$ Burgers vector, the flux in the irregular dislocation plaquette can be written in terms of the vector potential $A_{\text{defect}}$ on the defect lattice as follows:
\begin{equation}
\phi_{\text{irreg}}=A_{\text{defect},ij}+A_{\text{defect},jk}+A_{\text{defect},ki}.
\end{equation}
From Eq.~\eqref{eq:translateHop}, we see that the new hopping created in the defect lattice $tc_i^{\dagger} c_ke^{-iA_{\text{defect},ik}}$ can be calculated in terms of the vector potential on the clean lattice, i.e. $$A_{\text{defect},ik}=A_{ij}+\lambda_{j}.$$ 
Moreover, $ A_{\text{defect},ij} = A_{ij}$ and $ A_{\text{defect},jk} = A_{jk}$. Therefore 
\begin{equation}\label{eq:phi_irreg}
    \phi_{\text{irreg}}=A_{jk}-\lambda_j.
\end{equation}
In the main text we explicitly wrote $j = j_0, k = j_0 + \hat{y}$. 

\subsubsection{Change in $\phi_{\text{irreg}}$}

Next we prove the claims in Sec.~\ref{sec:c4_PO_DC} about how $\phi_{\text{irreg}}$ changes when either $ \lambda$ or $j_0$ is changed.

\underline{(1)}: Suppose we change $j_0 \rightarrow j_0 + (v_x,0)$, $v_x\in \Z$, keeping $A,\lambda$ fixed. Then the change in $\phi_{\text{irreg}}$ is 
\begin{align}
    \Delta \phi_{\text{irreg}} &= A_{j_0 + (v_x,0),j_0 + (v_x,0)+\hat{y}} \nonumber \\
    & \quad - \lambda_{j_0 + (v_x,0)} - (A_{j_0,j_0+\hat{y}} - \lambda_{j_0}).
\end{align}
Now we use Eq.~\eqref{eq:Ty-lambda}. On the infinite plane, Eq.~\eqref{eq:Ty-lambda} holds exactly, while on the torus, where it is not exact everywhere, we ensure that the dislocation is constructed within a region where this equality does hold. Then we get
\begin{equation}
    \lambda_{j_0} - \lambda_{j_0 + (v_x,0)} = A_{j_0,j_0 + (v_x,0)} - A_{j_0 + \hat{y},j_0 + (v_x,0) + \hat{y}}.
\end{equation}
Therefore
\begin{align}
    \Delta \phi_{\text{irreg}} &= A_{j_0 + (v_x,0),j_0 + (v_x,0)+\hat{y}} -A_{j_0,j_0+\hat{y}} \nonumber \\
    & + A_{j_0,j_0 + (v_x,0)} - A_{j_0 + \hat{y},j_0 + (v_x,0) + \hat{y}} \\
    &= +\phi v_x.
\end{align}
The last equality is because the sum over $A$ measures the flux in a rectangle of side lengths $v_x$ and $1$ in the $x$ and $y$ directions respectively.

\underline{(2)}: Suppose we change $\lambda_{j} \rightarrow \lambda_{j} + \chi$ for each $j$, and $\chi=\phi v_x$. $j_0$ and $A$ are fixed. Then from Eq.~\eqref{eq:phi_irreg}, $\Delta \phi_{\text{irreg}} = - \phi v_x$.

\subsubsection{Proof of Eq.~\eqref{eq:generalobar}}

With $\lambda_j$ as defined in Eq.~\eqref{eq:tyfullgaugeapp}, we calculate $\phi_{\text{irreg}}$ given that the dislocation is created at $x=j_{0,x}$ on the torus:
\begin{align}
    \phi_{\text{irreg}}=\begin{cases}
      (j_{0,x}-\overline{\OO}_x)\phi+\frac{\phi}{2}L_x\qquad   &j_{0,x}< \overline{\OO}_x\\
      (j_{0,x}-\overline{\OO}_x)\phi-\frac{\phi}{2}L_x\qquad &j_{0,x}> \overline{\OO}_x \\
      0\qquad &j_{0,x}=\overline{\OO}_x.
    \end{cases}
\end{align}

Since the sites have integer $x$ and $y$ coordinates, we use Eq.~\eqref{eq:lm_odef}, along with the following relation, which can be read off from Table~\ref{table:nuc}:
\begin{align}
  n_{\text{irreg},\OO}=- (\OO_x+\frac{1}{2})\mod 1.
\end{align}
This relation is defined mod 1 because an integer change of $n_{\text{irreg},\OO}$  does not change $\OO \mod \Z^2$.

From these two conditions we find that
$\OO_x=\overline{\OO}_x-\frac{1}{2}-\frac{L_x}{2}\mod 1$ whenever $j_{0,x} \ne \overline{\OO}_x$. This is the same as our numerical observation using Eq. \eqref{eq:generalobar}.

\begin{figure}[t]
    \centering
    \includegraphics[width=8cm]{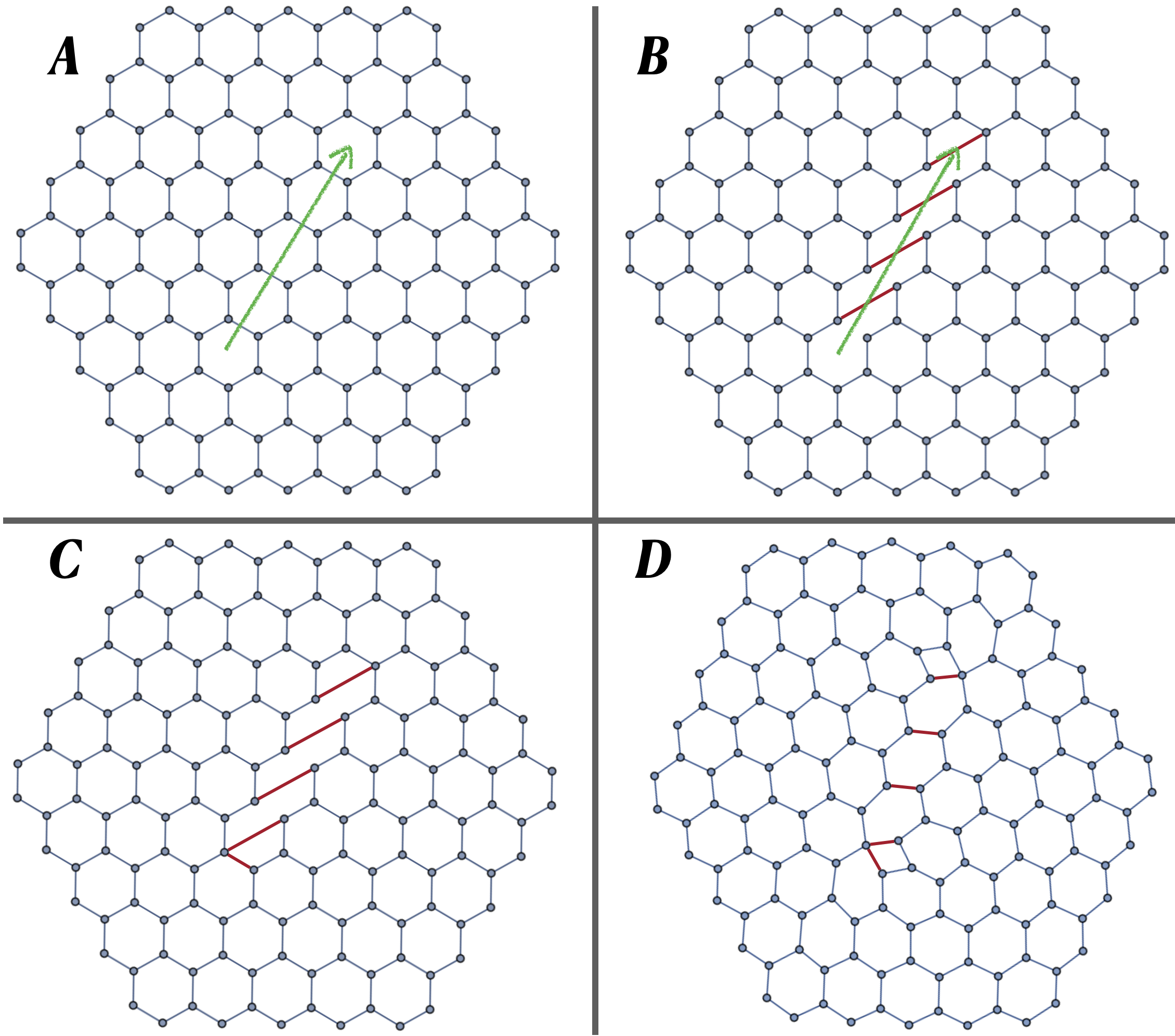}
    \caption{Dislocation construction procedure on a honeycomb lattice. \textbf{A.} Draw a cut in the $\hat{y}$ direction. \textbf{B.} Conjugate a set of hopping terms. \textbf{C.} Perform local moves. \textbf{D.} Reorganize}
    \label{fig:hexdislocation}
\end{figure}

\subsection{$C_3$, $C_6$ dislocation}

We can use a similar method to construct a $C_3$ or $C_6$ dislocation in the honeycomb lattice. Fig. \ref{fig:hexdislocation} shows the full procedure for constructing a $\vec{b}=(0,\pm 1)$ dislocation and anti-dislocation pair. 

With some fixed gauge origin $\overline{\text{o}}\equiv(\overline{\text{o}}_x,\overline{\text{o}}_y)$, we construct $H_{\text{defect}}$ from $H_{\text{clean}}$ following the same three steps as above. For consistency we choose the irregular unit cell to be quadrilateral (rather than, say, an octagon). As above, using $\delta\Phi_{W,\OO}=0$, we demand that $\tilde{T}_{\bf y}$ should create a dislocation with $\phi_{\text{irreg}}=\phi n_{\text{irreg}}$. Using $\lambda_j$ defined in \eqref{eq:tyfullgaugeappm3}, we again recover the relation between $\OO_x$ and $\overline{\OO}_x$ to be $\OO_x=\overline{\OO}_x-\frac{L_x}{2}-\frac{1}{2}\mod 1$.

\section{Independence of $\{\SO,\PO\}$ under shift of unit cell}\label{sec:trim}

In this section we describe the \textit{trimming} method which serves as a tool to prove that $\vec{\mathscr{P}}_{\text{o}}$ and $\mathscr{S}_{\text{o}}$ do not depend on the choice of unit cell $\Theta$. It is worth noting that the linear momentum calculation and the 1d polarization calculation are completely independent of $\Theta$. Only the charge response calculation depends on $\Theta$. We provide two examples to illustrate the method, for $\SO$ and $\PO$ respectively.

\subsection{$\SO$}

\begin{figure}[t]
    \centering
    \includegraphics[width=8cm]{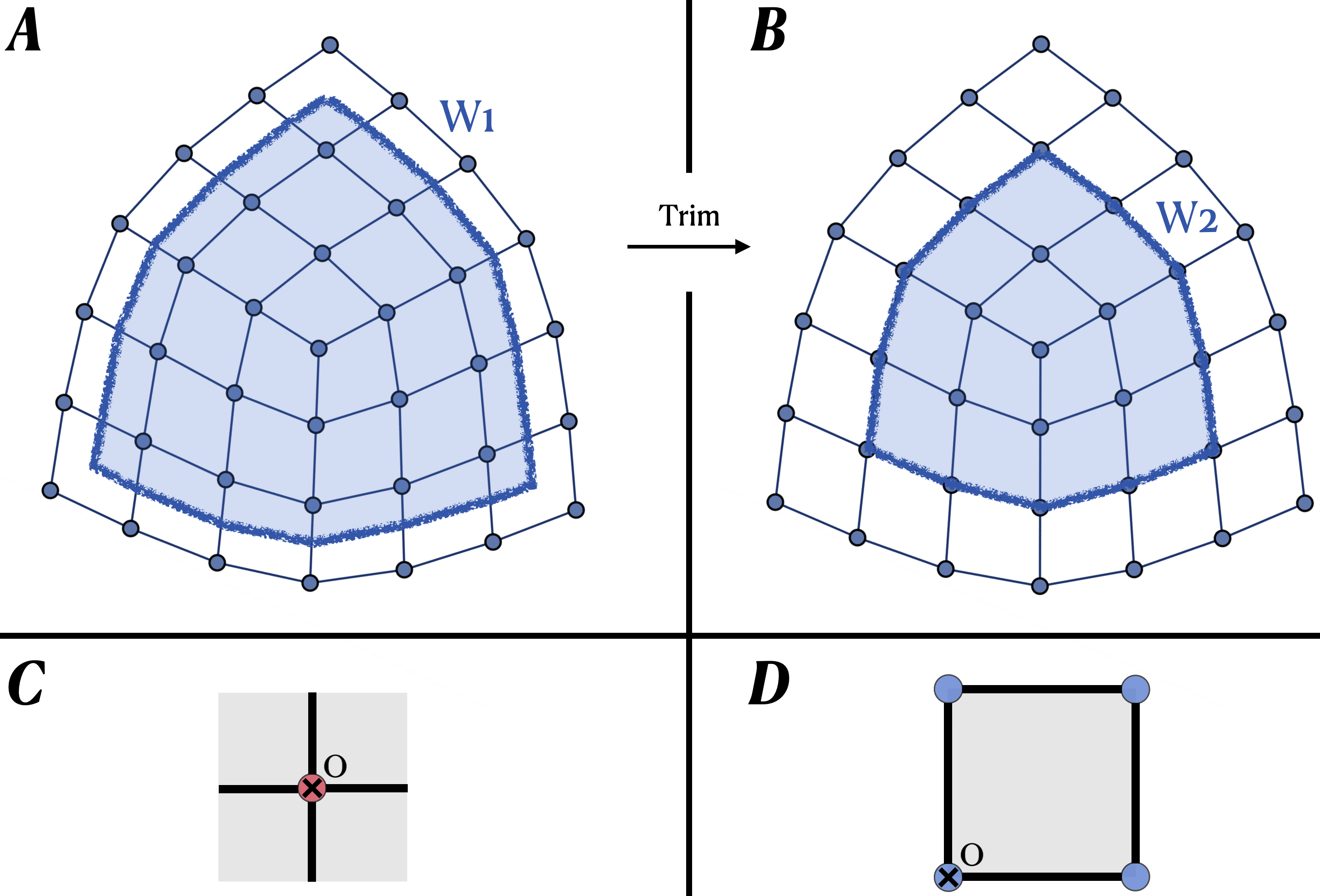}
    \caption{Trimming method for a pure disclination on the square lattice. We consider two regions \textbf{A.} $W_1$ and \textbf{B.} $W_2$. $W_1, W_2$ correspond to choices of the unit cell shown in \textbf{C.} and \textbf{D.} respectively. The origin $\text{o}$ is defined to be at a site. Black solid lines represent hoppings. Here $W_1$ is trimmed into $W_2$ by excluding subcells at the boundary.}
    \label{fig:squaretrim}
\end{figure}

Consider a pure disclination on the square lattice with the $C_4$ symmetric origin $\text{o}$ at a site, as shown in Fig.\ref{fig:squaretrim} \textbf{A}. We consider the two choices of unit cell $\Theta_1, \Theta_2$ shown in Fig.\ref{fig:squaretrim} \textbf{C} and \textbf{D}. We also define two regions $W_1, W_2$ whose boundaries are aligned with those of $\Theta_1, \Theta_2$ respectively.

Following Eq.\eqref{eq:generalpuredisclinationcharge}, the charge response for $W_1$ can be written as
\begin{equation}
    Q_{W1}=\frac{\mathscr{S}_\text{o}}{4}+(k_1+\frac{3}{4})\nu+\frac{C\delta{\Phi}_{W1}}{2\pi} \mod 1,
\end{equation}
where $k_1$ is the integer part of $n_{W_1}$. Since $\text{o}$ is at the center of the unit cell, we have $n_{\OO,\text{irreg},\Omega}=1-\frac{\Omega}{2\pi}=\frac{3}{4}$ as defined in Eq.~\eqref{eq:uco}.
The charge response for $W_2$ can be written as
\begin{equation}
 Q_{W2}=\frac{\mathscr{S}_\text{o}}{4}+k_2\nu+\frac{C\delta{\Phi}_{W2}}{2\pi} \mod 1.
\end{equation}
The extra flux $\delta{\Phi}_{W}$ is only inserted near the defect, and $W_1$ and $W_2$ only differ at their boundary. Therefore $\delta{\Phi}_{W1}=\delta{\Phi}_{W2}$.

In order to prove that $\mathscr{S}_{\text{o}}$ extracted in the two equations above is the same, we need to show that $Q_{W1}-Q_{W2}=(k_3+\frac{3}{4})\nu$. Since we can arbitrarily add a full unit cell on the boundary of $W_1$ or $W_2$, $k_3$ can vary freely without changing $\SO$, but the fractional part in parentheses needs to be $\frac{3}{4}$.

In order to compare $Q_{W1}$ and $Q_{W2}$, we \textit{trim} $W_1$ into $W_2$. This only requires cutting out \textit{subcells} on the boundary, which is assumed to be far away from the defect. Thus, on the boundary we expect the charge in each unit cell to be $\nu$, and the charge in each \textit{subcell} to be $\frac{1}{4}\nu$. With this information, we can explicitly calculate $Q_{W1}-Q_{W2}$.

For example, in Fig.\ref{fig:squaretrim} \textbf{A} and \textbf{B}, $Q_{W1}-Q_{W2}=(6+\frac{3}{4})\nu$, which satisfies the aforementioned condition.
The above process is repeated for disclinations with $M=2,3,4,6$ lattice with arbitrary $\text{o}$ and $\Omega$. In all cases, the value of $Q_{W_1}-Q_{W_2}$ obtained from the trimming method is consistent with a fixed value for $\SO$. This allows us to check that the extracted $\mathscr{S}_{\text{o}}$ does not depend on $\Theta$.

\subsection{$\PO$}

\begin{figure}[t]
    \centering
    \includegraphics[width=8cm]{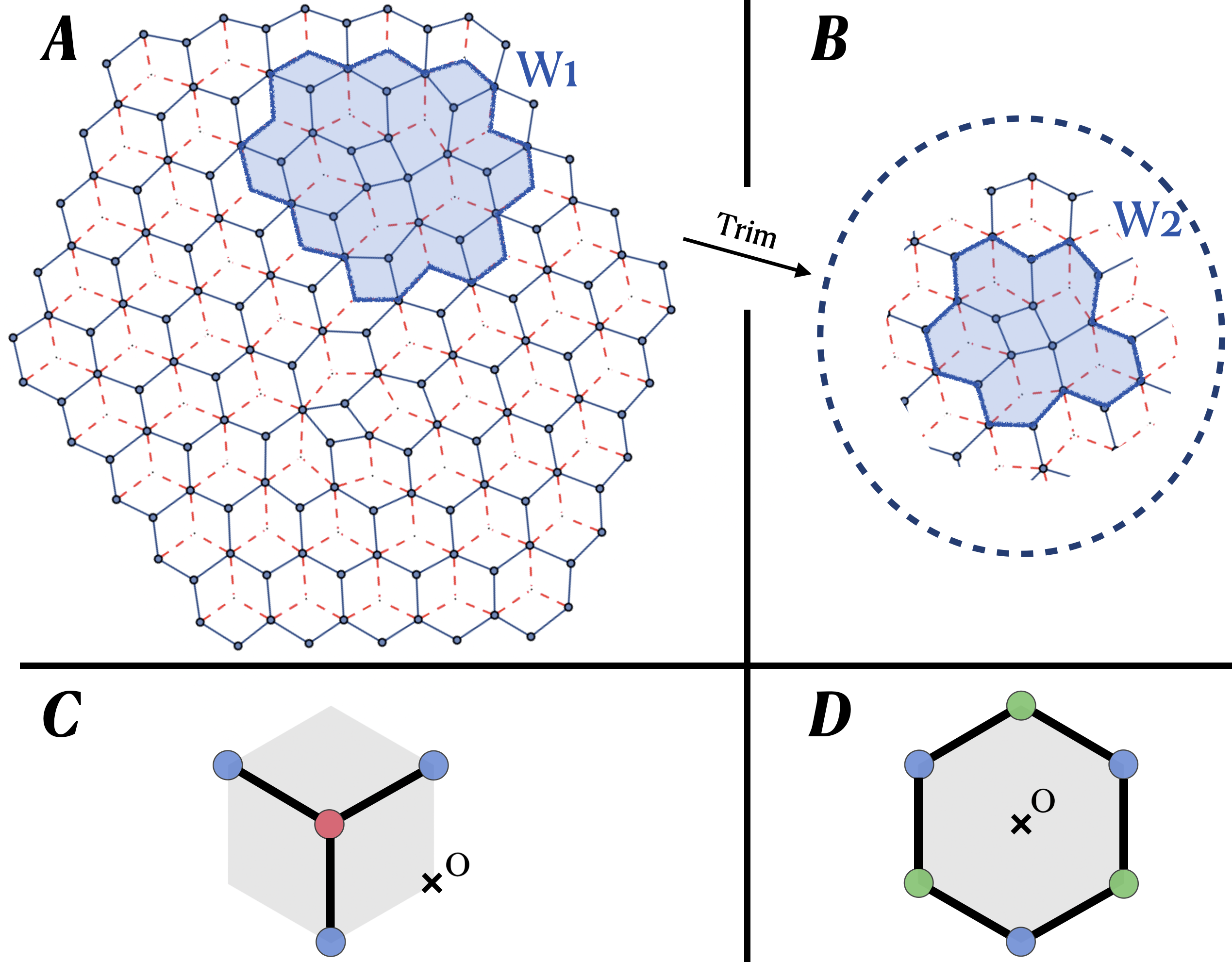}
    \caption{Trimming method for a dislocation. \textbf{A.} Honeycomb dislocation with region $W_1$ covering the $\vec{b}=(0,-1)$ dislocation. The boundary of $W$ is aligned with unitcell boundary of \textbf{C.}, shown as red dashed line.
    \textbf{B.} The region $W_2$ after trimming. Now the boundary of $W_2$ is aligned with the hopping, which is also the unit cell boundary of \textbf{D.}, shown as black solid line.}
    \label{fig:tileNtrim}
\end{figure}

We use a similar procedure to prove that $\vec{\mathscr{P}}_{\text{o}}$ does not depend on the choice of $\Theta$.
Consider the honeycomb latice dislocation shown in Fig.\ref{fig:hexdislocation} \textbf{D} with Burgers vector $\vec{b}=(0,-1)$. Let us consider $\text{o}$ to be at a plaquette center. Two choices of $\Theta$ are shown in Fig. \ref{fig:tileNtrim} \textbf{C} and \textbf{D}. 

The charge response with unit cell as in Fig. \ref{fig:tileNtrim} \textbf{C} can be written as
\begin{equation}
Q_{W1}=\vec{\mathscr{P}}_{\text{o}}\cdot\vec{b}+(k_1+\frac{2}{3})\nu+\frac{C\delta{\Phi}_{W1}}{2\pi} \mod 1.
\end{equation}
Again, $k_1$ is the integer part of the number of unit cells enclosed by $W_1$. Since the position of the origin relative to the center of the unit cell is $(\frac{2}{3},-\frac{1}{3})$, the fractional value of $n_{\text{irreg}}$ can be read off from Table \ref{table:nuc}. Similarly, the charge response with $\Theta_2$ as in Fig. \ref{fig:tileNtrim} \textbf{D} can be written as
\begin{equation}
Q_{W2}=\vec{\mathscr{P}}_{\text{o}}\cdot\vec{b}+k_2\nu+\frac{C\delta{\Phi}_{W2}}{2\pi} \mod 1.
\end{equation}
Similar to the previous example, $\delta{\Phi}_{W1}=\delta{\Phi}_{W2}$, and we need to show that $Q_{W1}-Q_{W2}=(k_3+\frac{2}{3})\nu$. 

In the example shown in Fig.\ref{fig:tileNtrim} \textbf{A} and \textbf{B}, each subcell contributes $\frac{1}{3}\nu$. We have $Q_{W1}-Q_{W2}=\frac{14}{3}\nu=(4+\frac{2}{3})\nu$, giving the expected answer.

This procedure can be extended to $M=2,3,4,6$ lattices with arbitrary $\text{o}$ and $\vec{b}$. With this we can verify that $\vec{\mathscr{P}}_{\text{o}}$ does not depend on $\Theta$.

\section{Further details in linear momentum calculation}\label{app:linmom_details}

\begin{figure}[t]
    \centering
    \includegraphics[width=8.7cm]{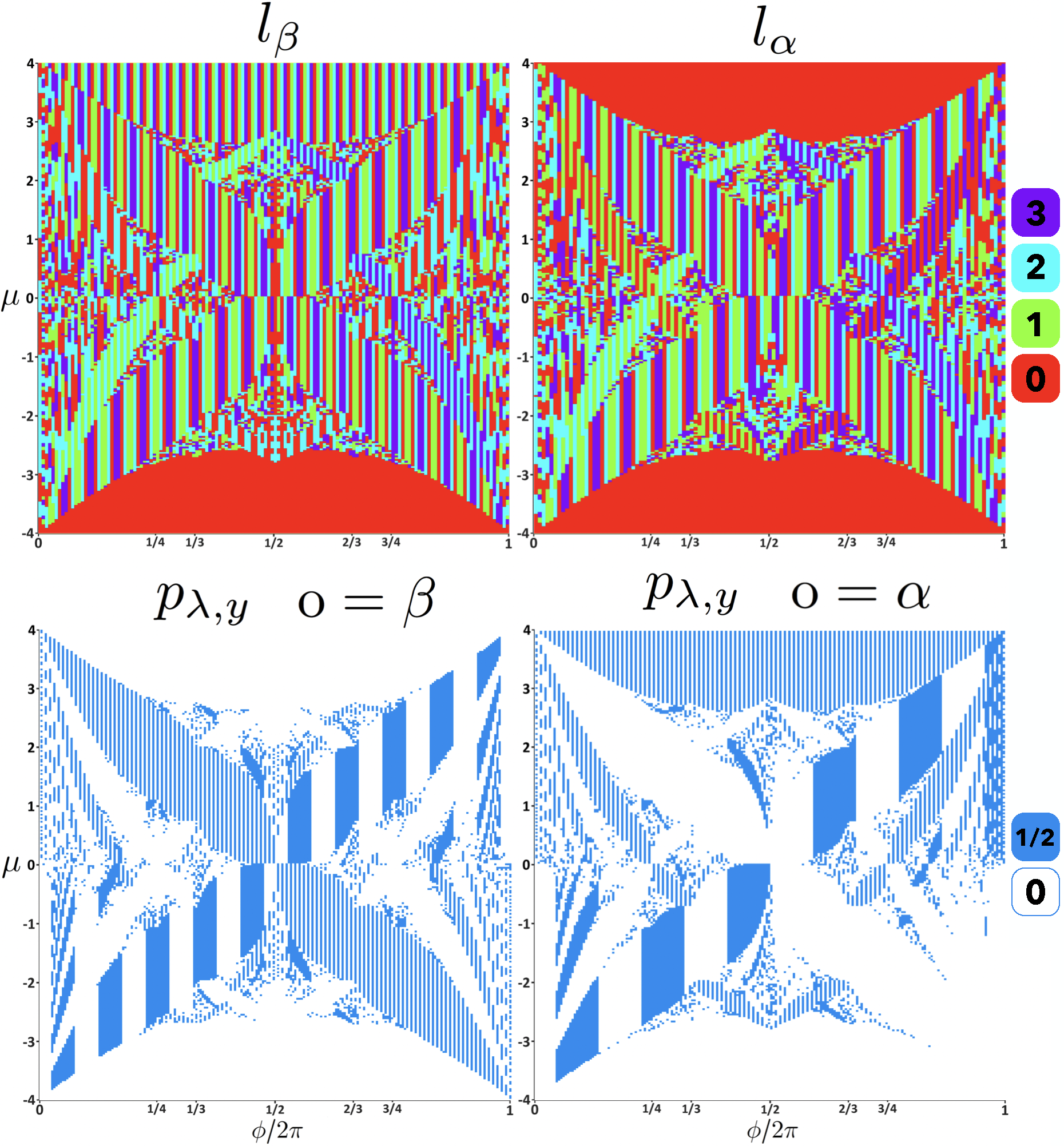}
    \caption{Bare numerical result of \textbf{Top:} angular momentum $l_{\OO}$ which follows Eq.~\eqref{eq:angularMomentum}; both calculated on an $L_x\times L_y=12 \times 12$ torus. \textbf{Bottom:} linear momentum $p_{\lambda,y}$ which follows Eq.~\eqref{eq:linearmomentum}; calculated on the $L_x\times L_y=19\times 11$ and $L_x\times L_y=20\times 11$ torus respectively. All sub-figure are plotted in the $0<\phi<2\pi$ range.}
    \label{fig:bare_momentums}
\end{figure}

In this appendix, we first state some numerical observations on how the quantization of the linear momentum changes upon starting with the operator $\tilde{T}_{\bf y}$ as in Eq.~\eqref{eq:tyfullgaugeapp} but then changing the location of the strip that violates translation symmetry. Then we discuss some technical details regarding the partial translation method. The numerical results of linear momentum can be seen in Fig.~\ref{fig:bare_momentums}.

We note that in the main text, the origin $\OO$ for the charge polarization was defined in terms of the `gauge origin' $\overline{\OO}$. This is however not the only way to define $\OO$. For a given $A,\lambda$ where $\lambda$ is associated to the operator $\tilde{T}_{\bf y}$, there is a point $\tilde{\OO}$ which satisfies 
\begin{equation}
    \lambda_{\tilde{\OO}} - A_{\tilde{\OO},\tilde{\OO} + \hat{y}} = 0.
\end{equation}
$\tilde{\OO}$ does not change under gauge transformations, which take $A_{ij} \rightarrow A_{ij} + f_j - f_i$ and $\lambda_j \rightarrow \lambda_j - f_j + f_{j + \hat{y}}$. This transformation of $\lambda_j$ is obtained by solving Eq.~\eqref{eq:Ty-lambda} after gauge transforming $A$. Therefore $\tilde{\OO}$ is a gauge-invariant point, and we can in principle define $\OO$ in terms of $\tilde{\OO}$ by constraining the flux at a dislocation defect. This involves arguments similar to the ones in App.~\ref{sec:dislocationConstruction}. However, we do not use $\tilde{\OO}$ in the present work.

\subsection{Freedom in changing location of strip}

Consider the choice of $\lambda$ in Eq.~\eqref{eq:tyfullgaugeapp}. Apart from global $U(1)$ transformations, we can consider other choices of $\lambda_j$ which serve equally well as approximate symmetries. For example, we can consider 
\begin{align}
\label{eq:tyfullgaugeapp_delta}
    \tilde{T}'_{\bf{y}}=\begin{cases}
      \hat{T}_{\bf{y}}e^{i\sum_j -\frac{\pi m}{L_y} c_j^{\dagger}c_j}\qquad   &j_x< \overline{\OO}_x + \delta\\
      \hat{T}_{\bf{y}}e^{i\sum_j \frac{\pi m}{L_y} c_j^{\dagger}c_j}\qquad   &j_x> \overline{\OO}_x + \delta\\
      \hat{T}_{\bf{y}}\qquad &j_x=\overline{\OO}_x + \delta,
    \end{cases}
\end{align}
for some real number $\delta \ne 0$. The vector potential is again almost symmetric with respect to $\tilde{T}'_{\bf y}$; the strip where $\tilde{T}'_{\bf y}$ is locally not the symmetry of the Hamiltonian is however shifted from $\overline{\OO}$ by the amount $\delta$ in the $x$ direction. However, the expectation value of $\tilde{T}'_{\bf y}$ does not yield the desired invariant $\mathscr{P}_{\Oy}$. We have numerically checked that for any $\delta \neq 0$, the resulting values of $\mathscr{P}_{\OO,y}$ are not even quantized throughout each gapped phase of the Hofstadter model when $C \ne 0$. And when $C=0$, choosing $\delta\ne 0$ continues to give a quantized result, but certain choices of $\delta$ will give $\mathscr{P}_{\OO+(1/2,0),y}$ instead of $\mathscr{P}_{\Oy}$.  

These observations can be understood as follows. Suppose we calculate $\mathscr{P}_{\Oy}$ using $\tilde{T}_{\bf y}$ for a given choice of strip, and then change the strip position by $ \delta \hat{x}$ as in Eq.~\eqref{eq:tyfullgaugeapp_delta}. In general, if $\delta \ne 0$, $\lambda_j$ changes by $k_j \pi m/L_y$ for each $j$ that was crossed by the strip as it was shifted. $k_j = \pm 1$ if $j$ lies exactly on the initial or the final position of the strip, and $k_j= \pm 2$ if $j$ lies in between the initial and final positions of the strip. Let $S$ be the set of points crossed by the strip. Then
\begin{equation}
    \tilde{T}_{\bf y} \rightarrow \tilde{T}_{\bf y} e^{i \sum_{j \in S} \frac{k_j\pi m}{L_y} \hat{n}_j}.
\end{equation}
When $C=0$, we verified numerically that the state $\ket{\Psi}$ is an almost exact eigenstate of $\hat{n}_j$, with expectation value $\kappa$ (the statement being exact for the Hofstadter model). Thus we obtain
\begin{equation}
    \tilde{T}_{\bf y} \rightarrow \tilde{T}_{\bf y} e^{i \pi m \kappa  \sum_{j \in S}\frac{k_j}{L_y}}.
\end{equation}
Since the number of points $j$ corresponding to a fixed $k_j$ is always a multiple of $L_y$, the extra factor is of the form $(-1)^{k' m}$, where $k' = 0$ or 1, and is fixed by the actual values of $k_j$ and $\kappa$. If $k'=0$, $\mathscr{P}_{\Oy}$ is invariant. But if $k'=1$, $\mathscr{P}_{\Oy} \rightarrow \mathscr{P}_{\Oy} + 1/2$.

Note that this reasoning breaks down when $C \ne 0$: $\ket{\Psi(m)}$ is \textit{not} an eigenstate of $\sum_{j \in S} \hat{n}_j$, and $\mathscr{P}_{\Oy}$ hence can be expected to depend sensitively on the strip position. As mentioned in the main text, empirically we have found that the strip needs to be aligned with the `gauge origin' $\overline{\OO}$ in order to obtain consistent results for $\mathscr{P}_{\Oy}$.

\subsection{Partial translation}

\begin{figure}[t]
    \centering
    \includegraphics[width=9cm]{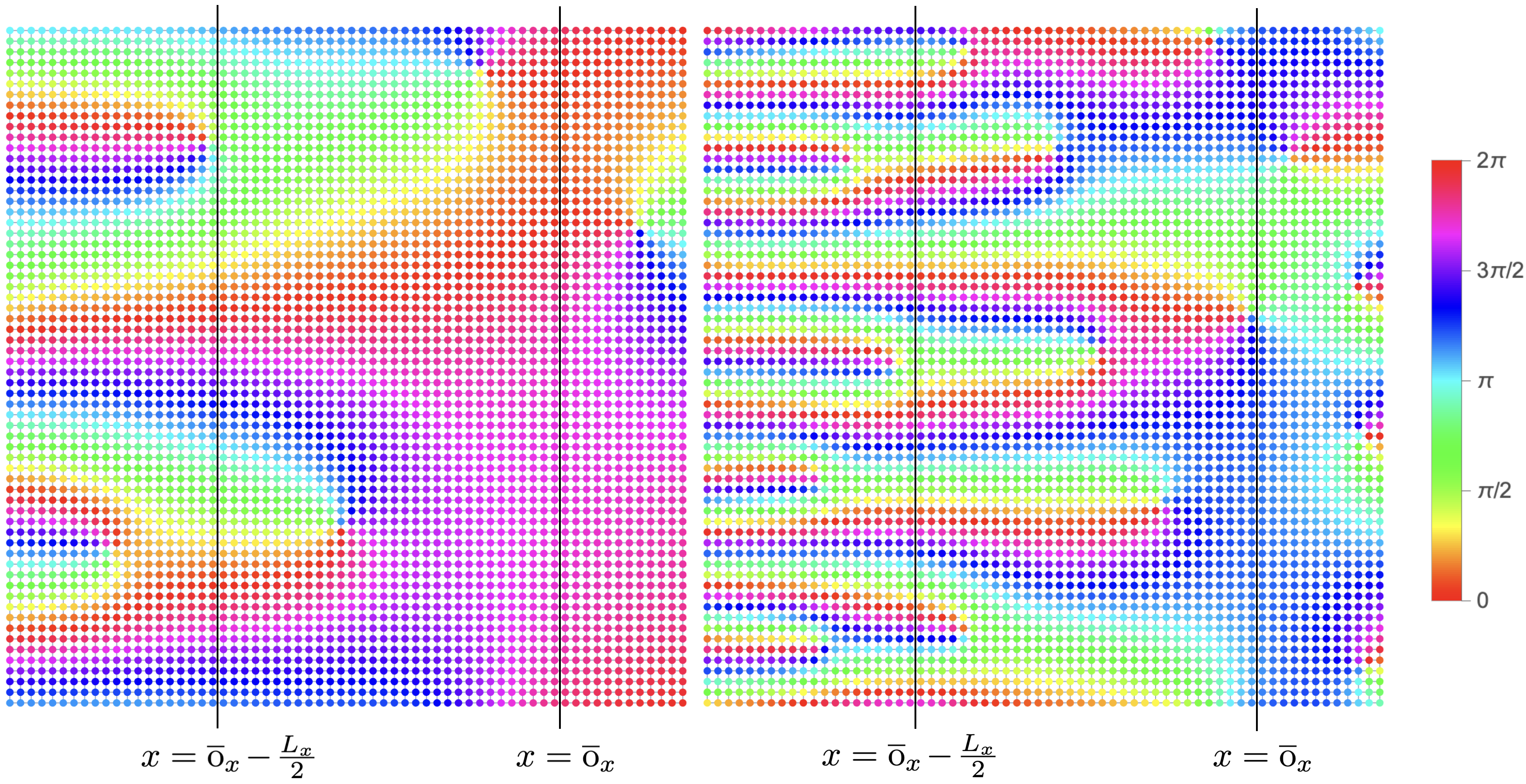}
    \caption{The $U(1)$ phase of a randomly chosen low-energy single particle eigenstate on a $64\times64$ torus. The gauge origin is at $\overline{\OO}=(51+\frac{1}{2},0)$. The parameters are \textbf{Left.} $m=5, C=1$ \textbf{Right.} $m=15, C=1$. The variation in phase along $y$ is lowest around $x = \overline{\OO}_x$.}
    \label{fig:phaseplot}
\end{figure}

In the partial translation calculation we defined the operator $\tilde{T}_{\textbf{y}}|_D$ and found that we need to pick a particular choice of $D$, so that $D$ is a cylinder defined mirror symmetrically about the cycle $x=\overline{\OO}_x-\frac{L_x}{2}$ (with holonomy $(-1)^m$). This set of choices recovers $\PO$ that is consistent with the dislocation charge calculation.
There is another natural choice of parameters: we can choose $D$ to be defined  around $x=\overline{\OO}_x$, and choose $\tilde{T}_{\textbf{y}}|_D=\hat{T}_{\textbf{y}}|_D$. We numerically found that this choice gives $\PO=0$ everywhere in the butterfly.
We do not have a complete explanation for why the first set of choices is required. However, we find a resemblance between this result and the calculation of angular momentum through partial rotations discussed previously in Ref. \cite{zhang2022fractional}.

In that calculation we need to choose a disk $D$ centered around a special point $o_2$, which is one of the two vertices invariant under the chosen rotation operator at a vertex $v$, $\hat{C}_{4,v}$. The holonomy of $A$ around either cycle which crosses $o_2$ is $e^{i \pi m}$ in a system with $m$ flux quanta. But if we choose a disk centered around the other fixed point $o_1$, with respect to which $A$ has trivial holonomy, the resulting value of the discrete shift varies irregularly within the same Hofstadter lobe and is not meaningful. Thus if we consider partial spatial symmetries, the calculation of both $\SO$ and $\PO$ involves the part of the system with $e^{i\pi m}$ holonomy rather than the part of the system with trivial holonomy.

We can provide some more heuristic numerical evidence explaining this choice. We consider a system at fixed $C$ and plot the phase of a randomly chosen low-energy single particle eigenstate for different choices of $m$ (see Fig.~\ref{fig:phaseplot}). We empirically find that the phase variation in the $y$ direction nearly vanishes around the cycle $x=\overline{\OO}_x$. Note that our choice of $D$ excludes the region around $x=\overline{\OO}_x$. The phase of $\bra{\Psi}\tilde{T}_{\textbf{y}}|_D\ket{\Psi}$ is nearly unaffected by this choice of $D$. However, choosing $D$ to exclude the region around $x=\overline{\OO}_x-\frac{L_x}{2}$ will significantly change $\bra{\Psi}\tilde{T}_{\textbf{y}}|_D\ket{\Psi}$. The above discussion suggests that the region around $x=\overline{\OO}_x-\frac{L_x}{2}$ is in a sense more important than the region around $x=\overline{\OO}_x$ in order to extract $\PO$.

\section{Derivation of topological response theory}\label{sec:field}
Here we discuss the construction of the topological response theory for invertible fermionic phases with the symmetry group $G_f = U(1)^f \times_{\phi} [\Z^2 \rtimes \Z_M]$. This construction was first developed for bosonic topological phases in \cite{manjunath2021cgt}, and a partial extension to invertible fermionic phases focussing on $\SO$ was discussed in Ref.~\cite{zhang2022fractional}. In this section, after reviewing the definition of crystalline gauge fields, we will present two computations that are new: a derivation of the complete response theory using results from Ref.~\cite{barkeshli2021invertible}, and a derivation of the quantization of $\PO$ from field theory.

\subsection{Review of crystalline gauge fields}

Consider a 3-manifold $\mathcal{M}^3$. We define real-valued differential 1-form gauge fields $\delta A, X, Y, \omega \in \Omega^1(\mathcal{M}^3,\mathbb{R})$. $\delta A$ denotes the part of the vector potential in excess of the uniform background, and we identify $\delta A \sim \delta A + 2\pi$. (Note that the quantity $A$ is taken to mean the entire vector potential.) $\vec{R} = (X,Y)$ represents the $\Z^2$ translation component. $\omega$ is a $\Z_M$ rotation gauge field. The full gauge field is denoted as $B = (\delta A, \vec{R},\omega )$. It is nonabelian and composes according to the group law of $G_f$. 

The flux of $B$ has three components $(F, \vec{T}, d\omega)$. $F := d\delta A + \phi A_{XY}$ is the density of magnetic flux. It consists of a background contribution $\phi A_{XY}$ proportional to the area element $A_{XY}$, and an excess contribution which is measured by $d \delta A$. We can equivalently write $F = dA$ where $A$ is the full vector potential.

$A_{XY}$ depends on $(X,Y)$, and its precise form also depends on the value of $M$, but when $\omega = 0$, $A_{XY} = X \cup Y$. 

$\vec{T}$ is the torsion; the quantity $\frac{1}{2\pi} \int_{D^2} \vec{T}$ measures the total dislocation Burgers vector within a 2d region $D^2$. When we use orthogonal coordinates, for example when $M=4$, we can write $\vec{T} := d\vec{R} + i \sigma_y \omega \wedge \vec{R}$ where $i\sigma_y = \begin{pmatrix} 0 & 1 \\ -1 & 0\end{pmatrix}$. The general definition of $\vec{T}$ depends on how we fix our coordinate basis. We note, however, that in the original simplicial formulation of the response theory, we can simply define $\vec{T} = d\vec{R}$ for any value of $M$, because the second term can be removed by a suitable gauge transformation. \cite{manjunath2021cgt}

$d\omega$ is the disclination density; the quantity $\frac{1}{2\pi} \int_{D^2} d\omega$ measures the total disclination angle within $D^2$. In order to correctly encode the curvature and torsion of $\mathcal{M}^3$, we set $\vec{R},\omega$ equal to the $SO(2)$ components of the coframe fields and the spin connection on $\mathcal{M}^3$.

There is a very important issue that was not commented upon in Refs.~\cite{manjunath2021cgt,zhang2022fractional}. Since the dislocation Burgers vector can in general be origin-dependent, the translation gauge fields must themselves be origin-dependent. This means that quantities in the action constructed out of the crystalline gauge fields may all in principle carry an origin-dependence. We return to this point in Appendix ~\ref{app:ftcomps}.

We impose the condition that $B$ is a $G_f$ gauge field through quantization conditions on the total flux of $B$ through closed and open 2d submanifolds of $\mathcal{M}^3$. If $D^2 \subset \mathcal{M}^3$ is a closed 2D submanifold, then the gauge field on $D^2$ must be flat, i.e.
\begin{align}
    \vec{T}(\vec{x}) &= 2\pi (1-U(\frac{2\pi}{M})) \sum_j \vec{l}_j \delta^{(2)}(\vec{x}-\vec{x}_j) \\
    \delta A(\vec{x}) &= 2\pi\sum_j m_j \delta^{(2)}(\vec{x}-\vec{x}_j) \\
    \omega(\vec{x}) &= 2\pi\sum_j n_j \delta^{(2)}(\vec{x}-\vec{x}_j)
\end{align}
within $D^2$, where $\vec{l}_j \in \Z^2$ and $m_j, n_j \in \Z$. This physically corresponds to saying that the only sources of flux in $D^2$ are points at the positions $\vec{x}_j$, which each carry an integer number of flux quanta. Note that the factor $(1-U(\frac{2\pi}{M}))$ appears in the condition for $\vec
T$, because a Burgers vector of the form $(1-U(\frac{2\pi}{M})) \vec{r}, \vec{r} \in \Z^2$, is in the trivial class (see App.~\ref{sec:bgd}) and therefore corresponds to a trivial quantum of flux. 

On open manifolds, we can have non-flat gauge field configurations, which can assign fractions of a flux quantum. The following conditions ensure that the net flux of the crystalline gauge fields in a region, although fractional, always corresponds to an element of $G_{\text{space}}$:
\begin{align}
    \vec{T}(\vec{x}) &= 2\pi \sum_j \vec{l}_j \delta^{(2)}(\vec{x}-\vec{x}_j) \\
    \omega(\vec{x}) &= \frac{2\pi}{M} \sum_j n_j \delta^{(2)}(\vec{x}-\vec{x}_j).
\end{align}
Even when we allow non-flat gauge field configurations on an open manifold, the integral through any closed $2$d submanifold $D^2$ must still be appropriately quantized:
\begin{align}
    \frac{1}{2\pi}\int_{D^2} \vec{T} &=  \sum_j \vec{l}_j = (1-U(\frac{2\pi}{M})) \vec{l}_{\text{tot}};
    \nonumber \\
    \frac{1}{2\pi}\int_{D^2} dA  &= \frac{1}{M}\sum_j m_j = m_{\text{tot}}
    \nonumber \\ 
    \frac{1}{2\pi}\int_{D^2} d\omega &= n_{\text{tot}}, 
\end{align}
where $m_{\text{tot}}, n_{\text{tot}} \in \Z$ and $\vec{l}_{\text{tot}} \in \Z^2$. Physically, $n_{\text{tot}}$ is the Euler characteristic of $D^2$.

\subsection{Response theory}

\subsubsection{General derivation}
The general construction of a response theory for invertible fermionic topological phases with symmetry group $G_f = U(1)^f\times_{\phi}[\Z^2 \rtimes \Z_M]$ was outlined in Ref.~\cite{zhang2022fractional}. In that work, specific quantization conditions on the discrete shift were derived by ignoring translations. Here we will recall the main steps in the derivation, but also consider translations, with a view to obtaining the quantization of $\PO$. 

First note that the quantity $[\omega_2] \in \H^2(G_b,\Z_2)$ specifies $G_f$ as a group extension of $G_b$ by $\Z_2^f$. To express $\omega_2$ in terms of the gauge fields, we define
\begin{equation}\label{eq:pb_om2}
    B_2^{(2)} := \frac{1}{2} (2F + k d\omega) = \pi B^* \omega_2,
\end{equation}
where $B^*$ refers to the pull-back via $B$ (meaning $B^* \omega_2(x) = \omega_2(B(x))$ for $x \in \mathcal{M}^3$)\footnote{In Ref.~\cite{zhang2022fractional} we set $\phi=0$ and wrote $ B_2^{(2)} := \frac{1}{2} (dA_b + k d\omega)$. $A_b$ is a gauge field for the `bosonic' $U(1)^f/\Z_2^f$ symmetry. But here $A$ always refers to the original $U(1)$ gauge field, which equals $A_b/2$. Thus $dA_b = 2 dA$. To account for $\phi \ne 0$, it is sufficient to replace $2dA$ with $2F$.}. As mentioned in Ref.~\cite{zhang2022fractional}, if the chosen $2\pi$ rotation operator acts as the identity, we should set $k=1$ in the above definition; if it acts as the fermion parity $(-1)^F$, we should set $k=0$. Accordingly, here we use $k=1$.

Note that the choice of $k$ in the definition of $\omega_2$ and hence $B_2^{(1)}$ specifies the angular momentum mod 1 of the fermion. In general, $\omega_2$ specifies the fractional $G_b$ quantum numbers of the fermion \cite{barkeshli2021invertible}.

Next, we define the quantity $B_2^{(1)} =\pi B^* n_2$, where $n_2 \in Z^2(G_b,\Z_2)$ is a general 2-cocycle which is also one piece of the data in the classification of Ref.~\cite{barkeshli2021invertible}. In our case, $n_2$ specifies the fermion parity change in the ground state after introducing a $G_b$ flux quantum. It turns out that the most general possible parametrization of $B_2^{(1)}$ is as follows:
\begin{align}
  B_2^{(1)} &:= \frac{1}{2} (2k_F F + k_{\omega} d\omega + (1-U(2\pi/M))^{-1}\vec{k}_T \cdot \vec{T} \nonumber \\
  & + k_A A_{XY}),
\end{align}
where $k_F,k_{\omega},k_A \in \Z$ and $\vec{k}_T \in \Z^2$. For $M=3,6$, we require $\vec{k}_T = (0,0)$. This parametrization of $n_2$ can be found, for example, in App.D of Ref.~\cite{manjunath2021cgt}. We will also use the fact that $n_2 \simeq n_2 + \omega_2$ (this is proven in Ref.~\cite{barkeshli2021invertible} and is an equivalence due to relabelling fermion parity defects in the invertible phase). We will use this equivalence to set $k_F = 0$.

The three remaining terms have the following interpretation: 1) $k_{\omega}$ is the fermion parity change upon inserting a set of disclinations with $\Omega = 2\pi$; 2) $\vec{k}_T \cdot \vec{\Lambda}$ specifies the fermion parity change in a region upon introducing dislocations with total Burgers vector $(1-U(2\pi/M)) \vec{\Lambda}$; 3) $k_A$ specifies the fermion parity per unit cell.

Then, the main result (obtained using the general theory developed in Ref.~\cite{barkeshli2021invertible}) is that the Lagrangian density $\mathcal{L}$ which gives the topological response theory for an invertible fermionic phase with symmetry $G_f$ and chiral central charge $c_-$ must satisfy
\begin{align}\label{eq:dL}
     d\mathcal{L} &= \frac{2}{\pi}\left( \frac{1}{2} B_2^{(1)} \wedge (B_2^{(1)} + B_2^{(2)}) + \frac{c_-}{8} B_2^{(2)} \wedge B_2^{(2)}\right) \nonumber \\ &\mod 2\pi.
\end{align}
Its meaning is the following. The 2-cocycles $n_2,\omega_2$ (alternatively, the quantities $B_2^{(1)},B_2^{(2)}$) together specify the $G_b$ quantum numbers of the fermions and the fermion parity defects in the system. But in order to ensure that the $G_b$ quantum numbers of all symmetry defects are well-defined in (2+1)D, we need an extra condition given by Eq.~\eqref{eq:dL}. Note that if the rhs of this equation corresponds to a nontrivial class in the group $\H^4(G_b,U(1))$, there is no solution for $L$, and the system can only live on the boundary of a (3+1)D SPT whose effective action is defined by the quantity $d\mathcal{L}$. 

For free fermion phases, we can set $C = c_-$; in general, $C = c_- \mod 8$. 

The response theory finally obtained from this condition, by substituting $B_2^{(1)}, B_2^{(2)}$ and integrating, takes the form
\begin{align}\label{eq:L}
  \mathcal{L} &=  \frac{C}{4\pi} A \wedge dA + \frac{\mathscr{S}_{\text{o}}}{2\pi} A \wedge d\omega + \frac{\vec{\mathscr{P}}_{\text{o}}}{2\pi} \cdot A \wedge \vec{T} + \frac{\kappa}{2\pi}A \wedge A_{XY} \nonumber \\
   &+ \frac{\tilde{\ell}_s}{4\pi} \omega \wedge d\omega + \frac{\vec{\mathscr{P}}_{s}}{2\pi} \cdot \omega \wedge \vec{T} +\frac{\nu_s}{2\pi} \omega \wedge A_{XY} + \dots
\end{align}
where the $\dots$ refer to additional terms that can be nonzero even if $\delta A, \omega$ are both zero. We note that because $d\mathcal{L}$ only involves $F$, $\mathcal{L}$ is written entirely in terms of $A$, even though one might expect that it should be written in terms of $\delta A$. The coefficients in this equation can all be expressed in terms of $k_{\omega}, k_T, k_A$, as we will now show for $\PO$.

\subsubsection{Quantization of $\PO$ using response theory}
The quantization of $\SO$ and its dependence on $c_-$ was discussed in Ref.~\cite{zhang2022fractional}. Below we only discuss the quantization of $\PO$. In the next section we discuss the origin dependence of $\SO$ and $\PO$. 

Note that the derivative of the term with $\PO$ in Eq.~\eqref{eq:L} can be written as $\frac{\PO}{2\pi} \cdot dA \wedge \vec{T} =  \frac{\PO}{2\pi} \cdot F \wedge \vec{T} + \dots$. 
When $k_F = 0$, such a term can only come from the $B_2^{(1)} \wedge B_2^{(2)}$ term in Eq.~\eqref{eq:dL}. Indeed, by comparing the two equations we must have
\begin{align}
    & \frac{\PO}{2\pi} F \wedge \vec{T} \nonumber \\
    &= \frac{1}{\pi} F \wedge \frac{1}{2}(1-U(2\pi/M))^{-1}\vec{k}_T \cdot \vec{T} \mod 2\pi.
\end{align}
The above equation implies that 
\begin{equation}
    \PO = (1-U(2\pi/M))^{-1}(\vec{k}_T + 2 \vec{k}_{T,SPT}) \mod \Z^2.
\end{equation}
The term with $\vec{k}_{T,SPT} \in \Z^2$ arises as a constant of integration. It does not contribute to $d\mathcal{L}$, and can physically be thought of as the polarization in a bosonic SPT phase which is stacked onto the original fermionic phase. We now use the conditions $\vec{k}_T,\vec{k}_{T,SPT} \in \Z^2$, along with the fact that integer choices of $\PO$ do not contribute to the partition function. With this we obtain the quantization result claimed in the main text. In particular, the classification of $\PO$ for invertible bosonic and fermionic systems is the same, in contrast to the results for the Hall conductance and $\SO$. 

There is no dependence of $\PO$ on $c_-$ because in Eq.~\eqref{eq:dL}, the terms proportional to $c_-$ only depend on $F$ and $d\omega$, and not on $\vec{T}$. The nontrivial values of $\PO$ are entirely due to $\vec{k}_T$ when $M=2,4$, and entirely due to $\vec{k}_{T,SPT}$ when $M=3$.

Finally, we note that the first four terms in Eq.~\eqref{eq:L} can be rewritten in a perhaps more familiar form, using the relations $\nu = \kappa + \frac{C \phi}{2\pi}$ and $F = d \delta A + \phi A_{XY}$:
\begin{align}
   \mathcal{L} &=  \frac{C}{4\pi} \delta A \wedge d\delta A + \frac{\mathscr{S}_{\text{o}}}{2\pi} \omega \wedge d\delta A + \frac{\vec{\mathscr{P}}_{\text{o}}}{2\pi} \cdot \vec{R} \wedge d\delta A \nonumber \\ &+ \frac{\nu}{2\pi}d\delta A \wedge A_{XY} + \dots
\end{align}
The advantage of this representation is that it makes the physical invariant $\nu$ explicit, and is also closer in spirit to the Chern-Simons response theory of the continuum quantum Hall effect, which only includes $\delta A$. However, it is a weaker form of the field theory, because it leaves out several terms that were explicit in Eq.~\eqref{eq:L}. For the derivations in the main text, we write the field theory in terms of $A$ and not $\delta A$. Remarkably, our empirical findings are easier to explain using Eq.~\eqref{eq:L} than with the above Lagrangian.

\section{Computations for Section \ref{app:ftrelabel}}\label{app:ftcomps}

Here we will use the simplicial formulation of the topological response theory \cite{manjunath2021cgt} to derive the quantities $\rho_M, \vec{\mu}_M, \vec{\tau}_M$ defined in Eqs.~\eqref{eq:k1},~\eqref{eq:k2},~\eqref{eq:k3}.
\subsubsection{Fractional gauge transformations}
The definition of each group element ${\bf g}_i = ({\bf r}_i,h_i) \in G_{\text{space}}$ implicitly depends on the choice of origin $\text{o}$. For example, ${\bf h} = ({\bf 0}, 1 \mod M)$ is the element in $G_{\text{space}}$ corresponding to a rotation by angle $2\pi/M$ (about $\OO$). Now suppose $\OO' = \OO + \vec{v}$, as before. The group element ${\bf g}_i'$ which implements the same transformation but with respect to the new origin $\OO'$ can be expressed as follows:
\begin{equation}\label{eq:fracGT}
    {\bf g}_i' = (\vec{v},0){\bf g} (-\vec{v},0),
\end{equation}
which corresponds to translating from $\text{o}'$ to $\text{o}$, applying ${\bf g}$, and translating back. Now we use the group multiplication law in $G_\text{space}$,
\begin{equation}
    ({\bf r}_1,h_1)({\bf r}_2,h_2) = ({\bf r}_1+U(2\pi h_1/M){\bf r}_2,h_1+h_2),
\end{equation}
to find that
\begin{equation}\label{eq:grouptr}
{\bf g}_i' = ({\bf r}_i - U(2\pi h_i/M)\vec{v}+\vec{v},h_i).
\end{equation}

\subsubsection{Change in $\mathscr{S}_{\text{o}}$}
First we study the transformation of $d\omega$. A cocycle representative for $d\omega$ can be explicitly written on a 2-simplex with the group elements ${\bf g}_1,{\bf g}_2,({\bf g}_1{\bf g}_2)^{-1}$:
\begin{equation}\label{eq:domDef}
    d\omega({\bf g}_1,{\bf g}_2) = \frac{[h_1]_M + [h_2]_M - [h_1 + h_2]_M}{M}
\end{equation}
where $[a]_M := a \mod M$. Since $d\omega$ does not depend on the translation group variables, it is unaffected by the transformation in Eq.~\eqref{eq:grouptr}. Therefore we have
\begin{equation}
    d\omega' = d\omega.
\end{equation}

Next we study the quantity $d\vec{R}$. An explicit cocycle representative is
\begin{align}
    d\vec{R}({\bf g}_1,{\bf g}_2) &= {\bf r}_1 + {\bf r}_2 - ({\bf r}_1 + U(2\pi h_1/M) {\bf r}_2) \nonumber \\
    &= (1-U(2\pi h_1/M)){\bf r}_2.
\end{align}
This implies that
\begin{align}
    d\vec{R}'({\bf g}_1,{\bf g}_2) &:= d\vec{R}({\bf g}'_1, {\bf g}'_2) \nonumber \\
    &= (1-U(2\pi h_1/M))({\bf r}_2 + (1-U(2\pi h_2/M))\vec{v}) \\
    &= d\vec{R} + {(1-U(2\pi h_1/M))(1-U(2\pi h_2/M))}\vec{v}.
\end{align}
The second term only depends on $h_1,h_2$, just like $d\omega$. In fact, it is coboundary equivalent to a multiple of $d\omega$. This can be seen by computing a cohomology invariant we denote $I_{\Omega}$, which for a general cocycle $f_2 \in Z^2(G_{\text{space}},\Z)$ is defined as
\begin{equation}
    I_{\Omega}[f_2] := \sum\limits_{j=0}^{M-1} f_2({\bf h}^j,{\bf h}) \mod M,
\end{equation}
where we recall ${\bf h} = ({\bf 0},1)$. We can show that $I_{\Omega}(d\omega) = 1$ using Eq.~\eqref{eq:domDef}, and also that $I_{\Omega}(d\vec{R}) = I_{\Omega}(A_{XY}) = 0$. Thus $I_{\Omega}$ is an invariant corresponding to the $\Z_M$ subgroup of $\H^2(G_{\text{space}},\Z)$. 

Up to coboundaries, we have by assumption
\begin{equation}
    d\vec{R}' = d\vec{R} + \vec{\tau}_M d\omega; 
\end{equation}
therefore, for any integer vector $\vec{u}$,
\begin{align}\label{eq:IOmega1}
     I_{\Omega}(\vec{u} \cdot d\vec{R}') &=  I_{\Omega}(\vec{u} \cdot d\vec{R}) +  I_{\Omega}(\vec{u} \cdot \vec{\tau}_M d\omega) \nonumber \\
     &= I_{\Omega}(\vec{u} \cdot d\vec{R}) + \vec{u} \cdot \vec{\tau}_M \mod M.
\end{align}
But we can compute
\begin{align}
     I_{\Omega}(\vec{u} \cdot d\vec{R}') &= I_{\Omega}(\vec{u} \cdot d\vec{R}) \nonumber \\ &+ \vec{u} \cdot \sum\limits_{j=0}^{M-1} (1-U(2\pi j/M))(1-U(2\pi/M))\vec{v} \nonumber \\
     &=I_{\Omega}(d\vec{R}) + \vec{u} \cdot M(1-U(2\pi/M))\vec{v} \mod M. \label{eq:IOmega2}
\end{align}
By comparing Eqs.~\eqref{eq:IOmega1},~\eqref{eq:IOmega2}, and using the fact that $\vec{u}$ can be arbitrary, we see that 
\begin{equation}\label{eq:tau}
    \vec{\tau}_M = M(1-U(2\pi/M))\vec{v}.
\end{equation}
The components of $\vec{\tau}_M$ are ambiguous up to multiples of $M$, since $I_{\Omega}$ is a $\Z_M$ invariant. This only means that $\mathscr{S}_{\text{o}'}$ is determined up to multiples of $M$, as we already know. 

Finally, we study the transformation of the area element $A_{XY}$. We know that $A_{XY}$ should be quadratic in the translation gauge fields if it corresponds to an area element. Consider the ansatz 
\begin{equation}
    A_{XY}({\bf g}_1,{\bf g}_2) = {\bf r}_1^T Q_M U(2\pi h_1/M){\bf r}_2
\end{equation}
where $Q_M$ is a $2\times 2$ matrix. The 2-cocycle condition for $A_{XY}$ is
\begin{align}
    & A_{XY}({\bf g}_1,{\bf g}_2) + A_{XY}({\bf g}_1{\bf g}_2,{\bf g}_3) \nonumber \\
    = & A_{XY}({\bf g}_2,{\bf g}_3) + A_{XY}({\bf g}_1,{\bf g}_2{\bf g}_3).
\end{align}
Upon substituting in the ansatz for $A_{XY}$, we find that
\begin{equation}
   {\bf r}_1^T (U(2\pi/M)^T Q_M U(2\pi/M)-Q_M) {\bf r}_2 = 0.
\end{equation}
This implies that
\begin{equation} \label{eq:uquq}
    U(2\pi/M)^T Q_M U(2\pi/M) = Q_M.
\end{equation}
Thus the particular form we chose for $A_{XY}$ ensures that $Q_M$ is a constant matrix with no dependence on group variables. We also demand that 
$$A_{XY}({\bf x},{\bf y})-A_{XY}({\bf y},{\bf x}) = 1,$$
which is a normalization condition. Solving these two conditions, we get the following possibilities for $Q_M$:
\begin{align}
    Q_{2} &= \begin{pmatrix}
        Q_{xx} & 1 \\ 0 & Q_{yy}    \end{pmatrix}, Q_{xx},Q_{yy} \in \Z \\
    Q_{4} &= \frac{1}{2}\begin{pmatrix}
        2Q_{0} & 1 \\ -1 & 2Q_{0}   \end{pmatrix} \quad Q_{0} \in \Z\\
        Q_{3,6} &= \begin{pmatrix}
        -1 & 1 \\ 0 & -1    \end{pmatrix}.
\end{align}
Note that for $M=2,4$, $Q_M$ is not uniquely fixed by Eq.~\eqref{eq:uquq}. This will be important in what follows. Upon shifting the origin, we obtain
\begin{equation}\label{eq:AprimeXY}
    A'_{XY} = ({\bf r}_1 - (U(h_1)-1)\vec{v})^T Q_M U(h_1)({\bf r}_1 - (U(h_2)-1)\vec{v}).
\end{equation}
We now write 
$$A_{XY}' = A_{XY} + \rho_M d\omega + \vec{\mu}_M \cdot d\vec{R}$$
for some $\rho_M, \vec{\mu}_M$, up to coboundaries. Evaluating the invariant $I_{\Omega}$, we find that
\begin{equation}
    I_{\Omega}(A'_{XY}) = \rho_M = -M ((U(2\pi/M)-1) \vec{v})^T Q_M U(2\pi/M)\vec{v}.
\end{equation}
After expanding out this expression for each $M$ using Table~\ref{tab:U_mats}, we find that
\begin{align}
    \rho_2 &= 4(v_x v_y + v_x^2 Q_{xx} + v_y^2 Q_{yy}) \mod 2 \\
    \rho_3 &= -3(v_x^2 + v_y^2 + v_x v_y) \mod 3 \\
    \rho_4 &= -2(v_x^2 + v_y^2)(1+2Q_0) \mod 4\\
    \rho_6 &= 0 \mod 6.
\end{align}
Note that the transformation rule depends on the choice of $Q_{xx},Q_{yy}$ for $M=2$ and on $Q_0$ for $M=4$. We find that numerical results are matched only if we set $Q_{xx}=Q_{yy}=1$ and $Q_0=0$. Upon doing so, we obtain
\begin{align}\label{eq:rho}
    \rho_2 &= 4(v_x v_y + v_x^2 + v_y^2 ) \mod 2 \nonumber \\
    \rho_3 &= -3(v_x^2 + v_y^2 + v_x v_y) \mod 3 \nonumber\\
    \rho_4 &= -2(v_x^2 + v_y^2) \mod 4 \nonumber\\
    \rho_6 &= 0 \mod 6.
\end{align}
Knowing $\vec{\tau}_M, \rho_M$ allows us to conclude that
\begin{equation}\label{eq:Sshiftfinal}
     \mathscr{S}_{\text{o}'} = \mathscr{S}_{\text{o}} + M \vec{\mathscr{P}}_{\text{o}} \cdot (1-U(2\pi/M))\vec{v} + \rho_M,
\end{equation}
which gives the complete expression for the transformation of $\mathscr{S}_{\OO}$. Thus the field theory does not uniquely fix the transformation rules for $M=2,4$; we needed to make some additional choices guided by numerics.

\subsubsection{Change in $\PO$}\label{app:change_PO}

Finally we find $\vec{\mu}_M$, which fixes the transformation of $\vec{\mathscr{P}}_{\text{o}}$. 
The calculation is given below. We discuss the different values of $M$ separately. We assume
\begin{equation}
    A_{XY}' = A_{XY} + \rho_M d\omega + \vec{\mu}_M \cdot d\vec{R}.
\end{equation}

\underline{M=2}: The invariants which characterize the cohomology class of $\vec{\mu}_2 \cdot d\vec{R}$ are $\mathcal{I}_X, \mathcal{I}_Y$, which are defined as
\begin{align}
    \mathcal{I}_X[f_2] &:= f_2({\bf xh},{\bf xh}) + f_2({\bf h},{\bf h}) \mod 2 \\
     \mathcal{I}_Y[f_2] &:= f_2({\bf yh},{\bf yh}) + f_2({\bf h},{\bf h}) \mod 2.
\end{align}
Here ${\bf x} = ((1,0),0),{\bf y} = ((0,1),0), {\bf h} = ((0,0),1)$.
We can check that $I_X(d\omega) = I_Y(d\omega) = 0 \mod 2$.
We can also show that
\begin{align}
    \mathcal{I}_X[\vec{\mu}_2 \cdot d\vec{R}]& = 2 \mu_{2,x} \mod 2 \\
    \mathcal{I}_Y[\vec{\mu}_2 \cdot d\vec{R}]& = 2 \mu_{2,y} \mod 2.
\end{align}
Therefore,
\begin{equation}
    I_{X(Y)}(A'_{XY}-A_{XY}) = 2\mu_{2,x(y)}.
\end{equation}
But when $M=2$, $(1-U(h_1))\vec{v} = 2 h_1 \vec{v}$ for any $\vec{v}$ (where $h_1$ is just an integer mod 2). Using this in Eq.~\eqref{eq:AprimeXY}, we obtain
$$A'_{XY}({\bf g}_1,{\bf g}_2) = ({\bf r}_1+2  \vec{v})^T Q_2 ({\bf r}_2 + 2 h_2 \vec{v}).$$
Then, by direct calculation we get 
\begin{align}
    \mathcal{I}_X[A'_{XY}-A_{XY}] &= (1+2v_x,2v_y) Q_2 (1+2v_x,2v_y)^T \nonumber \\ &- (2v_x,2v_y) Q_2  (2v_x,2v_y)^T - Q_{xx}\\
    &= 2 v_y \mod 2; \\
    \mathcal{I}_Y[A'_{XY}-A_{XY}] &= (2v_x,1+2v_y) Q_2 (2v_x,1+2v_y)^T \nonumber \\& - (2v_x,2v_y) Q_2 (2v_x,2v_y)^T- Q_{yy} \\
    &= 2 v_x \mod 2.
\end{align}
This means that $2\vec{\mu}_2 = 2(v_y,v_x) \mod \Z^2$. Putting this back into the transformation rule for $\PO$, we get the following result (mod 1):
\begin{align}
   2\mathscr{P}_{\OO',x} &:= 2\mathscr{P}_{\OO,x} + \kappa\mathcal{I}_X[A'_{XY}-A_{XY}] = 2\mathscr{P}_{\OO,x} + 2 \kappa v_y \\
   2\mathscr{P}_{\OO',y} &:= 2\mathscr{P}_{\OO,y} + \kappa\mathcal{I}_Y[A'_{XY}-A_{XY}]= 2\mathscr{P}_{\OO,y} + 2 \kappa v_x.
\end{align} 
Note that this result is independent of the choice for $Q_{xx},Q_{yy}$ made above, as these quantities cancel out of the final expression.

\underline{M=4}: We can directly see the result by setting $\PO = \frac{\overline{\mathscr{P}}_{\text{o}}}{2}(1,1)$ and $v_x = v_y = \frac{v_0}{2} \mod 1$ in the above calculation for $M=2$. The result is 
\begin{equation}
    \overline{\mathscr{P}}_{\text{o}'} = \overline{\mathscr{P}}_{\text{o}} +\kappa v_0 \mod 2.
\end{equation}
The results for $M=2,4$ are in fact all equivalent to Eq.~\eqref{eq:Pshift}.

\underline{M=3}: We wish to prove Eq.~\eqref{eq:Pshift}. If we parametrize $\vec{v} = \frac{v_0}{3}(1,1)$ and $\PO = \frac{\overline{\mathscr{P}}_{\text{o}}}{3}(1,2)$, this requires us to show that 
\begin{equation}\label{eq:Po'}
\overline{\mathscr{P}}_{\text{o}'} = \overline{\mathscr{P}}_{\text{o}} + 2 \kappa v_0 \mod 3.
\end{equation}
The invariant giving the cohomology class of $\vec{\mu}_3 \cdot d\vec{R}$ is $\mathcal{I}_X$, defined as
\begin{align}
    \mathcal{I}_X[f_2] &:= f_2({\bf xh},{\bf xh}) + f_2(({\bf xh})^2,{\bf xh}) \nonumber \\ &\quad - f_2({\bf h},{\bf h}) - f_2({\bf h}^2,{\bf h})  \mod 3.
\end{align}
By direct calculation, we can show that
\begin{align}
    \mathcal{I}_X[\vec{\mu}_3 \cdot d\vec{R}]& = 3 \mu_{3,x} = \overline{\mathscr{P}}_{\text{o}} \mod 3.
\end{align}
Again we use $A'_{XY}({\bf g}_1,{\bf g}_2) = ({\bf r}_1-(1-U(h_1)\vec{v}))^T Q_3 U(h_1)({\bf r}_2 -(1-U(h_2)\vec{v}))$. We use the same procedure discussed for $M=2$. After some (tedious) algebra, we indeed obtain Eq.~\eqref{eq:Po'}, or equivalently Eq.~\eqref{eq:Pshift}.

\subsection{Example: verifying the transformation rules for $\SO$}\label{app:SOExample}

\begin{figure}[t]
    \centering
    \includegraphics[width=7.5cm]{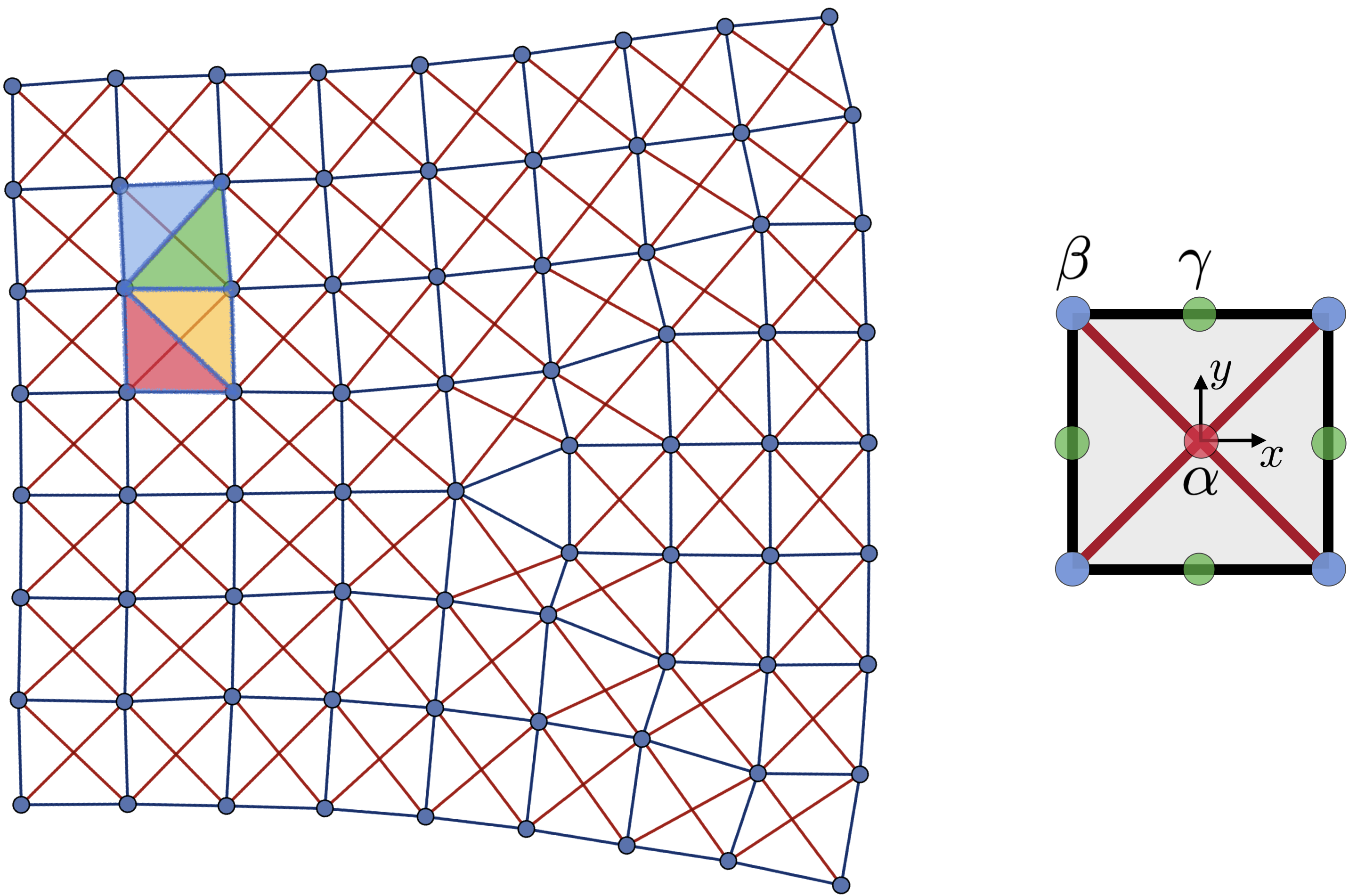}
    \caption{\textbf{Left.} $\vec{b}=(0,1)$ dislocation of a square lattice with next nearest neighbour hopping (red links). The $C_4$ symmetry requires all 4 colored triangles with different orientations having same flux $\phi/2$. \textbf{Right.} The choice of unit cell. There are \textit{no} sites at the MWP $\alpha$ and $\gamma$, and there is only one site at $\beta$. The next nearest neighbour hoppings cross each other.}
    \label{fig:nnlattice}
\end{figure}

\begin{figure*}[t]
    \centering
    \includegraphics[width=16cm]{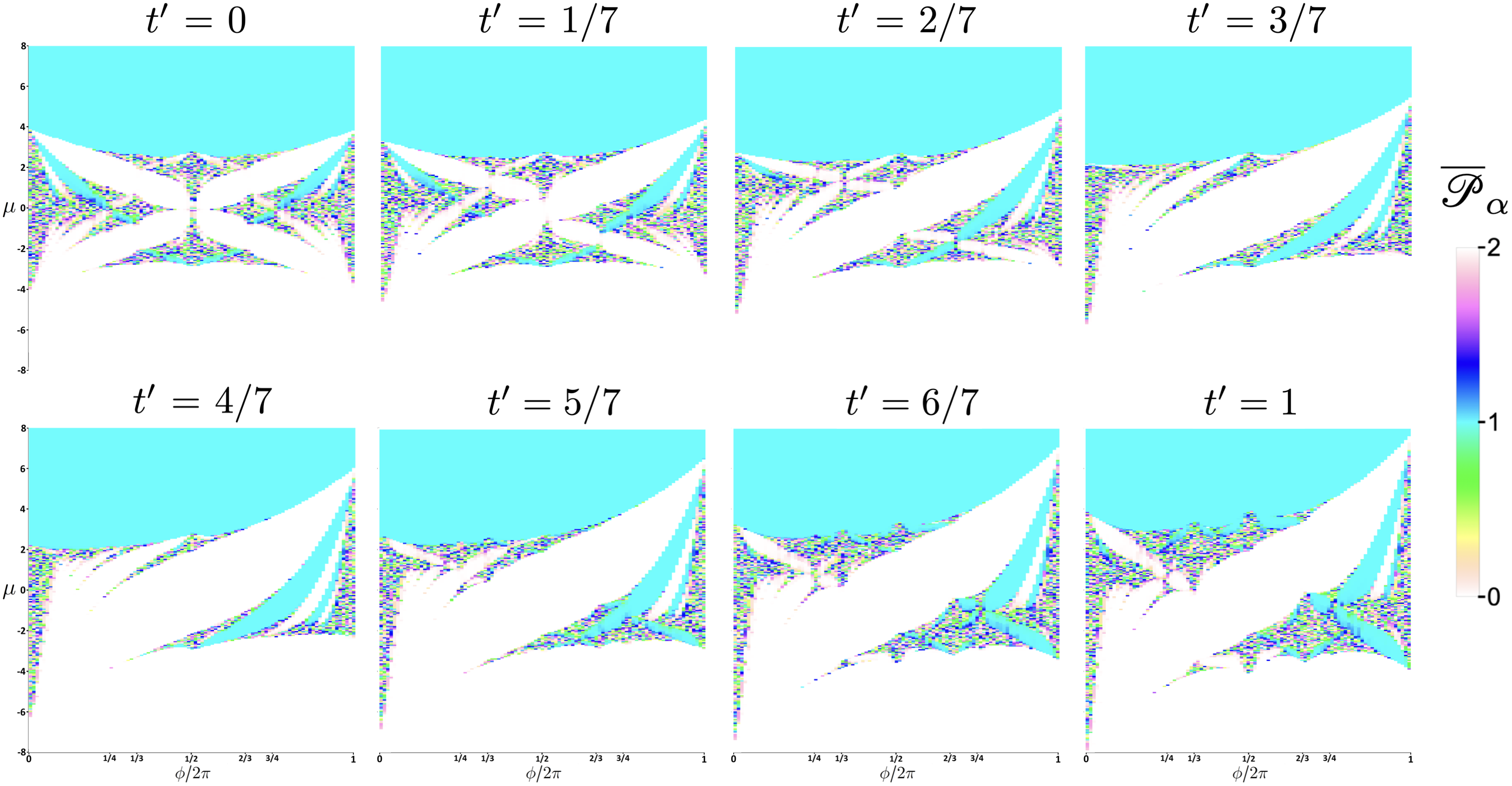}
    \caption{$\overline{\mathscr{P}}_{\alpha}$ as a function of the next neighbour hopping amplitude $t'$; the nearest neighbor hopping amplitude is $t=1$. The calculation is done on a $L_x\times L_y=41\times41$ open disk, with a $\vec{b}=(0,1)$ dislocation placed at the center. The defects are located at the center of the disk. The noisy features appear since the butterfly is numerically calculated on a finite size system rather than analytically derived from an empirical formula as in Fig.\ref{fig:SandP}.}
    \label{fig:nnbutterflies}
\end{figure*}

In this section we use the honeycomb lattice as an example to verify that the field theory formulas obtained above agree with numerical results.

Consider the honeycomb lattice with a disclination whose disclination angle is $\Omega=\frac{\pi}{3}$, as shown in Fig. \ref{fig:exampleLattice} \textbf{A}. We define the unit cell $\Theta$ so that the plaquette center is the center of the unit cell (i.e the point $\alpha$). Let us consider three choices of $\text{o}$, shown in Fig. \ref{fig:exampleLattice} \textbf{B,C,D}, which leads to different values of $\vec{b}_{\OO}$ and $n_{\text{irreg}}$, as summarized in Table~\ref{table:nuc}:

\begin{enumerate}
    \item Setting $\text{o}=\alpha$, we get a \textit{pure} disclination with $\vec{b}_{\OO}\in[(0,0)]$ and $n_{\text{irreg},\OO,\Omega}=\frac{2}{3}$ (Fig.~\ref{fig:exampleLattice} \textbf{B}).
    \item Setting $\text{o}=\beta_1$, we get an \textit{impure} disclination with $\vec{b}_{\OO}\in[(0,-1)]$ and $n_{\text{irreg},\OO,\vec{b}}=\frac{1}{3}$ (Fig.~\ref{fig:exampleLattice} \textbf{C}).
    \item Setting $\text{o}=\beta_2$, we obtain a different impure disclination with $\vec{b}_{\OO}\in[(-1,0)]$ and $n_{\text{irreg},\OO,\vec{b}}=\frac{1}{3}$ (Fig.~\ref{fig:exampleLattice} \textbf{D}).
\end{enumerate}

Without loss of generality we can assume the flux through the quadrilateral plaquette to be $\frac{2}{3}\phi$. Since $W$ is defined to align with the boundary of the unit cell, $W$ is the same for all three choices above. This means that $Q_W$ is also the same, but is described by three different equations:
\begin{equation}\label{eq:foureqs}
    \begin{aligned}
    Q_W&=\frac{\mathscr{S}_{\alpha}}{3}+(k+2/3)\nu\\
    Q_W&=\frac{\mathscr{S}_{\beta_1}}{3}+(k+1/3)\nu+\vec{\mathscr{P}}_{\beta_1}\cdot(0,-1)+\frac{C\phi}{6\pi}\\
    Q_W&=\frac{\mathscr{S}_{\beta_2}}{3}+(k+1/3)\nu+\vec{\mathscr{P}}_{\beta_2}\cdot(-1,0)+\frac{C\phi}{6\pi},\\
    \end{aligned}
\end{equation}
where $k$ is the integer part of $n_W$. 
We then equate the three expressions and cancel out the $\phi$ dependence to derive a relation between $\mathscr{S}_{\alpha}$ and $\mathscr{S}_{\beta}$. Equating the first two gives us
\begin{equation}\label{eq:SCSD}
\begin{aligned}
\mathscr{S}_{\alpha}+2\kappa&=\mathscr{S}_{\beta_1}+\kappa+\overline{\mathscr{P}}_{\beta(\frac{1}{3},\frac{1}{3})}(1,2)\cdot(0,-1) \mod 3\\
\mathscr{S}_{\alpha}&=\mathscr{S}_{\beta_1} +\overline{\mathscr{P}}_{\beta(\frac{1}{3},\frac{1}{3})}-\kappa\mod 3.
\end{aligned}
\end{equation}
where we have used $\kappa\equiv \nu -\frac{C\phi}{2\pi}$.
On the other hand, setting $\OO=\beta_1$, $\OO+(-\frac{1}{3},-\frac{1}{3})=\alpha$, the field theory (Eq.\eqref{eq:Sshiftfinal}) predicts:
\begin{equation}
\begin{aligned}
    \mathscr{S}_{\alpha}&=\mathscr{S}_{\beta_1}+3\vec{\mathscr{P}}_{\beta(\frac{1}{3},\frac{1}{3})}\cdot(1,0)-\kappa\mod 3\\
    \mathscr{S}_{\alpha}&=\mathscr{S}_{\beta_1}+\overline{\mathscr{P}}_{\beta(\frac{1}{3},\frac{1}{3})}-\kappa\mod 3,
\end{aligned}
\end{equation}
which is exactly the same equation. Eq. \eqref{eq:SCSD} can also be confirmed numerically (the numerical raw data is given in Fig.~\ref{fig:hexnum}). This verifies the field theory prediction Eq. \eqref{eq:Sshiftfinal}.

We remark that we can use the same honeycomb lattice Hamiltonian $H_{\text{clean}}$ to construct defects with different disclination angles. We can then apply similar procedures to verify the $M = 2,3,6$ transformation rules for $\mathscr{S}_{\OO+\vec{v}}$ and $\vec{\mathscr{P}}_{\OO+\vec{v}}$.

\section{Generalization to models with next nearest neighbour hopping}\label{app:nnnhopping}

In the main text, we argued that our method of extracting $\SO$ and $\PO$ works for an arbitrarily complicated unit cell with further neighbor hopping and interaction terms, as long as it is $C_M$ symmetric. In order to apply our dislocation or disclination charge calculation to measure $\SO,\PO$ in these generalized Hamiltonians, we need to construct a defect Hamiltonian $H_{\text{defect}}$, measure the charge $Q_W$ in a region $W$, and then specify the quantities $n_{\text{irreg},\OO}, \delta \Phi_{W,\OO}$. We argued that our numerical procedure allows us to specify these quantities as long as $H_{\text{clean}}$ is local and has a gapped, symmetry-preserving ground state, irrespective of its other properties.

We now provide numerical evidence that our procedure gives the expected quantized $\PO$ when we add next neighbor hopping (numerically it is much harder to verify this for interaction terms, and we do not pursue this here). Consider a Hofstadter model with the unit cell shown in Fig.\ref{fig:nnlattice}. The Hamiltonian with this unit cell has next neighbour hopping. We set the nearest neighbour hopping amplitude equal to $t=1$, and tune the next neighbour (diagonal) hopping amplitude $t'$ between 0 and 1. There is a flux $\phi$ through each unit cell. Since we require the unit cell to be $C_4$ symmetric, we also require that there is a flux $\frac{\phi}{2}$ through each triangle in the unit cell, as shown in Fig.\ref{fig:nnlattice}.

We extract the $\Z_2$ invariant $\overline{\mathscr{P}}_{\alpha}$ from a dislocation charge computation. The respective butterflies as a function of $t'$ are shown in fig.~\ref{fig:nnbutterflies}. We see that the invariant is indeed quantized and well-defined mod 2 throughout the main Hofstadter lobes. In the $t'=0$ limit, the butterfly reduces to the one for $\overline{\mathscr{P}}_{\alpha}$ in Fig.~\ref{fig:SandP}.

\bibliography{hofstadter_refs}
\end{document}